\documentclass[12pt]{article}

\usepackage{amsfonts,amssymb,bm,cite,amsmath}
\usepackage[dvips]{graphicx}
\usepackage[dvips]{psfrag}

\newcommand{\cL}{{\cal L}}
\newcommand{\cZ}{{\cal Z}}

\newcommand{\cg}[3]{C^{#1}_{#2\;#3}}

\newcommand {\beq}{\begin{equation}}
\newcommand {\eeq}{\end{equation}}
\newcommand {\beqa}{\begin{eqnarray}}
\newcommand {\eeqa}{\end{eqnarray}}
\newcommand {\n}{\nonumber \\}
\newcommand {\tr}{\mbox{tr}}
\newcommand {\Tr}{\mbox{Tr}}

\renewcommand{\theequation}{\thesection.\arabic{equation}}

\allowdisplaybreaks

\begin{document}
\setlength{\oddsidemargin}{0cm}
\setlength{\baselineskip}{7mm}

\begin{titlepage}
\renewcommand{\thefootnote}{\fnsymbol{footnote}}
\begin{normalsize}
\begin{flushright}
\begin{tabular}{l}
OU-HET 567\\
October 2006
\end{tabular}
\end{flushright}
  \end{normalsize}

~~\\

\vspace*{0cm}
    \begin{Large}
    \begin{bf}
       \begin{center}
         {Embedding of theories with ${\bf SU(2|4)}$ symmetry into \\
          the plane wave matrix model}
       \end{center}
    \end{bf}   
    \end{Large}
\vspace{1cm}

\begin{center}
           Goro I{\sc shiki}$^{1)}$\footnote
            {
e-mail address : 
ishiki@het.phys.sci.osaka-u.ac.jp},
Shinji S{\sc himasaki}$^{1)}$\footnote
            {
e-mail address : 
shinji@het.phys.sci.osaka-u.ac.jp},
           Yastoshi T{\sc akayama}$^{2)}$\footnote
            {
e-mail address : 
takayama@nbi.dk}
           {\sc and}
           Asato T{\sc suchiya}$^{1)}$\footnote
           {
e-mail address : tsuchiya@phys.sci.osaka-u.ac.jp}\\
      \vspace{1cm}
                    
       $^{1)}$ {\it Department of Physics, Graduate School of  
                     Science}\\
               {\it Osaka University, Toyonaka, Osaka 560-0043, Japan}\\
      \vspace{0.5cm}
       $^{2)}$ {\it The Niels Bohr Institute, Copenhagen University}\\
               {\it Blegdamsvej 17, DK-2100 Copenhagen \O , Denmark}
               
\end{center}

\vspace{1cm}

\begin{abstract}
\noindent
We study theories with $SU(2|4)$ symmetry, which include
the plane wave matrix model, $2+1$ SYM on $R \times S^2$ and ${\cal N}=4$
SYM on $R\times S^3/Z_k$. All these theories possess many vacua. From 
Lin-Maldacena's method which gives the gravity dual 
of each vacuum, it is  predicted that the theory around 
each vacuum of $2+1$ SYM on $R\times S^2$
and ${\cal N}=4$ SYM on $R\times S^3/Z_k$
is embedded in the plane wave matrix model. 
We show this directly on the gauge theory side.
We clearly reveal relationships among the spherical harmonics on $S^3$,
the monopole harmonics
and the harmonics on fuzzy spheres.
We extend the compactification (the T-duality) in matrix models a la Taylor
to that on spheres.
\end{abstract}
\vfill
\end{titlepage}
\vfil\eject

\setcounter{footnote}{0}

\tableofcontents

\section{Introduction}
\setcounter{equation}{0}
\renewcommand{\thefootnote}{\arabic{footnote}} 
The gauge/gravity (string) correspondence is
one of the most important concepts
in studying nonperturbative aspects of string theory and gauge theories.
An exhaustively investigated example
is the AdS/CFT correspondence \cite{Maldacena,GKP,Witten}.
Recently, Lin and Maldacena proposed the gauge/gravity correspondence
for theories with
$SU(2|4)$ symmetry \cite{Lin:2005nh}, which include on the gauge theory side
the plane wave matrix model (PWMM) \cite{Berenstein:2002jq},
$2+1$ super Yang Mills on $R\times S^2$ ($\mbox{SYM}_{R\times S^2}$)
\cite{Maldacena:2002rb} and ${\cal N}=4$ super Yang Mills on
$R\times S^3/Z_k$ ($\mbox{SYM}_{R\times S^3/Z_k}$).
These theories share the common feature that they have 
many vacua, a mass gap and a discrete energy spectrum.
Lin and Maldacena developed a unified method for providing the gravity
dual of each vacuum of these theories. This method is an extension of 
the so-called bubbling AdS geometry \cite{LLM}.

$\mbox{}$From Lin-Maldacena's method, it is predicted
that the theory around each vacuum of 
$\mbox{SYM}_{R\times S^2}$ and $\mbox{SYM}_{R\times S^3/Z_k}$
is embedded in PWMM.
In this paper, we prove this prediction for every vacuum of 
$\mbox{SYM}_{R\times S^2}$ and the trivial vacuum of 
$\mbox{SYM}_{R\times S^3/Z_k}$.
Our results do not only serve as a nontrivial check of the gauge/gravity
correspondence for the theories with $SU(2|4)$ symmetry, but they are
also interesting in the following aspects. 
First, we extend the compactification (the T-duality) in
matrix models a la Taylor \cite{Taylor:1996ik} to that on spheres.
We realize $S^3/Z_k$ as a $U(1)$ bundle on $S^2$ in matrices.
Second, we clearly reveal relationships among various spherical harmonics:
the spherical harmonics on $S^3$, 
the monopole harmonics developed by Wu, Yang and others 
\cite{Wu:1976ge,Kazama:1976fm,Wu:1977qk,Olsen:1990jm} and the harmonics
on a set of concentric fuzzy spheres with different radii 
\cite{Grosse:1995jt,Baez:1998he,Dasgupta:2002hx}. 
We give an alternative understanding and a generalization of
topologically nontrivial configurations and their topological charges
on fuzzy spheres
studied
in \cite{Aoki:2002fq,Ydri:2002nt,Balachandran:2003ay,Aoki:2003ye,Aoki:2004sd}.
Our results would shed light on problems of
describing curved space \cite{Hanada:2005vr} 
and topological invariants in matrix models \cite{BFSS,IKKT,DVV}.
In what follows, we review known facts on the gauge theory side and the gravity
side of the theories with $SU(2|4)$ symmetry as well as describe our strategy
and the organization of this paper.

In \cite{Lin:2005nh}, PWMM, $\mbox{SYM}_{R\times S^2}$ and 
$\mbox{SYM}_{R\times S^3/Z_k}$ were defined by truncations of 
${\cal N}=4$ SYM on $R\times S^3$ ($\mbox{SYM}_{R\times S^3}$) as follows.
$\mbox{SYM}_{R\times S^3}$ has the superconformal symmetry
$SU(2,2|4)$, whose bosonic subgroup is $SO(2,4)\times SO(6)$, where
$SO(2,4)$ is the conformal group in 4 dimensions and $SO(6)$ is the R-symmetry.
$SO(2,4)$ has a subgroup $SO(4)$ that is the isometry of the $S^3$ on 
which the theory is defined. $SO(4)$ is identified with 
$SU(2)\times \tilde{SU}(2)$, where we have marked one of two $SU(2)$'s with a
tilde to focus on it. The above theories are obtained by dividing the 
original $\mbox{SYM}_{R\times S^3}$ by subgroups of $\tilde{SU}(2)$.
Dividing it by full $\tilde{SU}(2)$ gives rise to PWMM. Indeed this fact
was first found in \cite{Kim:2003rz}.\footnote{We make a remark on a relation
of PWMM with a supersymmetric quantum mechanics
that is given by the dimensional reduction of
10D ${\cal N}=1$ SYM to $1+0$ dimensions. General 
mass deformation of this quantum mechanics which preserves all supersymmetries
was studied in \cite{Bonelli:2002mb}, and it was recently shown 
in \cite{Kim:2006wg}
that the deformation is unique and gives PWMM.}
Dividing $\mbox{SYM}_{R\times S^3}$
by $Z_k$ gives rise to $\mbox{SYM}_{R\times S^3/Z_k}$.
In a coordinate system of $S^3$ defined in appendix A, 
this corresponds to making an identification 
$(\theta,\phi,\psi)\sim (\theta,\phi, \psi+\frac{4\pi}{k})$.
The $k \rightarrow\infty$ limit of $\mbox{SYM}_{R\times S^3/Z_k}$ is 
nothing but $\mbox{SYM}_{R\times S^2}$. That is, $\mbox{SYM}_{R\times S^2}$
is obtained by dividing $\mbox{SYM}_{R\times S^3}$ by $U(1)$, in
other words, 
by dimensionally reducing $\mbox{SYM}_{R\times S^3}$ or 
$\mbox{SYM}_{R\times S^3/Z_k}$
in the $\psi$ direction.
In \cite{Maldacena:2002rb}, the trivial vacuum of
$\mbox{SYM}_{R\times S^2}$ was obtained by removing fuzziness of fuzzy spheres
in a vacuum of PWMM. By viewing this procedure inversely, one finds that
PWMM is obtained as a dimensional reduction of $\mbox{SYM}_{R\times S^2}$.
It can be said that we achieve `inverse' of these dimensional 
reductions in this paper, keeping the philosophy of \cite{EK}
in mind: we obtain $\mbox{SYM}_{R\times S^3/Z_k}$ from 
$\mbox{SYM}_{R\times S^2}$
and $\mbox{SYM}_{R\times S^2}$ from PWMM.
In section 2.1, we review these dimensional reductions. 

The vacua of PWMM are characterized by configuration of concentric membrane
fuzzy spheres \cite{Berenstein:2002jq}. The vacua of $\mbox{SYM}_{R\times S^2}$
are labeled by monopole charges and unbroken gauge group
\cite{Maldacena:2002rb,Lin:2005nh}.
The vacua of $\mbox{SYM}_{R\times S^3/Z_k}$ are parameterized by the
holonomy along nontrivial generator of $\pi_1(S^3/Z_k)$ \cite{Lin:2005nh}.
In section 2.2, we review these facts, and we clarify
correspondence between the holonomy parameterizing the vacua of 
$\mbox{SYM}_{R\times S^3/Z_k}$ with $k\rightarrow\infty$ and the monopole
charges and the unbroken gauge group
labeling the vacua of $\mbox{SYM}_{R\times S^2}$.

On the gravity side, Lin and Maldacena reduced the problem
of finding a supergravity solution dual to each vacuum of
the above theories to the problem of finding
an axially symmetric solution 
to the 3-dimensional Laplace equation for the electrostatic potential, 
where the boundary condition involves
charged conducting disks and a background potential.
Each theory is specified by a background potential and each vacuum
is specified by a configuration of charged conducting disks.
In section 3.1, we review Lin-Maldacena's method and 
the one-to-one correspondences between
the configurations of charged conducting disks and the vacua.
In particular, by using the correspondence described in section 2.2,
we clarify the one-to-one correspondence between
the configurations of charged conducting disks and 
the monopole charges and the unbroken gauge group labeling the vacua
of $\mbox{SYM}_{R\times S^2}$.

In section 3.2, from the one-to-one correspondences between 
the configurations of
charged conducting disks and the vacua, we obtain the following two predictions
about relations between the vacua of different gauge theories: 
if the gauge/gravity
correspondence for the theories with $SU(2|4)$ symmetry is valid,
1) the theory around each vacuum of $\mbox{SYM}_{R\times S^2}$ is embedded 
in PWMM and 
2) the theory around each vacuum of $\mbox{SYM}_{R\times S^3/Z_k}$ 
is embedded in $\mbox{SYM}_{R\times S^2}$. More precisely, 
1) the theory around each vacuum
of $\mbox{SYM}_{R\times S^2}$ is equivalent to the theory around a certain
vacuum of PWMM and 2) the theory around each vacuum 
of $\mbox{SYM}_{R\times S^3/Z_k}$
is equivalent to the theory around a certain 
vacuum of $\mbox{SYM}_{R\times S^2}$ with a periodicity imposed.
In \cite{Maldacena:2002rb}, the prediction 1) for the trivial vacuum of 
$\mbox{SYM}_{R\times S^2}$ was already shown as mentioned above, and
its consistency with the gravity duals was recently shown 
in \cite{Ling:2006up}.
The prediction 1) for some nontrivial vacua of 
$\mbox{SYM}_{R\times S^2}$ was also
suggested in \cite{Maldacena:2002rb,Lin}. 
We give a complete proof of the prediction 1) for generic nontrivial vacua
of $\mbox{SYM}_{R\times S^2}$ in this paper.
Combining the predictions 1) and 2) leads to a remarkable statement
that the theory around every vacuum of $\mbox{SYM}_{R\times S^3/Z_k}$ and
$\mbox{SYM}_{R\times S^2}$ is embedded in PWMM.

In order to prove the predictions, we make harmonic expansions for the theories
around various vacua. We use the spherical harmonics on $S^3$, the
monopole harmonics on $S^2$ and the harmonics on a set of fuzzy spheres with
different radii, which we call the fuzzy sphere harmonics.
In section 4, as a preparation for the proofs, we describe properties of 
these harmonics.
In section 4.1, we recall the properties of the spherical harmonics
on $S^3$ summarized in \cite{ITT} and add some new results.
In section 4.2, we generalize the results on the monopole harmonics
in \cite{Wu:1976ge,Kazama:1976fm,Wu:1977qk,Olsen:1990jm} and 
reveal relationship between
the monopole harmonics and the spherical harmonics on $S^3$.
In section 4.3, we study the fuzzy sphere harmonics, which is an appropriate
basis for the vector space of rectangular matrices 
\cite{Grosse:1995jt,Baez:1998he,Dasgupta:2002hx}.
We further develop the works 
\cite{Grosse:1995jt,Baez:1998he,Dasgupta:2002hx}: we consider general
spin $S$ fuzzy sphere harmonics and derive various formula about them,
and furthermore we clearly reveal their relationship with
the monopole harmonics.
It is well known \cite{Hoppe,deWit:1988ig,Hoppe:1988gk} 
that a basis for the vector space of square matrices is
the harmonics on a fuzzy sphere and is regarded as a regularization
of the ordinary spherical harmonics on $S^2$,
where the size of matrices plays a role
of an ultraviolet cut-off for the angular momentum.
Analogously, a basis for the vector space of rectangular matrices is
the fuzzy sphere harmonics 
and is regarded
as a regularization of the monopole harmonics, where the size of matrices
plays a role of an ultraviolet cut-off while
a half of the difference between
the numbers of raws and columns is fixed and 
identified with the monopole charge.

By using the results in sections 4.2 and 4.3, 
we prove the prediction 1) in section 5.1.
In section 5.2, we comment on a relation of our result in section 5.1 with
the works \cite{Aoki:2003ye,Aoki:2004sd}.
In section 6.1, by using the results in sections 4.1 and 4.2 and
the mode expansion around the trivial vacuum of $\mbox{SYM}_{R\times S^3/Z_k}$
performed in \cite{ITT},
we prove the prediction 2) for the trivial vacuum of
$\mbox{SYM}_{R\times S^3/Z_k}$. Following the suggestion given by
the gravity side, we consider a configuration of matrices in 
$\mbox{SYM}_{R\times S^2}$ with a periodicity and recover the 
$\psi$ direction by `T-duality'.
This is an extension of the compactification (the T-duality) in matrix models
a la Taylor to that on spheres, where $S^3/Z_k$ is realized as a nontrivial
$S^1$ fibration over $S^2$ in matrices rather than a direct product.
In section 6.2, we combine the predictions 1) and 2) and make some comments
on construction of $S^3$ in terms of three matrices.

Section 7 is devoted to summary and discussion. Some details are gathered
in appendices.

\section{Theories with $SU(2|4)$ symmetry}
\setcounter{equation}{0}
In this section, we review the gauge theory side of the theories with
$SU(2|4)$ symmetry with some new insights. 
In section 2.1, starting with $\mbox{SYM}_{R\times S^3}$ or 
$\mbox{SYM}_{R\times S^3/Z_k}$, we first obtain
$\mbox{SYM}_{R\times S^2}$ by a dimensional reduction.
After rewriting it using a 3-dimensional notation, we again make 
a dimensional reduction for it to obtain PWMM. We fix our notation in the
above process. In section 2.2, we classify vacua of the theories with
$SU(2|4)$ symmetry. In particular, we clarify correspondence between the  
vacua of $\mbox{SYM}_{R\times S^2}$ and the vacua of
$\mbox{SYM}_{R\times S^3/Z_k}$ with the $k\rightarrow\infty$ limit.

\subsection{Dimensional reductions from ${\cal N}=4$ SYM on $R\times S^3$}
We start with $\mbox{SYM}_{R\times S^3}$ \cite{Breitenlohner:1982jf,
Nicolai:1988ek,Bergshoeff:1988jx,Okuyama:2002zn}.
Here the gauge group is $U(N)$ and the radius of $S^3$ is fixed 
to $\frac{2}{\mu}$.
Borrowing the ten-dimensional notation, we can write down the action 
as follows:
\beqa
&&S_{R\times S^3}=\frac{1}{g_{R\times S^3}^2}
\int dt\frac{d\Omega_3}{(\mu/2)^3} 
\mbox{Tr}\left(
-\frac{1}{4}F_{ab}F^{ab}-\frac{1}{2}D_aX_mD^aX_m-\frac{1}{12}\hat{R}X_m^2 
\right.\n
&&\hspace{5cm}\;\;\left.-\frac{i}{2}\bar{\lambda}\Gamma^aD_a\lambda
-\frac{1}{2}\bar{\lambda}\Gamma^m[X_m,\lambda]+\frac{1}{4}[X_m,X_n]^2\right),
\label{action of N=4 SYM on R times S^3}
\eeqa
where $a$ and $b$ are the (3+1)-dimensional local Lorentz indices 
and run from $0$ to $3$,
and $m$ runs from $4$ to $9$.
$\Gamma^a$ and $\Gamma^m$ are the 10-dimensional gamma matrices, which satisfy
\beqa
\{\Gamma^a,\Gamma^b\}=2\eta^{ab},\;\;\;\;\; \{\Gamma^m,\Gamma^n\}=2\delta^{mn},
\eeqa 
where $\eta^{ab}=\mbox{diag}(-1,1,1,1)$.
$\lambda$ is the Majorana-Weyl spinor in 10 dimensions, which satisfies
\beqa
C_{10}\bar{\lambda}^T=\lambda,\;\;\; \Gamma^{11}\lambda=\lambda,
\eeqa
where $C_{10}$ is the charge conjugation matrix.
$\hat{R}$ is the scalar curvature of $S^3$ which is equal to 
$\frac{3\mu^2}{2}$.
The field strength and the covariant derivatives take the form
\beqa
&&F_{ab}=\nabla_aA_b-\nabla_bA_a-i[A_a,A_b], \n
&&D_aX_m=\nabla_aX_m-i[A_a,X_m], \;\;\;
D_a\lambda=\nabla_a\lambda-i[A_a,\lambda],
\eeqa
where 
\beqa
\nabla_aA_b=e^{\mu}_a(\partial_{\mu}A_b+\omega_{\mu b}^{\;\;\;\;\;c}A_c),\;\;\;
\nabla_aX_m=e^{\mu}_a\partial_{\mu}X_m, \;\;\;
\nabla_a\lambda=e^{\mu}_a(\partial_{\mu}\lambda+\frac{1}{4}\omega_{\mu}^{bc}
\Gamma_{bc}\lambda).
\eeqa
In appendix A, we list the metric, the vierbeins and the spin connections 
for $R\times S^3$ used in this paper. In this metric, 
\beqa
\int d\Omega_3
=\frac{1}{8}\int^{\pi}_0d\theta\int^{2\pi}_0d\phi\int^{4\pi}_0d\psi\sin\theta,
\eeqa
so that $\int d\Omega_3 1=2\pi^2$.

$\mbox{SYM}_{R\times S^3/Z_k}$ is obtained by identifying the value
at $(\theta,\phi,\psi)$ with that at $(\theta,\phi,\psi+\frac{4\pi}{k})$
for all the fields in $\mbox{SYM}_{R\times S^3}$. 
The relation between the coupling constant of $\mbox{SYM}_{R\times S^3/Z_k}$ 
and that of $\mbox{SYM}_{R\times S^3}$ is given by
\beqa
g_{R\times S^3}^2=kg_{R\times S^3/Z_k}^2.
\eeqa
The $k\rightarrow\infty$ limit of this procedure can be regarded as
a dimensional reduction. This dimensional reduction with 
a redefinition
of the gauge fields gives rise to $\mbox{SYM}_{R\times S^2}$.

In order to obtain $\mbox{SYM}_{R\times S^2}$,
we make following replacements:
\beqa
A=A_0dt+A_{\theta}d\theta+A_{\phi}d\phi+A_{\psi}d\psi \:\rightarrow
A_0dt+A_{\theta}d\theta+(A_{\phi}+\frac{1}{\mu}\cos\theta \Phi)d\phi
+\frac{1}{\mu}\Phi d\psi,
\label{replacement}
\eeqa
We also assume that all the fields are independent of $\psi$.
Then, using the metric, the dreibeins and the spin connections 
for $R\times S^2$
listed in appendix A, it is easy to see that
(\ref{action of N=4 SYM on R times S^3}) is reduced 
to an action on $R\times S^2$.
For instance, the space components of
the gauge field strength are reduced to quantities on $R\times S^2$ as
\beqa
F_{12}\:\rightarrow\: F_{12}-\mu\Phi,\;\;\;
F_{13}\:\rightarrow\: D_1\Phi, \;\;\;
F_{23}\:\rightarrow\: D_2\Phi.
\label{reduction of field strength}
\eeqa
The final result is
\beqa
&&S_{R\times S^2}
=\frac{1}{g_{R\times S^2}^2}\int dt\frac{d\Omega_2}{\mu^2} \mbox{Tr}\left(
-\frac{1}{4}F_{a'b'}F^{a'b'}-\frac{1}{2}D_{a'}\Phi D^{a'}\Phi
-\frac{\mu^2}{2}\Phi^2+\mu F_{12}\Phi \right.\n
&&\hspace{3.5cm}-\frac{1}{2}D_{a'}X_mD^{a'}X_m-\frac{\mu^2}{8}X_m^2 
+\frac{1}{4}[X_{m},X_{n}]^2+\frac{1}{2}[\Phi,X_{m}]^2 \n
&&\hspace{3.5cm}\left.-\frac{i}{2}\bar{\lambda}\Gamma^{a'}D_{a'}\lambda
+\frac{i\mu}{8}\bar{\lambda}\Gamma^{123}\lambda
-\frac{1}{2}\bar{\lambda}\Gamma^3[\Phi,\lambda]
-\frac{1}{2}\bar{\lambda}\Gamma^m[X_m,\lambda]\right),
\label{action of 2+1 SYM on R times S^2}
\eeqa
where $a'$ and $b'$ are the $(2+1)$-dimensional local Lorentz indices and 
run from 0 to 2.
The radius of $S^2$ is fixed to $\frac{1}{\mu}$ and 
\beqa
\int d\Omega_2=\int^{\pi}_0d\theta\int^{2\pi}_0d\phi\sin\theta,
\eeqa
so that $\int d\Omega_2 1=4\pi$. 
When $\mbox{SYM}_{R\times S^2}$ is identified with the 
$k\rightarrow\infty$ limit of 
$\mbox{SYM}_{R\times S^3/Z_k}$, the coupling constant $g_{R\times S^2}$
is expressed as
\beqa
g_{R\times S^2}^2
=\lim_{k\rightarrow\infty}\frac{k\mu g_{R\times S^3/Z_k}^2}{4\pi},
\label{coupling relation}
\eeqa
so that $kg_{R\times S^3/Z_k}^2$ must be fixed in the $k\rightarrow\infty$ 
limit. This relation will be used in comparison with the gravity duals in
section 3.1.
(\ref{action of 2+1 SYM on R times S^2}) is 
$\mbox{SYM}_{R\times S^2}$ obtained in \cite{Maldacena:2002rb}.

For later convenience, we rewrite (\ref{action of 2+1 SYM on R times S^2})
using the 3-dimensional flat space notation, which is represented by
the orthogonal coordinates system $(x_1,x_2,x_3)$ or 
the polar coordinates system
$(r,\theta,\phi)$.
We introduce the flat space nabla
\beqa
\vec{\partial}
=\vec{e}_i\partial_i
=\vec{e}_r\partial_r +\vec{e}_{\theta}\frac{1}{r}\partial_{\theta}
             +\vec{e}_{\phi}\frac{1}{r\sin\theta}\partial_{\phi},
\eeqa
where $\vec{e}_i\;\;(i=1,2,3)$ are the unit vectors of $x_i$ directions,
and $\vec{e}_r$, $\vec{e}_{\theta}$ and $\vec{e}_{\phi}$ are the unit vectors
of the $r$, $\theta$ and $\phi$ directions, respectively.
In the followings, the $r$-derivative in $\vec{\partial}$ does not contribute 
and $r$ in $\vec{\partial}$ is fixed to $\frac{1}{\mu}$.
We construct a 3-dimensional vector from $A_{\theta}$ and $A_{\phi}$ as 
\beqa
\vec{A}=\mu A_{\theta}\vec{e}_{\theta}
        +\frac{\mu}{\sin\theta}A_{\phi}\vec{e}_{\phi},
\eeqa
and define a vector,
\beqa
\vec{\Gamma}=\Gamma^i\vec{e}_i.
\eeqa
We make a unitary transformation for the fermion,
\beqa
\lambda \: \rightarrow \: 
e^{\frac{\pi}{4}\Gamma_{12}}e^{\frac{\theta}{2}\Gamma_{31}}
e^{\frac{\phi}{2}\Gamma_{12}}\lambda.
\eeqa
Then, it is easy to see the transformation of the following two terms:
\beqa
&&\Tr\left(-\frac{i}{2}\bar{\lambda}\Gamma^{a'}D_{a'}\lambda\right) 
\: \rightarrow \:
\Tr\left(-\frac{i}{2}\bar{\lambda}\Gamma^{0}D_{0}\lambda 
-\frac{i}{2}\bar{\lambda}\vec{\Gamma}\cdot(\vec{e}_r\times\vec{D})\lambda
-\frac{i\mu}{2}\bar{\lambda}\Gamma^{123}\lambda\right), 
\label{fermion transformation 1}\\
&&\Tr\left(-\frac{1}{2}\bar{\lambda}\Gamma^3[\Phi,\lambda]\right)
\: \rightarrow \:
\Tr\left(-\frac{1}{2}\bar{\lambda}\vec{\Gamma}\cdot\vec{e}_r[\Phi,\lambda]
\right).
\label{fermion transformation 2}
\eeqa
where $\vec{D}=\vec{\partial}-i[\vec{A},\;]$.
The other terms including the fermion are unchanged.
Note that the last term on the righthand side of 
(\ref{fermion transformation 1})
shifts the coefficient of the fermion mass term.
In order to rewrite the bosonic part, we define the following quantities:
\beqa
&&\vec{Y}=\vec{e}_r\Phi+\vec{e}_r\times\vec{A}, \n
&&\vec{L}^{(0)}=-i\mu^{-1}\vec{e}_r\times\vec{\partial}, \n
&&\vec{{\cal Z}}=\mu\vec{Y}+i(\mu\vec{L}^{(0)}
                 \times\vec{Y}-\vec{Y}\times\vec{Y}), \n
&&\vec{{\cal L}}=\mu\vec{L}^{(0)}-[\vec{Y},\;].
\label{vecY and so on}
\eeqa
$\vec{{\cal Z}}$ is evaluated as 
\beqa
\vec{{\cal Z}}=(-\mu\Phi+F_{12})\vec{e}_r+D_1\Phi\vec{e}_{\theta}
               +D_2\Phi\vec{e}_{\phi}.
\label{cal Z}
\eeqa
\newpage
Finally, we obtain
\beqa
&&S_{R\times S^2}
=\frac{1}{g_{R\times S^2}^2}\int dt\frac{d\Omega_2}{\mu^2} \mbox{Tr}\left(
\frac{1}{2}(D_0\vec{Y}-i\mu\vec{L}^{(0)}A_0)^2-\frac{1}{2}\vec{{\cal Z}}^2
+\frac{1}{2}(D_0X_m)^2+\frac{1}{2}(\vec{{\cal L}}X_m)^2-\frac{\mu^2}{8}X_m^2
\right. \n
&&\hspace{2.5cm}\left.+\frac{1}{4}[X_m,X_n]^2 
-\frac{i}{2}\bar{\lambda}\Gamma^{0}D_{0}\lambda
+\frac{1}{2}\bar{\lambda}\vec{\Gamma}\cdot\vec{{\cal L}}\lambda
-\frac{3i\mu}{8}\bar{\lambda}\Gamma^{123}\lambda
-\frac{1}{2}\bar{\lambda}\Gamma^m[X_m,\lambda] \right).
\label{2+1 SYM convenient form}
\eeqa

It is now easy to obtain PWMM.
We dimensionally reduce (\ref{2+1 SYM convenient form}) to $1+0$ dimensions
by dropping $\vec{\partial}$.
The result is
\beqa
&&S_{PW}=\frac{1}{g_{PW}^2}\int \frac{dt}{\mu^2}\: \Tr \left(
\frac{1}{2}(D_0Y_i)^2-\frac{1}{2}(\mu Y_i-\frac{i}{2}\epsilon_{ijk}[Y_j,Y_k])^2
+\frac{1}{2}(D_0X_m)^2-\frac{\mu^2}{8}X_m^2 \right.\n
&&\qquad\qquad\left. 
+\frac{1}{2}[Y_i,X_m]^2+\frac{1}{4}[X_m,X_n]^2 
-\frac{i}{2}\bar{\lambda}\Gamma^0D_0\lambda
-\frac{3i\mu}{8}\bar{\lambda}\Gamma^{123}\lambda
-\frac{1}{2}\bar{\lambda}\Gamma^{i}[Y_i,\lambda] 
-\frac{1}{2}\bar{\lambda}\Gamma^{m}[X_m,\lambda]\right), \n
\label{action of PWMM}
\eeqa
where $4\pi g_{PW}^2=g_{R\times S^2}^2$.
In appendix B, we show that this is indeed equivalent to
the action of PWMM used in the literature.

In appendix C, we describe the supersymmetry transformations of all the 
theories. In appendix A, we rewrite the actions 
(\ref{action of N=4 SYM on R times S^3}), (\ref{2+1 SYM convenient form})
and (\ref{action of PWMM}) in terms of the $SU(4)$ symmetric notation.
We will make mode expansions for these $SU(4)$ symmetric forms of the actions 
in sections 5 and 6.
In the remaining of the present paper, 
it is convenient to assume that the gauge groups of 
PWMM, $\mbox{SYM}_{R\times S^2}$ and $\mbox{SYM}_{R\times S^3/Z_k}$
are $U(\hat{N})$, $U(\tilde{N})$ and $U(N)$, respectively.

\subsection{Nontrivial vacua}
While $\mbox{SYM}_{R\times S^3}$ has the unique trivial vacuum,
$\mbox{SYM}_{R\times S^3/Z_k}$ has many vacua. Those vacua are given
by the space of flat connections on $S^3/Z_k$.
The space is parameterized by the holonomy $U$ along
nontrivial generator of $\pi_1(S^3/Z_k)=Z_k$ up to gauge transformations.
$U$ satisfies $U^k=1$, so that 
$U$ can be diagonalized as 
\beqa
U=\mbox{diag}(\underbrace{e^{i\frac{2\pi}{k}\beta_1},e^{i\frac{2\pi}{k}\beta_1},\cdots,e^{i\frac{2\pi}{k}\beta_1}}_{N_1},
\underbrace{e^{i\frac{2\pi}{k}\beta_2},e^{i\frac{2\pi}{k}\beta_2},\cdots,
e^{i\frac{2\pi}{k}\beta_2}}_{N_2},\cdots,
\underbrace{e^{i\frac{2\pi}{k}\beta_T},e^{i\frac{2\pi}{k}\beta_T},\cdots,
e^{i\frac{2\pi}{k}\beta_T}}_{N_T}), \n
\label{vacuum for S^3/Z_k}
\eeqa
where all $\beta_s \;\;(s=1,\cdots,T,\;T\leq k)$ are different integers 
mod $k$, 
and $N_1+\cdots+N_T=N$.
The vacua of $\mbox{SYM}_{R\times S^3/Z_k}$ are parameterized by 
$U$ in (\ref{vacuum for S^3/Z_k}). By applying the flat connection condition
to the supersymmetry transformation (\ref{susy transf for S^3}), 
it is easy to see
that these vacua preserve all 16 supercharges.
In the vacuum (\ref{vacuum for S^3/Z_k}), the gauge symmetry $U(N)$ is 
spontaneously broken to 
$U(N_1)\times U(N_2)\times \cdots \times U(N_T)$.

Next, let us discuss the vacua of $\mbox{SYM}_{R\times S^2}$. 
The condition for the 
vacua of $\mbox{SYM}_{R\times S^2}$ is obtained from 
the $k\rightarrow\infty$ limit of the condition 
for the vacua of $\mbox{SYM}_{R\times S^3/Z_k}$, which are given by the space
of the flat connections on $R\times S^3/Z_k$.
Then, it is seen from (\ref{reduction of field strength}) 
that the condition for the vacua of $\mbox{SYM}_{R\times S^2}$ is
\beqa
&&F_{12}-\mu\Phi=0, \n
&&D_1\Phi=D_2\Phi=0.
\label{vacuum condition for 2+1 SYM}
\eeqa
On the other hand, the condition for vacua derived from
(\ref{2+1 SYM convenient form}) is 
\beqa
\vec{{\cal Z}}=0,
\label{cal Z=0}
\eeqa
which is indeed equivalent to (\ref{vacuum condition for 2+1 SYM}) as
seen from (\ref{cal Z}).
In order to solve the equations (\ref{vacuum condition for 2+1 SYM}),
we take a gauge in which $\Phi$ is diagonal. Then, the second equation in
(\ref{vacuum condition for 2+1 SYM}) implies that $\Phi$ is constant.
We parameterize $\Phi$ as
\beqa
\Phi=\frac{\mu}{2}\mbox{diag}
(\underbrace{\alpha_1,\alpha_1,\cdots,\alpha_1}_{N_1},
\underbrace{\alpha_2,\alpha_2,\cdots,\alpha_2}_{N_2},\cdots,
\underbrace{\alpha_T,\alpha_T,\cdots,\alpha_T}_{N_T}),
\label{background Phi}
\eeqa
where all $\alpha_s$'s $(s=1,\cdots,T)$ are different, and
$N_1+\cdots+N_T=\tilde{N}$.
Then, it is seen from the second equation in
(\ref{vacuum condition for 2+1 SYM}) that $A_1$ and $A_2$ are
block-diagonal, where the sizes of the blocks are $N_1,N_2,\cdots,N_T$.
Using the remaining $U(N_1)\times U(N_2)\times \cdots \times U(N_T)$,
we take a gauge in which $A_1=0$.
Then, the first equation reduces to
\beqa
\nabla_1A_2+\mu\cot\theta A_2=\mu\Phi.
\eeqa
This equation can be easily solved by introducing patches on $S^2$ as
\beqa
A_2=\left\{\begin{array}{ll}
            \tan\frac{\theta}{2}\:\Phi  & \mbox{in region I} \\
            -\cot\frac{\theta}{2}\:\Phi & \mbox{in region II}
           \end{array} \right. ,
\eeqa
where the region I corresponds to $0 \leq \theta < \frac{\pi}{2}+\varepsilon$
while the region II corresponds to 
$\frac{\pi}{2}-\varepsilon < \theta \leq \pi$.
To summarize, the solution to (\ref{vacuum condition for 2+1 SYM}) is 
\beqa
&&\hat{\Phi}=\frac{\mu}{2}\mbox{diag}
(\underbrace{\alpha_1,\alpha_1,\cdots,\alpha_1}_{N_1},
\underbrace{\alpha_2,\alpha_2,\cdots,\alpha_2}_{N_2},\cdots,
\underbrace{\alpha_T,\alpha_T,\cdots,\alpha_T}_{N_T}), \n
&&\hat{A}_1=0, \n
&&\hat{A}_2=\left\{\begin{array}{ll}
            \tan\frac{\theta}{2}\:\hat{\Phi}  & \mbox{in region I} \\
            -\cot\frac{\theta}{2}\:\hat{\Phi} & \mbox{in region II}
           \end{array} \right. 
\label{vacuum for S^2}
\eeqa
Each diagonal element of $\hat{A}_1$ and $\hat{A}_2$ is 
the configuration of a monopole with magnetic charge $q_s=\frac{\alpha_s}{2}$.
In the overlap of the regions I and II, the configurations in both patches
are transformed
each other by the gauge transformation given by
\beqa
V_{I\rightarrow II}=\mbox{exp}\left(i\frac{2}{\mu}\hat{\Phi}\phi\right).
\label{V from I to II}
\eeqa
It follows from the single-valuedness of $V_{I\rightarrow II}$
that all $\alpha_s$'s 
$(s=1,\cdots,T)$ in (\ref{vacuum for S^2}) are integers.
This is nothing but Dirac's quantization condition for the monopole charges.
One can understand this condition from a different point of view as follows.
In the $k\rightarrow\infty$ limit, each vacuum of 
$\mbox{SYM}_{R\times S^3/Z_k}$ would reduce to a vacuum of 
$\mbox{SYM}_{R\times S^2}$.
As mentioned in the previous subsection, 
$S^3/Z_k$ is obtained by making an identification on $S^3$, 
$(\theta,\phi,\psi)\;\sim\;(\theta,\phi,\psi+\frac{4\pi}{k})$.
A generator of $\pi_1(S^3/Z_k)$ is a non-contractible loop,
$C:(\frac{\pi}{2},0,\psi) \;\;\psi\in[0,\frac{4\pi}{k}]$. The holonomy along
this loop is 
\beqa
U=P\exp\left[i\int_0^{\frac{4\pi}{k}}A_{\psi}d\psi\right].
\eeqa
In the $k\rightarrow\infty$ limit, from (\ref{replacement}), this reduces to
\beqa
U=\exp\left[i\frac{4\pi}{k}\frac{1}{\mu}\Phi(\theta,\phi)\right].
\label{reduced U}
\eeqa
Substituting (\ref{background Phi}) into (\ref{reduced U}) yields
\beqa
U=\mbox{diag}
(\underbrace{e^{i\frac{2\pi}{k}\alpha_1},e^{i\frac{2\pi}{k}\alpha_1},\cdots,
e^{i\frac{2\pi}{k}\alpha_1}}_{N_1},
\underbrace{e^{i\frac{2\pi}{k}\alpha_2},e^{i\frac{2\pi}{k}\alpha_2},\cdots,
e^{i\frac{2\pi}{k}\alpha_2}}_{N_2},\cdots,
\underbrace{e^{i\frac{2\pi}{k}\alpha_T},e^{i\frac{2\pi}{k}\alpha_T},\cdots,
e^{i\frac{2\pi}{k}\alpha_T}}_{N_T}). \n
\label{vacuum for S^2 2}
\eeqa
The condition $U^k=1$ indeed implies that all $\alpha_s$'s $(s=1,\cdots,T)$ are
integers. This consideration also clarifies correspondence between
the vacua of $\mbox{SYM}_{R\times S^3/Z_k}$ with the $k\rightarrow\infty$ limit
and the vacua of $\mbox{SYM}_{R\times S^2}$. Using (\ref{susy transf for S^2}),
it is easy to show that the vacua (\ref{vacuum for S^2}) preserve all
16 supercharges.
In the vacuum (\ref{vacuum for S^2}), the gauge group $U(\tilde{N})$ is
spontaneously broken to $U(N_1)\times U(N_2)\times \cdots \times U(N_T)$.

Finally, we discuss the vacua of PWMM.
The condition for the vacua would be obtained by dropping the derivative
in (\ref{cal Z=0}). The result is
\beqa
\mu Y_i-\frac{i}{2}\epsilon_{ijk}[Y_j,Y_k]=0.
\label{vacuum condition for PWMM}
\eeqa
This condition is also read off directly from (\ref{action of PWMM}).
The general solution to the equation (\ref{vacuum condition for PWMM}) is
\beqa
Y_i=-\mu L_i,
\eeqa
where $L_i$ is a representation matrix for a
$\hat{N}$-dimensional representation of $SU(2)$, which is in general 
reducible, and satisfies $[L_i,L_j]=i\epsilon_{ijk}L_k$.
One can decompose it into irreducible pieces as
\begin{align}
L_i=
 \begin{pmatrix}
  \rotatebox[origin=tl]{-35}
  {$
  \overbrace{\rotatebox[origin=c]{35}{$L_{i}^{[j_{1}]}$} \;
  \cdots \;
  \rotatebox[origin=c]{35}{$L_{i}^{[j_{1}]}$}}^{\rotatebox{35}{$N_1$}}
  \;\;\;
  \overbrace{\rotatebox[origin=c]{35}{$L_i^{[j_2]}$} \;
  \cdots \;
  \rotatebox[origin=c]{35}{$L_i^{[j_2]}$}}^{\rotatebox{35}{$N_2$}}
  \;\;\; \cdots \;\;\;
  \overbrace{\rotatebox[origin=c]{35}{$L_{i}^{[j_{T}]}$} \;
  \cdots \;
  \rotatebox[origin=c]{35}{$L_{i}^{[j_{T}]}$}}^{\rotatebox{35}{$N_T$}}
  $}
 \end{pmatrix}
 \label{vacuum for PWMM}
\end{align}
where $L_i^{[j_s]}\;\;(s=1,\cdots,T)$ stands for 
the $(2j_s+1)\times (2j_s+1)$ representation 
matrix for the spin $j_s$ representation of $SU(2)$ and satisfies
\beqa
&&[L_i^{[j_s]},L_j^{[j_s]}]=i\epsilon_{ijk}L_k^{[j_s]},\n
&&(L_i^{[j_s]})^2=j_s(j_s+1)1_{2j_s+1},
\eeqa
and 
\beqa
(2j_1+1)N_1+(2j_2+1)N_2+\cdots+(2j_T+1)N_T=\hat{N}.
\eeqa
The vacuum (\ref{vacuum for PWMM}) can be interpreted as
a set of coincident $N_s$ fuzzy spheres
with the radius $\mu\sqrt{j_s(j_s+1)}$ $(s=1,\cdots,T)$,
where all the fuzzy spheres are
concentric. One can see from (\ref{susy transf for PWMM}) that this vacuum
preserves all 16 supercharges.
In this vacuum, the gauge symmetry $U(\hat{N})$ is spontaneously broken to 
$U(N_1)\times U(N_2)\times \cdots \times U(N_T)$.

\section{Gravity duals}
\setcounter{equation}{0}
In this section, we consider the gravity duals of the theories with
$SU(2|4)$ symmetry. In section 3.1, we review the electrostatics problem
that gives the gravity dual of each vacuum of these theories.
In section 3.2, from relations between 
the configurations of conducting disks for the vacua,
we obtain two predictions on relations between the vacua of different theories.
\subsection{Electrostatics problem}
It was shown in \cite{Lin:2005nh} that a general smooth 
solution of type IIA supergravity
that preserves the $SU(2|4)$ symmetry is characterized by a single 
function $V(\rho,\eta)$ and  takes the form
\begin{eqnarray}
&&ds_{10}^2=\left( \frac{\ddot{V}-2\dot{V}}{-V''}\right) \left\{
-4\frac{\ddot{V}}{\ddot{V}-2\dot{V}}dt^2+\frac{-2V''}{\dot{V}}(d\rho^2+d\eta^2)
+4d\Omega_5^2+2\frac{V''\dot{V}}{\Delta}d\Omega_2^2 \right\}, \n
&&e^{4\phi}=\frac{4(\ddot{V}-2\dot{V})^3}{-V''\dot{V}^2\Delta^2},\n
&&C_1=-\frac{2\dot{V}'\dot{V}}{\ddot{V}-2\dot{V}}dt, \n
&&F_4=dC_3,\;\;\;\; 
C_3=-4\frac{\dot{V}^2V''}{\Delta}dt\wedge d^2\Omega, \n
&&H_3=dB_2,\;\;\;\; 
B_2=\left(\frac{\dot{V}\dot{V}'}{\Delta}+\eta\right)d^2\Omega, \n
&&\Delta=(\ddot{V}-2\dot{V})V''-(\dot{V}')^2,
\end{eqnarray}
where the dot and the prime stands for the derivatives with respect to 
$\log\rho$ and $\eta$, respectively. 
$V$ can be regarded as an electrostatic potential
for an axially symmetric system with 
conducting disks and a background potential.
$\rho$ is the distance from the center axis and $\eta$ is the coordinate
in the direction along the center axis.
$V$ is decomposed as $V=V_b(\rho,\eta)+v(\rho,\eta)$, where
$V_b$ is the background potential, and
$v$ is determined by a configuration of conducting disks.
Each theory is specified by $V_b$ and each vacuum is specified by
a configuration of conducting disks. The distance $d$ between two disks
is proportional to the NS 5-brane charge, $d=\frac{\pi}{2}N_5$,
while the electric charge $Q$ on a disk is proportional to the D2-brane charge,
$Q=\frac{\pi^2}{8}N_2$.

\begin{figure}[htbp]
\begin{center}
\psfrag{aa}{$\eta$}
\psfrag{bb}{$\rho$}
\psfrag{cc}{$\pi k/2$}
\psfrag{dd}{$0$}
\psfrag{ee}{$\pi\beta_1/2$}
\psfrag{ii}{$\pi\beta_2/2$}
\psfrag{jj}{$\pi\beta_T/2$}
\psfrag{ff}{$\pi^2 N_1/8$}
\psfrag{gg}{$\pi^2 N_2/8$}
\psfrag{hh}{$\pi^2 N_T/8$}
\includegraphics[height=7cm, keepaspectratio, clip]{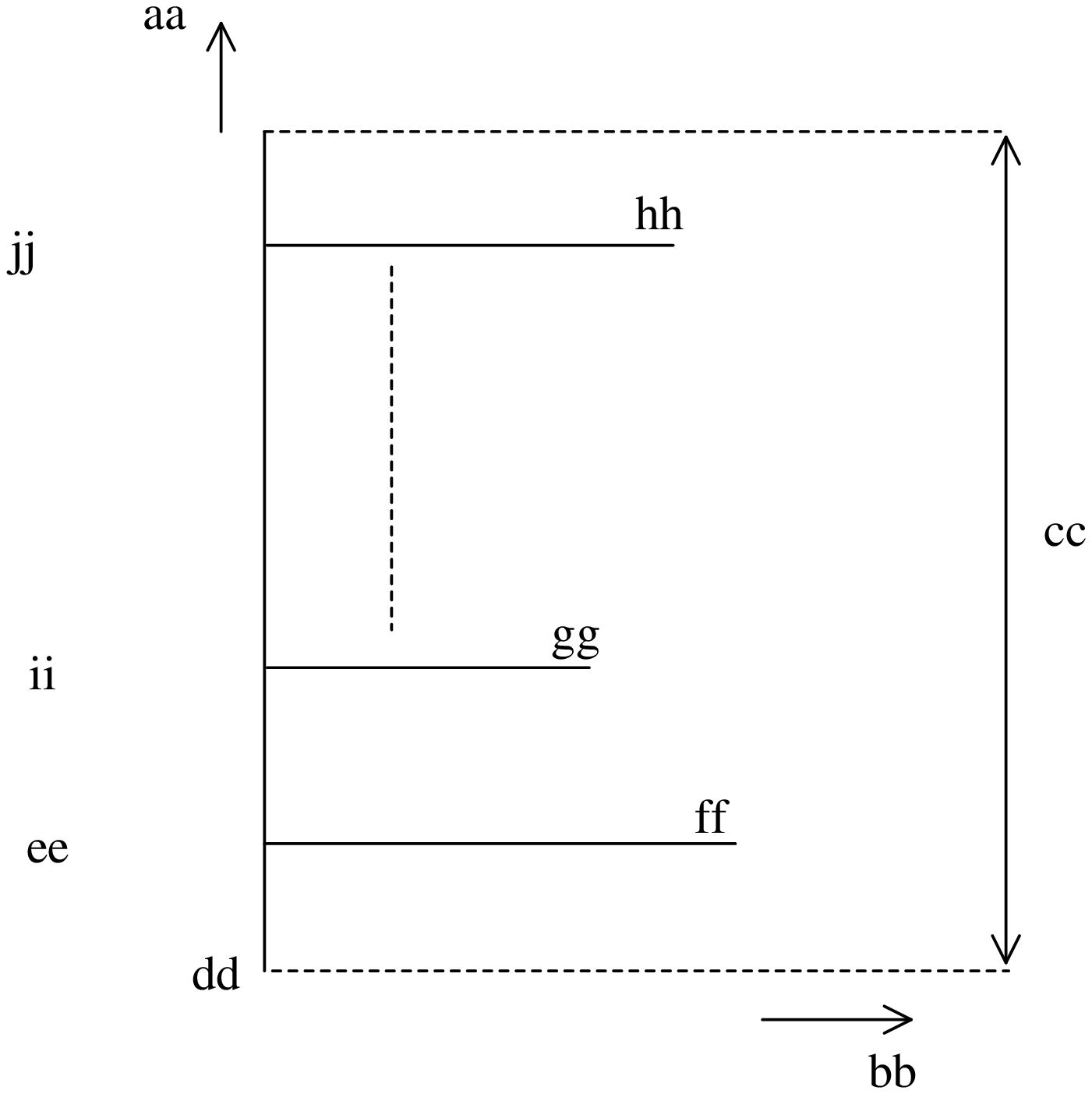}
\end{center}
\caption{Configuration of conducting disks for (\ref{vacuum for S^3/Z_k})}
\label{configuration of conducting disks for S^3/Z_k}
\end{figure}

The background potential for $\mbox{SYM}_{R\times S^3/Z_k}$ is
\beqa
V_b=W(\rho^2-2\eta^2),
\label{background potential for S^3/Z_k}
\eeqa
where $W=c/kg_{R\times S^3/Z_k}^2$ with $c$ a constant \cite{Lin:2005nh}.
In this case, the system is periodic with respect to $\eta$ 
with the period $\frac{\pi}{2}k$, and the total NS 5-brane charge is $k$.
One can concentrate a region $0\leq \eta \leq \frac{\pi}{2}k$, where
one can place conducting disks at 
$\eta=0,\frac{\pi}{2},\cdots,\frac{\pi}{2}(k-1)$.
For the vacuum (\ref{vacuum for S^3/Z_k}), $T$ disks are located at
$\eta_1=\frac{\pi}{2}\beta_1,\eta_2=\frac{\pi}{2}\beta_2,\cdots,
\eta_T=\frac{\pi}{2}\beta_T$.
The electric charges on these disks are equal to
$\frac{\pi^2}{8}N_1,\frac{\pi^2}{8}N_2,\cdots,\frac{\pi^2}{8}N_T$,
respectively. Fig.\ref{configuration of conducting disks for S^3/Z_k}
shows this configuration of conducting disks.

\begin{figure}[htbp]
\begin{center}
\psfrag{aa}{$\eta$}
\psfrag{bb}{$\rho$}
\psfrag{cc}{$\pi^2 N_1/8$}
\psfrag{dd}{$\pi^2 N_2/8$}
\psfrag{ee}{$\pi^2 N_T/8$}
\psfrag{ff}{$\pi\alpha_1/2$}
\psfrag{gg}{$\pi\alpha_2/2$}
\psfrag{hh}{$\pi\alpha_T/2$}
\includegraphics[height=7cm, keepaspectratio, clip]{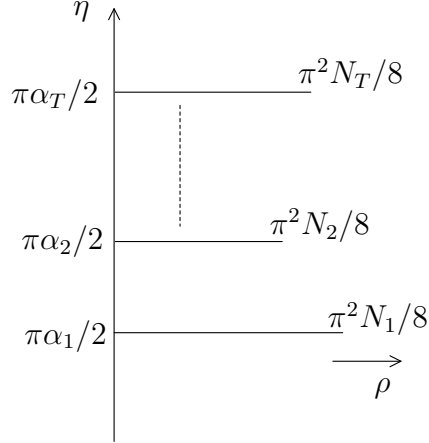}
\end{center}
\caption{Configuration of conducting disks for (\ref{vacuum for S^2})}
\label{configuration of conducting disks for S^2}
\end{figure}

$\mbox{SYM}_{R\times S^2}$ corresponds to the $k\rightarrow\infty$ limit
of $\mbox{SYM}_{R\times S^3/Z_k}$.
For $\mbox{SYM}_{R\times S^2}$, the region of $\eta$ becomes infinite.
The background potential for $\mbox{SYM}_{R\times S^2}$ is given by 
\beqa
V_b=\tilde{W}(\rho^2-2\eta^2),
\eeqa
where $\tilde{W}$ is given by the $k\rightarrow\infty$ limit of $W$,
so that $kg_{R\times S^3/Z_k}^2$ must be fixed. This is consistent with
the result in the gauge theory side, and from (\ref{coupling relation})
$\tilde{W}$ turns out to be
$c\mu/4\pi g_{R\times S^2}^2$.
By using the correspondence between the vacua of $\mbox{SYM}_{R\times S^3/Z_k}$
with the $k\rightarrow\infty$ limit 
and the vacua of $\mbox{SYM}_{R\times S^2}$ seen in the previous subsection, 
it is easy to construct
a configuration of conducting disks for each vacuum of 
$\mbox{SYM}_{R\times S^2}$.
For the vacuum (\ref{vacuum for S^2 2}), there are $T$ disks located
at $\eta_1=\frac{\pi}{2}\alpha_1,\eta_2=\frac{\pi}{2}\alpha_2,\cdots,
\eta_T=\frac{\pi}{2}\alpha_T$.
The electric charges on these disks are equal to
$\frac{\pi^2}{8}N_1,\frac{\pi^2}{8}N_2,\cdots,\frac{\pi^2}{8}N_T$, 
respectively. Fig.\ref{configuration of conducting disks for S^2}
shows this configuration of conducting disks.


\begin{figure}[htbp]
\begin{center}
\psfrag{aa}{$\eta$}
\psfrag{bb}{$\rho$}
\psfrag{cc}{$\pi(2j_1+1)/2$}
\psfrag{dd}{$\pi(2j_2+1)/2$}
\psfrag{ee}{$\pi(2j_T+1)/2$}
\psfrag{ff}{$\pi^2N_1/8$}
\psfrag{gg}{$\pi^2N_2/8$}
\psfrag{hh}{$\pi^2N_T/8$}
\psfrag{ii}{$0$}
\includegraphics[height=7cm, keepaspectratio, clip]{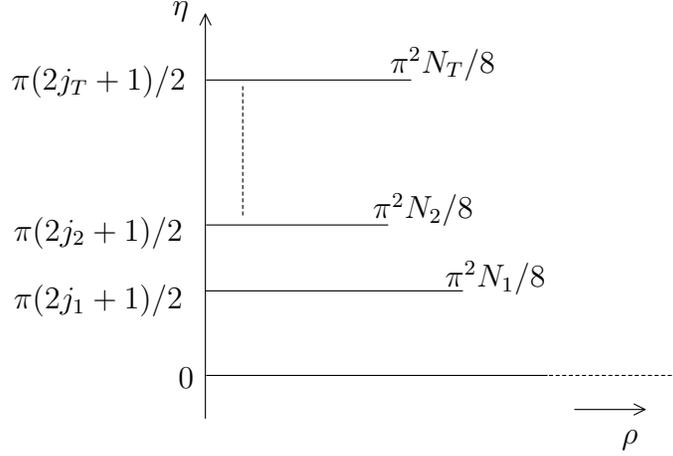}
\end{center}
\caption{Configuration of conducting disks for (\ref{vacuum for PWMM})}
\label{configuration of conducting disks for PWMM}
\end{figure}

The background potential for PWMM is
\beqa
V_b=\hat{W}(\rho^2\eta-\frac{2}{3}\eta^3),
\label{background potential for PWMM}
\eeqa
where $\hat{W}$ is represented in terms of a certain function $h$  
as \cite{Ling:2006up}
\beqa
\hat{W}=\frac{1}{g_{PW}^2}h(g_{PW}^2\hat{N}).
\eeqa
It was pointed out in \cite{Ling:2006up} 
that the correspondence between the trivial
vacuum of $\mbox{SYM}_{R\times S^2}$ and a certain vacuum of PWMM
shown in \cite{Maldacena:2002rb}
is consistent with the gravity side only if 
the function $h$
approaches some constant $h_{\infty}$ at large values of its argument.
Namely, this behavior of $h$ is true if the gauge/gravity correspondence
for the theories with $SU(2|4)$ symmetry is valid.
We assume this behavior, and we will use this assumption
to obtain the prediction 1). 
In the case of PWMM, only the region $\eta\geq 0$ is meaningful.
There is always a infinitely large disk sitting at $\eta=0$.
For the vacuum (\ref{vacuum for PWMM}), 
there are $T$ disks other than this disk.
They are located at $\eta_1=\frac{\pi}{2}(2j_1+1),\eta_2=\frac{\pi}{2}(2j_2+1),
\cdots,\eta_T=\frac{\pi}{2}(2j_T+1)$.
The electric charges on these disks are equal to
$\frac{\pi^2}{8}N_1,\frac{\pi^2}{8}N_2,\cdots,\frac{\pi^2}{8}N_T$, 
respectively. Fig.\ref{configuration of conducting disks for PWMM}
shows this configuration of conducting disks.

\subsection{Predictions on relations between vacua}
\begin{figure}[htbp]
\begin{center}
\psfrag{aa}{$\rightarrow\infty$}
\psfrag{bb}{$\rho$}
\psfrag{cc}{$\rho$}
\psfrag{dd}{$\eta$}
\psfrag{ee}{$\eta$}
\includegraphics[height=7cm, keepaspectratio, clip]{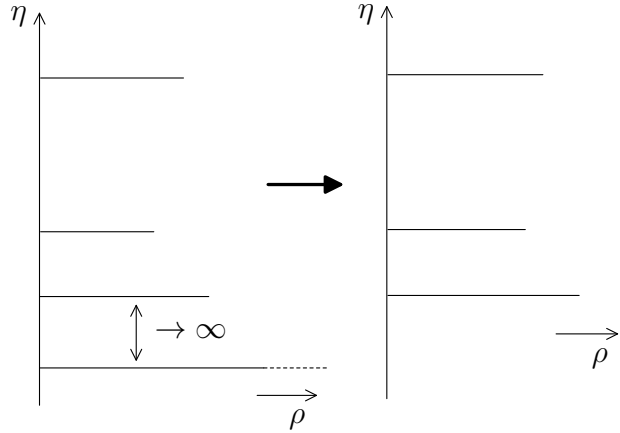}
\end{center}
\caption{From a vacuum of the plane wave matrix model to a vacuum of $2+1$ SYM
on $R\times S^2$}
\label{from PWMM to S^2}
\end{figure}

We first consider a limit that transforms
a vacuum of PWMM into a vacuum of $\mbox{SYM}_{R\times S^2}$.
Naively, by moving the infinitely large disk in
a configuration for a vacuum of PWMM away to infinity
as in Fig.\ref{from PWMM to S^2},
one obtains a configuration of disks for a vacuum of 
$\mbox{SYM}_{R\times S^2}$.
This motivates us to take the following limit.
We parameterize the positions of the disks for a vacuum of PWMM, 
which are proportional to 
the dimensions of representations of $SU(2)$ in the gauge theory, as
\beqa
&&2j_s+1=N_0+\zeta_s, \n
&&\eta_s=\eta_0+\tilde{\eta}_s, \n
&&\eta_0=\frac{\pi}{2}N_0, \;\;\;
\tilde{\eta}_s=\frac{\pi}{2}\zeta_s,
\eeqa
where $N_0$ and $\zeta_s$ are integers.
Under a shift $\eta \rightarrow \eta_0+\eta$, the background potential
(\ref{background potential for PWMM}) is transformed as 
\beqa
V_b \;\rightarrow\; -\frac{2}{3}\hat{W}\eta_0^3-2\hat{W}\eta_0^2\eta
+\hat{W}\eta_0(\rho^2-2\eta^2)+\hat{W}(\eta\rho^2-\frac{2}{3}\eta^3)
\eeqa
The first and second terms on the righthand side do not contribute
to the Laplace equation, the boundary condition for $V$ and
the geometry.
In the limit,
\beqa
\eta_0\rightarrow \infty, \;\;\; \hat{W}\rightarrow 0,\;\;\;
\hat{W}\eta_0=\tilde{W}=\mbox{fixed},
\label{limit}
\eeqa
the last term vanishes and only the third term survives resulting in
the background potential for $\mbox{SYM}_{R\times S^2}$.
In the $T=1$ case, it was explicitly shown in \cite{Ling:2006up} that
the charge $Q_1$ can be fixed in this limit. It is reasonable to expect that
all the charges $Q_s$'s $(s=1,\cdots,T)$ can be fixed in this limit 
for generic $T$.
Hence, the limit (\ref{limit}) indeed transforms the gravity dual of 
a vacuum of PWMM to the gravity dual of a vacuum of $\mbox{SYM}_{R\times S^2}$
(See Fig.\ref{from PWMM to S^2}).
This observation on the gravity side leads us to the prediction 1).
Indeed, by using the relation between $\tilde{W}$ and $g_{R\times S^2}$
and the behavior of $h$ in $\hat{W}$ discussed in the previous subsection,
we obtain the
prediction 1) that
on the gauge theory side the theory around
the vacuum (\ref{vacuum for PWMM}) of PWMM
coincides with the theory around
the vacuum (\ref{vacuum for S^2}) of $\mbox{SYM}_{R\times S^2}$ with
the identification $\zeta_s-\zeta_t=\alpha_s-\alpha_t \;\;(s,t=1,\cdots,T)$
in the limit
\beqa
N_0 \rightarrow \infty, \;\;\; 
\frac{N_0}{g^2_{PW}}=\mbox{fixed}\sim \frac{1}{g_{R\times S^2}^2}.
\label{coupling relation between S^2 and PWMM}
\eeqa
In section 5, we will prove the prediction 1).

\begin{figure}[htbp]
\begin{center}
\psfrag{aa}{$\pi k/2$}
\psfrag{bb}{$\eta$}
\psfrag{cc}{$\rho$}
\psfrag{dd}{$\pi^2N/8$}
\includegraphics[height=7cm, keepaspectratio, clip]{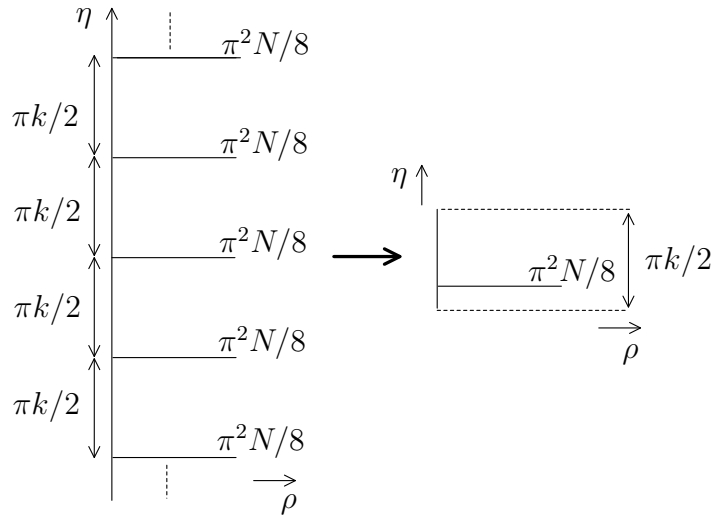}
\end{center}
\caption{From a vacuum of $2+1$ SYM on $R\times S^2$ to the trivial vacuum of 
${\cal N}=4$ SYM on $R\times S^3/Z_k$}
\label{periodic configuration}
\end{figure}

Next, let us discuss the prediction 2).
In the gravity dual of $\mbox{SYM}_{R\times S^2}$, we consider 
a configuration of disks which is periodic in the $\eta$ 
direction with period $\frac{\pi}{2}k$
and extract a single period. 
This procedure should yield the gravity dual of a theory around
a vacuum of $\mbox{SYM}_{R\times S^3/Z_k}$.
In the procedure, $W=\tilde{W}$, so that the coupling constant 
of the resultant theory around the 
vacuum of $\mbox{SYM}_{R\times S^3/Z_k}$ is given by a relation
\beqa
g_{R\times S^3/Z_k}^2=\frac{4\pi}{k\mu}g_{R\times S^2}^2.
\label{relation for coupling constant}
\eeqa
In particular, Fig.\ref{periodic configuration} shows 
the case in which the trivial vacuum of $\mbox{SYM}_{R\times S^3/Z_k}$
with the gauge group $U(N)$ is obtained.
The corresponding vacuum configuration of $\mbox{SYM}_{R\times S^2}$ is
\beqa
&&\hat{\Phi}=\frac{\mu}{2}\mbox{diag}
(\cdots,\underbrace{k(s-1),\cdots,k(s-1)}_{N},
\underbrace{ks,\cdots,ks}_{N},\underbrace{k(s+1),\cdots,k(s+1)}_{N},
\cdots), \n
&&\hat{A}_1=0, \n
&&\hat{A}_2=\left\{\begin{array}{ll}
            \tan\frac{\theta}{2}\:\hat{\Phi}  & \mbox{in region I} \\
            -\cot\frac{\theta}{2}\:\hat{\Phi} & \mbox{in region II}
           \end{array} \right. 
\label{periodic Phi}
\eeqa
where $s$ runs from $-\infty$ to $\infty$.
In section 6, we will show
that the theory around the trivial 
vacuum of $\mbox{SYM}_{R\times S^3/Z_k}$ with the gauge group $U(N)$
is obtained by the theory around the vacuum labeled
by (\ref{periodic Phi}) through the following procedure:
we impose a condition which
corresponds to the periodicity on the gravity side and
extract a single period, and input
the relation (\ref{relation for coupling constant}).
This is a proof of the prediction 2) for the trivial vacuum of 
$\mbox{SYN}_{R\times S^3/Z_k}$.

\section{Spherical harmonics}
\setcounter{equation}{0}
In this section, we consider various spherical harmonics:
the spherical harmonics on $S^3$ in section 4.1, the monopole harmonics
in section 4.2, and the fuzzy sphere harmonics
in section 4.3.
We reveal relationship between the spherical harmonics on $S^3$ and the
monopole harmonics in section 4.2,
and relationship between the monopole harmonics and 
the fuzzy sphere harmonics in section 4.3. 
The latter implies that the fuzzy sphere harmonics
can be regarded as a matrix regularization of the monopole harmonics.
In this section, we frequently use the formula 
for the representations of $SU(2)$ gathered in appendix D.
\subsection{Spherical harmonics on $S^3$}
In our previous publication \cite{ITT}, 
we summarized the properties of the spherical harmonics
based on \cite{Salam:1981xd,Cutkosky,Hamada} and found some new formula.
In this subsection, we recall the properties of the spherical harmonics
on $S^3$ based on \cite{ITT} and add some new formula.
We view $S^3$ as $G/H=SO(4)/SO(3)$, where
$G=SO(4)=SU(2)\times \tilde{SU}(2)$, and 
the subgroup $H=SO(3)$ is naturally identified with the local 
`Lorentz' group $SO(3)$ on $S^3$. We denote the generators of the $SU(2)$
in $G$ by $J_i$ and those of the $\tilde{SU}(2)$ in $G$
by $\tilde{J}_i$, where $i=1,2,3$. Then, the generators of $H$ are represented
by $S_i=J_i+\tilde{J}_i$. 

The irreducible representations of $G$ are labeled
by two spins, $J$ and $\tilde{J}$, which specify the irreducible 
representations of the $SU(2)$ and the $\tilde{SU}(2)$, respectively.
We denote the basis of the $(J,\tilde{J})$ representation by 
$|Jm\rangle |\tilde{J}\tilde{m}\rangle$. The basis of 
the spin $S$ representation of $H$ is constructed in terms of 
$|Jm\rangle |\tilde{J}\tilde{m}\rangle$:
\beqa
|Sn;J\tilde{J}\rangle\rangle
=\sum_{m\tilde{m}}C^{Sn}_{Jm \; \tilde{J}\tilde{m}}
|Jm\rangle |\tilde{J}\tilde{m}\rangle,
\eeqa
where $C^{Sn}_{Jm \; \tilde{J}\tilde{m}}$ is the Clebsch-Gordan coefficient
of $SU(2)$ and 
the triangular inequality, 
\beqa
|J-\tilde{J}|\leq S \leq J+\tilde{J},
\label{triangularinequality}
\eeqa
must be satisfied. 

A definite form of the representative element of $G/H$ is 
given by\footnote{We use the coordinate system given in appendix A, which
is different from the one in \cite{ITT}.}
\beqa
\Upsilon(\Omega)=e^{-i\phi J_3}e^{i\psi\tilde{J}_3}
e^{-i\frac{\theta}{2}(J_1-\tilde{J}_1)}.
\label{representative}
\eeqa
The spin $S$ spherical harmonics on $S^3$ is given by
\beqa
{\cal Y}^{Sn}_{Jm,\tilde{J}\tilde{m}}(\Omega)
=N^S_{J\tilde{J}}
\langle\langle Sn;J\tilde{J}|\Upsilon^{-1}(\Omega)
|Jm\rangle |\tilde{J}\tilde{m}\rangle,
\label{spinSsphericalharmonics}
\eeqa
where $N^S_{J\tilde{J}}$ is the normalization factor fixed as
\beqa
N^S_{J\tilde{J}}=\sqrt{\frac{(2J+1)(2\tilde{J}+1)}{2S+1}}.
\eeqa
The spherical harmonics (\ref{spinSsphericalharmonics}) satisfies the
orthonormal condition
\beqa
\int \frac{d\Omega_3}{2\pi^2} 
\:\sum_n({\cal Y}^{Sn}_{J_1m_1,\tilde{J}_1\tilde{m}_1})^* 
\:{\cal Y}^{Sn}_{J_2m_2,\tilde{J}_2\tilde{m}_2}
=\delta_{J_1J_2}\delta_{\tilde{J}_1\tilde{J}_2}\delta_{m_1m_2}
\delta_{\tilde{m}_1\tilde{m}_2}.
\label{orthonormality}
\eeqa
The complex conjugate of ${\cal Y}^{Ln}_{Jm,\tilde{J}\tilde{m}}$ is given by
\beqa
({\cal Y}^{Sn}_{Jm,\tilde{J}\tilde{m}})^*
=(-1)^{-J+\tilde{J}-S+m-\tilde{m}+n} \:
  {\cal Y}^{S\:-n}_{J\:-m,\tilde{J}\:-\tilde{m}}.
\label{complexconjugate}
\eeqa
The covariant derivative is understood as an algebraic manipulation:
\beqa
\nabla_i\:{\cal Y}^{Sn}_{Jm,\tilde{J}\tilde{m}}(\Omega)
=-iN^S_{J\tilde{J}}
 \langle\langle Sn;J\tilde{J}|(J_i-\tilde{J}_i)\Upsilon^{-1}(\Omega)
 |Jm\rangle |\tilde{J}\tilde{m}\rangle.
\label{covariantderivative}
\eeqa
Using this relation, it is easy to obtain the eigenvalue of the laplacian
for the spin $S$ spherical harmonics:
\beqa
\nabla^2{\cal Y}^{Sn}_{Jm,\tilde{J}\tilde{m}}
=-(2J(J+1)+2\tilde{J}(\tilde{J}+1)-S(S+1))
\:{\cal Y}^{Sn}_{Jm,\tilde{J}\tilde{m}}.
\label{laplacian}
\eeqa
Moreover, using (\ref{covariantderivative}) and (\ref{6j}),
we find a new formula
\beqa
&&C^{1r}_{S'n' \; Sn}\check{\nabla}_r{\cal Y}^{Sn}_{Jm,\tilde{J}\tilde{m}}
=-i(-1)^{J+\tilde{J}+S+S'-n'}\left(
\sqrt{3J(J+1)(2J+1)}\left\{
\begin{array}{ccc}
S & S' & 1 \\
J & J  & \tilde{J}
\end{array} \right\} \right.\n
&&\hspace{3.5cm}\left.
-(-1)^{S-S'}\sqrt{3\tilde{J}(\tilde{J}+1)(2\tilde{J}+1)}\left\{
\begin{array}{ccc}
S         & S'         & 1 \\
\tilde{J} & \tilde{J}  & J
\end{array} \right\}
\right) {\cal Y}^{S'-n'}_{Jm,\tilde{J}\tilde{m}},
\label{linear action}
\eeqa
where 
\beqa
\check{\nabla}_{\pm}=\mp\frac{1}{\sqrt{2}}(\nabla_1\pm i\nabla_2),\;\;\;
\check{\nabla}_0=\nabla_3.
\eeqa
In particular, when $S=S'$, this formula reduces to
\beqa
C^{1r}_{Sn' \; Sn}\check{\nabla}_r{\cal Y}^{Sn}_{Jm,\tilde{J}\tilde{m}}
=i(-1)^{S-n'}\sqrt{3}(J(J+1)-\tilde{J}(\tilde{J}+1))
{\cal Y}^{S-n'}_{Jm,\tilde{J}\tilde{m}}.
\eeqa
By using (\ref{CG symmetry}) and (\ref{complex conjugate of matrix element}),
we rewrite (\ref{spinSsphericalharmonics}) to an expression, in which
the connection to the monopole harmonics defined in the next 
subsection is clear:
\beqa
{\cal Y}^{Sn}_{Jm,\tilde{J}\tilde{m}}
={\cal K}_{Snn'}C^{Jm}_{\tilde{J}p \; Sn'}Y_{\tilde{J}p\tilde{m}},
\label{another expression}
\eeqa
where
\beqa
{\cal K}_{Snn'}=\langle Sn| e^{i\frac{\theta}{2}S_1}e^{i\phi S_3}|Sn'\rangle,
\eeqa
and $Y_{\tilde{J}p\tilde{m}}={\cal Y}^{00}_{\tilde{J}p,\tilde{J}\tilde{m}}$, 
which
is the scalar spherical harmonics. 
In \cite{ITT}, we found the compact formula for
the integral of the product of three spherical harmonics,
\beqa
&&\int \frac{d\Omega_3}{2\pi^2} \: \sum_{n_1n_2n_3}
                  ({\cal Y}^{S_1n_1}_{J_1m_1,\tilde{J}_1\tilde{m}_1})^* \: 
                  {\cal Y}^{S_2n_2}_{J_2m_2,\tilde{J}_2\tilde{m}_2}
                  \:{\cal Y}^{S_3n_3}_{J_3m_3,\tilde{J}_3\tilde{m}_3} \: 
                  C^{S_1n_1}_{S_2n_2 \; S_3n_3} \nonumber\\
&&=\sqrt{(2S_1+1)(2J_2+1)(2\tilde{J}_2+1)(2J_3+1)(2\tilde{J}_3+1)}\:
   \left\{   \begin{array}{ccc}
             J_1 & \tilde{J}_1 & S_1 \\
             J_2 & \tilde{J}_2 & S_2 \\
             J_3 & \tilde{J}_3 & S_3 
             \end{array}     \right\} \:
             C^{J_1m_1}_{J_2m_2 \; J_3m_3}
             C^{\tilde{J}_1\tilde{m}_1}_{\tilde{J}_2\tilde{m}_2 \; 
                \tilde{J}_3\tilde{m}_3} .  \n
\label{integralofthreeharmonics}
\eeqa
Here we rederive the formula in a different way, starting with a 
particular case of the formula,
\beqa
\int \frac{d\Omega_3}{2\pi^2}
(Y_{J_1m_1\tilde{m}_1})^*Y_{J_2m_2\tilde{m}_2}Y_{J_3m_3\tilde{m}_3}
=\sqrt{\frac{(2J_2+1)(2J_3+1)}{2J_1+1}}
C^{J_1m_1}_{J_2m_2 \; J_3m_3}
C^{J_1\tilde{m}_1}_{J_2\tilde{m}_2 \; J_3\tilde{m}_3}.
\label{integral of theree scalars}
\eeqa
By noting 
\beqa
\sum_{n_1n_2n_3}C^{S_1n_1}_{S_2n_2 \; S_3n_3}
({\cal K}_{S_1n_1{n_1}'})^*{\cal K}_{S_2n_2{n_2}'}{\cal K}_{S_3n_3{n_3}'}
=C^{S_1{n_1}'}_{S_2{n_2}' \; S_3{n_3}'},
\eeqa
we find that the lefthand side of (\ref{integralofthreeharmonics}) is equal to
\beqa
C^{S_1n_1}_{S_2n_2 \; S_3n_3}
C^{J_1m_1}_{\tilde{J}_1p_1 \; S_1n_1}
C^{J_2m_2}_{\tilde{J}_2p_2 \; S_2n_2}
C^{J_3m_3}_{\tilde{J}_3p_3 \; S_3n_3}
\int \frac{d\Omega_3}{2\pi^2}
(Y_{\tilde{J}_1p_1\tilde{m}_1})^*Y_{\tilde{J}_2p_2\tilde{m}_2}
Y_{\tilde{J}_3p_3\tilde{m}_3}.
\eeqa
Applying (\ref{integral of theree scalars}) and (\ref{9j}) to this expression 
leads to (\ref{integralofthreeharmonics}).

As an application of the above results, we
consider scalars, vectors and spinors on $S^3$.
The scalar corresponds to $S=0$. From the triangular inequality
(\ref{triangularinequality}), we see that
$(J,\tilde{J})=(J,J)$. We introduce a notation for the scalar:
\beqa
Y_{Jm\tilde{m}}\equiv {\cal Y}^{S=0,n=0}_{Jm,J\tilde{m}}.
\eeqa
The vector corresponds to $S=1$.
Then, the triangular inequality implies that
$(J,\tilde{J})$ takes $(J+1,J)$ or $(J,J+1)$ or $(J,J)$.
We assign $\rho=1$, $\rho=-1$ and $\rho=0$ to these three cases, respectively.
We make a change of basis from the basis of the $S_3$ eigenstates to
the vector basis:
\beqa
&&{\cal Y}^{1}_{Jm,\tilde{J}\tilde{m}}
=\frac{1}{\sqrt{2}}(-{\cal Y}^{11}_{Jm,\tilde{J}\tilde{m}}
                    +{\cal Y}^{1-1}_{Jm,\tilde{J}\tilde{m}}), \n
&&{\cal Y}^{2}_{Jm,\tilde{J}\tilde{m}}
=-\frac{i}{\sqrt{2}}({\cal Y}^{11}_{Jm,\tilde{J}\tilde{m}}
                     +{\cal Y}^{1-1}_{Jm,\tilde{J}\tilde{m}}), \n
&&{\cal Y}^{3}_{Jm,\tilde{J}\tilde{m}}
={\cal Y}^{10}_{Jm,\tilde{J}\tilde{m}}.
\label{change of basis}
\eeqa
We introduce a notation for the vector:
\beqa
Y^{\rho=1}_{Jm\tilde{m}i}=i{\cal Y}^{i}_{J+1\:m,J\tilde{m}}, \;\;\;
Y^{\rho=-1}_{Jm\tilde{m}i}=-i{\cal Y}^{i}_{Jm,J+1\:\tilde{m}}, \;\;\;
Y^{\rho=0}_{Jm\tilde{m}i}={\cal Y}^{i}_{Jm,J\tilde{m}}.
\eeqa
Here the factors $\pm i$ on the right-hand side are just a convention.
Note that $Y^{0}_{J=0\:M=(0,0)i}=0$.
The spinor corresponds to $S=\frac{1}{2}$.  The triangular inequality
implies that $(J,\tilde{J})$ takes $(J+\frac{1}{2},J)$ or $(J,J+\frac{1}{2})$.
We assign $\kappa=1$ to the former and $\kappa=-1$ to the latter.
We introduce a notation for the spinor:
\beqa
Y^{\kappa=1}_{Jm\tilde{m}\alpha}
={\cal Y}^{S=\frac{1}{2},\alpha}_{J+\frac{1}{2}\:m,J\tilde{m}}, \;\;\;
Y^{\kappa=-1}_{Jm\tilde{m}\alpha}
={\cal Y}^{S=\frac{1}{2},\alpha}_{Jm,J+\frac{1}{2}\:\tilde{m}}, 
\eeqa
where $\alpha$ takes $\frac{1}{2}$ and $-\frac{1}{2}$.
The orthonormality condition (\ref{orthonormality}) is translated to
the scalar, the vector and the spinor as
\beqa
&&\int \frac{d\Omega_3}{2\pi^2} \:(Y_{J_1m_1\tilde{m}_1})^* 
Y_{J_2m_2\tilde{m}_2}
=\delta_{J_1J_2}\delta_{m_1m2}\delta_{\tilde{m}_1\tilde{m}_2}, \n
&&\int \frac{d\Omega_3}{2\pi^2} \:(Y^{\rho_1}_{J_1m_1\tilde{m}_1i})^* 
Y^{\rho_2}_{J_2m_2\tilde{m}_2i}
=\delta_{\rho_1\rho_2}\delta_{J_1J_2}
\delta_{m_1m_2}\delta_{\tilde{m}_1\tilde{m}_2}, \n
&&\int \frac{d\Omega_3}{2\pi^2} \:(Y^{\kappa_1}_{J_1m_1\tilde{m}_1\alpha})^* 
Y^{\kappa_2}_{J_2m_2\tilde{m}_2\alpha}
=\delta_{\kappa_1\kappa_2}\delta_{J_1J_2}
\delta_{m_1m_2}\delta_{\tilde{m}_1\tilde{m}_2},
\label{orthnormality2}
\eeqa
while their complex conjugates are read off from (\ref{complexconjugate})
as
\beqa
&&(Y_{Jm\tilde{m}})^*=(-1)^{m-\tilde{m}}Y_{J-m-\tilde{m}}, \n
&&(Y^{\rho}_{Jm\tilde{m}i})^*=(-1)^{m-\tilde{m}+1}Y^{\rho}_{J-m-\tilde{m}i}, \n
&&(Y^{\kappa}_{Jm\tilde{m}\alpha})^*=(-1)^{m-\tilde{m}+\kappa\alpha+1}
                                   Y^{\kappa}_{J-m-\tilde{m}-\alpha}.
\label{complexconjugate2}
\eeqa 
The eigenvalues of the laplacian can be read off from (\ref{laplacian}):
\beqa
&&\nabla^2\:Y_{Jm\tilde{m}}=-4J(J+1)\:Y_{Jm\tilde{m}}, \n
&&\nabla^2\:Y^{\pm 1}_{Jm\tilde{m}i}
=-(4J(J+2)+2)\:Y^{\pm 1}_{Jm\tilde{m}i}, \n
&&\nabla^2\:Y^0_{Jm\tilde{m}i}=-(4J(J+1)-2)\:Y^0_{Jm\tilde{m}i}, \n
&&\nabla^2\:Y^{\kappa}_{Jm\tilde{m}\alpha}=-(2J(2J+3)+\frac{3}{4})\:
                                           Y^{\kappa}_{Jm\tilde{m}\alpha}.
\label{laplacian2}
\eeqa
Using (\ref{linear action}) yields identities,
\beqa
&&\nabla_i\:Y_{Jm\tilde{m}}=-2i\sqrt{J(J+1)}\:Y^0_{Jm\tilde{m}i}, \n
&&\nabla_i\:Y^{\rho}_{Jm\tilde{m}i}
=-2i\delta_{\rho 0}\sqrt{J(J+1)}Y_{Jm\tilde{m}}, \n
&&\epsilon_{ijk}\:\nabla_j\:Y^{\rho}_{Jm\tilde{m}k}
=-2\rho (J+1)\:Y^{\rho}_{Jm\tilde{m}i}, \n
&&\sigma^i_{\alpha\beta}\:\nabla_i\:Y^{\kappa}_{Jm\tilde{m}\beta}
=-i\kappa (2J+\frac{3}{2})\: Y^{\kappa}_{Jm\tilde{m}\alpha}.
\label{identities}
\eeqa
In \cite{ITT}, we defined various integrals of the product of
three scalar or spinor or vector harmonics, 
which we call vertex coefficients:
\beqa
&&{\cal C}^{J_1m_1\tilde{m}_1}_{J_2m_2\tilde{m}_2\;J_3m_3\tilde{m}_3}
\equiv\int \frac{d\Omega_3}{2\pi^2} 
\:(Y_{J_1m_1\tilde{m}_1})^*Y_{J_2m_2\tilde{m}_2}
Y_{J_3m_3\tilde{m}_3}. \n
&&{\cal D}^{Jm\tilde{m}}_{J_1m_1\tilde{m}_1\rho_1\;J_2m_2\tilde{m}_2\rho_2}
\equiv \int \frac{d\Omega_3}{2\pi^2}
\: (Y_{Jm\tilde{m}})^*Y^{\rho_1}_{J_1m_1\tilde{m}_1i}
Y^{\rho_2}_{J_2m_2\tilde{m}_2i}. \n
&&{\cal E}_{J_1m_1\tilde{m}_1\rho_1\;J_2m_2\tilde{m}_2\rho_2
\;J_3m_3\tilde{m}_3\rho_3}
\equiv\int \frac{d\Omega_3}{2\pi^2} \: \epsilon_{ijk}\: 
Y^{\rho_1}_{J_1m_1\tilde{m}_1i} Y^{\rho_2}_{J_2m_2\tilde{m}_2j} 
Y^{\rho_3}_{J_3m_3\tilde{m}_3k}. \n
&&{\cal F}^{J_1m_1\tilde{m}_1\kappa_1}_{J_2m_2\tilde{m}_2\kappa_2\;Jm\tilde{m}}
\equiv \int \frac{d\Omega_3}{2\pi^2} \:
           (Y^{\kappa_1}_{J_1m_1\tilde{m}_1\alpha})^*
           Y^{\kappa_2}_{J_2m_2\tilde{m}_2\alpha}Y_{Jm\tilde{m}}. \n
&&{\cal G}^{J_1m_1\tilde{m}_1\kappa_1}_{J_2m_2\tilde{m}_2\kappa_2\;
Jm\tilde{m}\rho}
\equiv \int \frac{d\Omega_3}{2\pi^2} \: 
         (Y^{\kappa_1}_{J_1m_1\tilde{m}_1\alpha})^*\sigma^i_{\alpha\beta}
          Y^{\kappa_2}_{J_2m_2\tilde{m}_2\beta}Y^{\rho}_{Jm\tilde{m}i}. 
\label{vertexcoefficients}
\eeqa
The expressions for the vertex coefficients are obtained by using the formula
(\ref{integralofthreeharmonics}) and given in appendix E.

\subsection{Monopole harmonics}
The angular momentum operator in the presence of a monopole with the magnetic
charge $q$ at the origin takes the form
\beqa
\vec{L}^{(q)}=\vec{x}\times (-i\vec{\partial}-\vec{A}^{(q)})
              -q\vec{e}_r,
\eeqa
where 
\beqa
\vec{A}^{(q)}
=\left\{\begin{array}{ll}
        \frac{q}{r}\tan\frac{\theta}{2}\vec{e}_{\phi} & \mbox{in region I}\\
        -\frac{q}{r}\cot\frac{\theta}{2}\vec{e}_{\phi}& \mbox{in region II}
        \end{array} \right.
\eeqa
The regions I and II are defined in section 2.2 and 
$q$ can take $0,\pm\frac{1}{2},\pm 1,\pm\frac{3}{2},\cdots$ due to 
Dirac's quantization condition, as explained in 2.2.
Noting $\vec{x}=r\vec{e}_r$, it is easy 
to see that neither $r$ nor the $r$-derivative appear in $\vec{L}^{(q)}$ in
the polar coordinates system. Note that $\vec{L}^{(0)}$ is nothing but
$\vec{L}^{(0)}$ in (\ref{vecY and so on}).
$\vec{L}^{(q)}$ satisfies the $SU(2)$ algebra:
\beqa
[L^{(q)}_i,L^{(q)}_j]=i\epsilon_{ijk}L^{(q)}_k.
\eeqa
The monopole harmonic function (section), $Y_{q,J,m}(\theta,\phi)$, 
was 
constructed by Wu and Yang \cite{Wu:1976ge}, 
where $J$ takes $|q|,|q|+1,|q|+2,\cdots$
and $m$ takes $-J,-J+1,\cdots,J-1,J$. The explicit expressions for 
$Y_{q,J,m}$ in the regions I and II are given in \cite{Wu:1976ge}.
It is convenient for us to multiply a phase and normalization factor:
\beqa
\tilde{Y}_{Jmq}=(-1)^J\sqrt{4\pi}Y_{q,J,m}
\eeqa
We see from \cite{Wu:1976ge,Wu:1977qk} that $\tilde{Y}_{Jmq}$ 
has the following properties.
\beqa
&&L^{(q)}_{\pm}\tilde{Y}_{Jmq}
=\sqrt{(J\mp m)(J\pm m+1)}\tilde{Y}_{Jm\pm 1q}, \n
&&L^{(q)}_3\tilde{Y}_{Jmq}=m\tilde{Y}_{Jmq}, \n
&&\vec{L}^{(q)2}\tilde{Y}_{Jmq}=J(J+1)\tilde{Y}_{Jmq}, \n
&&\int \frac{d\Omega_2}{4\pi} (\tilde{Y}_{Jmq})^*\tilde{Y}_{J'm'q}
=\delta_{JJ'}\delta_{mm'}, \n
&&(\tilde{Y}_{Jmq})^*=(-1)^{m-q}\tilde{Y}_{J-m-q}, \n
&&\int \frac{d\Omega_2}{4\pi} (\tilde{Y}_{J_1m_1q_1})^*\tilde{Y}_{J_2m_2q_2}
\tilde{Y}_{J_3m_3q_3}={\cal C}^{J_1m_1q_1}_{J_2m_2q_2 \; J_3m_3q_3} \;\;\;
\mbox{for} \;\;q_1=q_2+q_3,
\label{monopole harmonic function properties}
\eeqa
where ${\cal C}^{J_1m_1q_1}_{J_2m_2q_2 \; J_3m_3q_3}$ is the same as the
vertex coefficient defined in (\ref{vertexcoefficients}).
We emphasize that $J=|q|,|q|+1,|q|+2,\cdots$ and $q=0,\pm\frac{1}{2},
\pm 1,\pm\frac{3}{2},\cdots$.

The spin $S$ monopole harmonics is defined by
\beqa
\tilde{{\cal Y}}^{Sn}_{Jm,\tilde{J}q}
=C^{Jm}_{\tilde{J}p \; Sn}\tilde{Y}_{\tilde{J}pq}.
\label{spin S monopole spherical harmonics}
\eeqa
$\tilde{{\cal Y}}^{Sn}_{Jm,\tilde{J}q}$ possesses the properties
similar to the ones which ${\cal Y}^{Sn}_{Jm,\tilde{J}\tilde{m}}$ possesses 
with the identification $q=\tilde{m}$.
The counterparts of (\ref{orthonormality}) and (\ref{complexconjugate}) are
\beqa
&&\int \frac{d\Omega_2}{4\pi}
\:\sum_n(\tilde{{\cal Y}}^{Sn}_{J_1m_1,\tilde{J}_1q})^* 
\:\tilde{{\cal Y}}^{Sn}_{J_2m_2,\tilde{J}_2q}
=\delta_{J_1J_2}\delta_{\tilde{J}_1\tilde{J}_2}\delta_{m_1m_2},
\n
&&(\tilde{{\cal Y}}^{Sn}_{Jm,\tilde{J}q})^*
=(-1)^{-J+\tilde{J}-S+m-q+n} \:
 \tilde{{\cal Y}}^{S\:-n}_{J\:-m,\tilde{J}\:-q}.
\label{orthonormality and complex conjugate}
\eeqa
The counterpart of (\ref{linear action}) is 
\beqa
&&C^{1r}_{S'n' \; Sn}\check{L}^{(q)}_r 
\tilde{{\cal Y}}^{Sn}_{Jm,\tilde{J}q}
=(-1)^{-J-\tilde{J}+2S+n'+1}\sqrt{3\tilde{J}(\tilde{J}+1)(2\tilde{J}+1)}
\left\{\begin{array}{ccc}
       S         & S'        & 1 \\
       \tilde{J} & \tilde{J} & J 
       \end{array} \right\}
  \tilde{{\cal Y}}^{S'-n'}_{Jm,\tilde{J}q}, \n
\label{linear action 2}
\eeqa
where $\check{L}^{(q)}_{\pm}=\mp\frac{1}{\sqrt{2}}(L^{(q)}_1\pm iL^{(q)}_2), 
\;\;\;\check{L}^{(q)}_0=L^{(q)}_3$.
By comparing (\ref{another expression}) 
and (\ref{spin S monopole spherical harmonics})
and using the last identity in (\ref{monopole harmonic function properties}),
we can prove the counterpart of (\ref{integralofthreeharmonics})
in the same way:
\beqa
&&\int \frac{d\Omega_2}{4\pi} \: \sum_{n_1n_2n_3}
              (\tilde{{\cal Y}}^{S_1n_1}_{J_1m_1,\tilde{J}_1q_1})^* \: 
               \tilde{{\cal Y}}^{S_2n_2}_{J_2m_2,\tilde{J}_2q_2}
               \:\tilde{{\cal Y}}^{S_3n_3}_{J_3m_3,\tilde{J}_3q_3} \: 
               C^{S_1n_1}_{S_2n_2 \; S_3n_3} \nonumber\\
&&=\sqrt{(2S_1+1)(2J_2+1)(2\tilde{J}_2+1)(2J_3+1)(2\tilde{J}_3+1)}\:
   \left\{   \begin{array}{ccc}
             J_1 & \tilde{J}_1 & S_1 \\
             J_2 & \tilde{J}_2 & S_2 \\
             J_3 & \tilde{J}_3 & S_3 
             \end{array}     \right\} \:
             C^{J_1m_1}_{J_2m_2 \; J_3m_3}
             C^{\tilde{J}_1q_1}_{\tilde{J}_2q_2 \; \tilde{J}_3q_3},  \n
\label{integral of three harmonics 2}
\eeqa
where $q_1$ must be equal to $q_2+q_3$.

Here we make a remark. The similarity between the spherical harmonics
on $S^3$ and the monopole harmonics seen above can be understood 
through (\ref{another expression}), (\ref{spin S monopole spherical harmonics})
and the following equalities:
\beqa
&&Y_{Jm\tilde{m}}=(-1)^{J-m} \; \sqrt{2J+1}
\; d^{(J)}_{-m,\;\tilde{m}}(\theta)
e^{-i\tilde{m}(\psi-\pi/2)}e^{im(\phi+\pi/2)}, \n
&&\tilde{Y}_{Jmq}=\left\{
\begin{array}{ll}
(-1)^{J} \; \sqrt{2J+1} \;d^{(J)}_{-m,\; q}(\theta)
e^{i(q+m)\phi} & \mbox{in region I }\\
(-1)^{J} \; \sqrt{2J+1} \;d^{(J)}_{-m,\; q}(\theta)
e^{i(-q+m)\phi} & \mbox{in region I\hspace{-.1em}I}
\end{array}
\right.,
\eeqa
where 
\beqa 
\;d^{(J)}_{m,\; \tilde{m}}(\theta)
\equiv \langle  J m | \: e^{i\theta J_2} \: | J  \tilde{m}\rangle.
\eeqa

The monopole scalar harmonics, the monopole vector harmonics
and the monopole spinor harmonics are defined similarly: 
\beqa
&&\tilde{Y}_{Jmq}=\tilde{{\cal Y}}^{00}_{Jm,\tilde{J}q}, \n
&&\tilde{Y}^{\rho=1}_{Jmqi}=i\tilde{{\cal Y}}^{i}_{J+1\:m,Jq}, \;\;\;
\tilde{Y}^{\rho=-1}_{Jmqi}=-i\tilde{{\cal Y}}^{i}_{Jm,J+1\:q}, \;\;\;
\tilde{Y}^{\rho=0}_{Jmqi}=\tilde{{\cal Y}}^{i}_{Jm,Jq}, \n
&&\tilde{Y}^{\kappa=1}_{Jmq\alpha}
=\tilde{{\cal Y}}^{S=\frac{1}{2},\alpha}_{J+\frac{1}{2}\:m,Jq}, \;\;\;
\tilde{Y}^{\kappa=-1}_{Jmq\alpha}
=\tilde{{\cal Y}}^{S=\frac{1}{2},\alpha}_{Jm,J+\frac{1}{2}\:q},
\eeqa
where $\tilde{{\cal Y}}^{i}_{Jm,\tilde{J}q}$ is an analogue of
${\cal Y}^{i}_{Jm,\tilde{J}\tilde{m}}$ and defined in terms of 
$\tilde{{\cal Y}}^{1n}_{Jm,\tilde{J}q}$'s as in (\ref{change of basis}).
These harmonics are also orthonormal:
\beqa
&&\int \frac{d\Omega_2}{4\pi} 
\:(\tilde{Y}_{J_1m_1q})^* \tilde{Y}_{J_2m_2q}
=\delta_{J_1J_2}\delta_{m_1m_2}, \n
&&\int \frac{d\Omega_2}{4\pi} 
\:(\tilde{Y}^{\rho_1}_{J_1m_1qi})^* 
\tilde{Y}^{\rho_2}_{J_2m_2qi}
=\delta_{\rho_1\rho_2}\delta_{J_1J_2}\delta_{m_1m_2}, \n
&&\int \frac{d\Omega_2}{4\pi} 
\:(\tilde{Y}^{\kappa_1}_{J_1m_1q\alpha})^* 
\tilde{Y}^{\kappa_2}_{J_2m_2q\alpha}
=\delta_{\kappa_1\kappa_2}\delta_{J_1J_2}\delta_{m_1m_2}.
\label{orthnormality3}
\eeqa
Their complex conjugates are analogous to those of the spherical harmonics
on $S^3$:
\beqa
&&(\tilde{Y}_{Jmq})^*=(-1)^{m-q}\tilde{Y}_{J-m-q}, \;\;\;
(\tilde{Y}^{\rho}_{Jmqi})^*=(-1)^{m-q+1}\tilde{Y}^{\rho}_{J-m-qi}, \n
&&(\tilde{Y}^{\kappa}_{Jmq\alpha})^*=(-1)^{m-q+\kappa\alpha+1}
                                   \tilde{Y}^{\kappa}_{J-m-q-\alpha}.
\label{complexconjugate3}
\eeqa 
Using the formula (\ref{linear action 2}) yields the identities analogous
to (\ref{identities}):
\beqa
&&\vec{L}^{(q)}\tilde{Y}_{Jmq}=\sqrt{J(J+1)}\vec{\tilde{Y}}{}^{0}_{Jmq},\n
&&\vec{L}^{(q)}\cdot\vec{\tilde{Y}}{}^{\rho}_{Jmq}
=\sqrt{J(J+1)}\delta_{\rho 0}\tilde{Y}_{Jmq}, \n
&&i\vec{L}^{(q)}\times \vec{\tilde{Y}}{}^{\rho}_{Jmq}
+\vec{\tilde{Y}}{}^{\rho}_{Jmq}
=\rho (J+1)\vec{\tilde{Y}}{}^{\rho}_{Jmq}, \n
&&\left(\vec{\sigma}\cdot\vec{L}^{(q)}
+\frac{3}{4}\right)\tilde{Y}^{\kappa}_{Jmq}
=\kappa (J+\frac{3}{4})\tilde{Y}^{\kappa}_{Jmq}.
\label{identities2}
\eeqa
It follows from 
(\ref{integralofthreeharmonics}) and (\ref{integral of three harmonics 2})
that the integrals of various three monopole harmonics are equal to the 
corresponding integrals on $S^3$ (vertex coefficients) with the identification
$q=\tilde{m}$. Namely, the following identies hold.
\beqa
&&\int \frac{d\Omega_2}{4\pi} \:(\tilde{Y}_{J_1m_1q_1})^*
\tilde{Y}_{J_2m_2q_2}
\tilde{Y}_{J_3m_3q_3}
={\cal C}^{J_1m_1q_1}_{J_2m_2q_2\;J_3m_3q_3}. \n
&&\int \frac{d\Omega_2}{4\pi}\: (\tilde{Y}_{Jmq})^*
\tilde{Y}^{\rho_1}_{J_1m_1q_1i}
\tilde{Y}^{\rho_2}_{J_2m_2q_2i}
={\cal D}^{Jmq}_{J_1m_1q_1\rho_1\;J_2m_2q_2\rho_2}. \n
&&\int \frac{d\Omega_2}{4\pi} \: \epsilon_{ijk}\: 
\tilde{Y}^{\rho_1}_{J_1m_1q_1i} 
\tilde{Y}^{\rho_2}_{J_2m_2q_2j} 
\tilde{Y}^{\rho_3}_{J_3m_3q_3k}
={\cal E}_{J_1m_1q_1\rho_1\;J_2m_2q_2\rho_2\;J_3m_3q_3\rho_3}. \n
&&\int \frac{d\Omega_2}{4\pi} \:
           (\tilde{Y}^{\kappa_1}_{J_1m_1q_1\alpha})^*
           \tilde{Y}^{\kappa_2}_{J_2m_2q_2\alpha}
           \tilde{Y}_{Jmq}
={\cal F}^{J_1m_1q_1\kappa_1}_{J_2m_2q_2\kappa_2\;Jmq}. \n
&&\int \frac{d\Omega_2}{4\pi} \: 
         (\tilde{Y}^{\kappa_1}_{J_1m_1q_1\alpha})^*
          \sigma^i_{\alpha\beta}
          \tilde{Y}^{\kappa_2}_{J_2m_2q_2\beta}
          \tilde{Y}^{\rho}_{Jmqi}
={\cal G}^{J_1m_1q_1\kappa_1}_{J_2m_2q_2\kappa_2\;Jmq\rho},
\label{vertex coefficients 2}
\eeqa
where the monopoles charges must be conserved
as in the last equality in (\ref{monopole harmonic function properties}).

\subsection{Fuzzy sphere harmonics}
Let us consider the set of linear maps from a $(2j'+1)$-dimensional 
complex vector space $V_{j'}$ 
to a $(2j+1)$-dimensional complex vector space $V_{j}$, where
$j$ and $j'$ are non-negative half-integers. 
We denote the set by ${\cal M}_{jj'}$.
${\cal M}_{jj'}$ is identified with the set of 
$(2j+1)\times (2j'+1)$ rectangular complex matrices and is 
a $((2j+1)\times (2j'+1))$-dimensional complex vector space.
It is convenient for us to consider the basis of the spin $j$ and $j'$
representations of $SU(2)$
as a basis of $V_{j}$ and $V_{j'}$, respectively, and to 
construct a basis of ${\cal M}_{jj'}$ as 
\beqa
|jr\rangle\langle j'r'|, \;\;\;
(r=-j,-j+1,\cdots,j-1,j; \; r'=-j',-j'+1,\cdots,j'-1,j').
\eeqa
Then, an arbitrary element of ${\cal M}_{jj'}$, $M$, is expressed as
\beqa
M=\sum_{r,r'}M_{rr'}\:|jr\rangle\langle j'r'|.
\eeqa
One can define linear maps from ${\cal M}_{jj'}$ to ${\cal M}_{jj'}$ 
by its operation on the basis:
\beqa
L_i \circ |jr\rangle\langle j'r'|
=L_i|jr\rangle\langle j'r'|-|jr\rangle\langle j'r'|L_i,
\eeqa
where $L_i$ is a generator of $SU(2)$.
The matrix element $M_{rr'}$ is transformed under these maps as
\beqa
(L_i\circ M)_{rr'}=(L_i^{[j]})_{rp}M_{pr'}-M_{rp'}(L_i^{[j']})_{p'r'},
\eeqa
where $L_i^{[j]}$ is the $(2j+1)\times (2j+1)$ representation matrix of
the spin $j$ representation of $SU(2)$.
These maps form a $((2j+1)\times (2j'+1))$-dimensional representation of 
$SU(2)$, which is in general reducible, because the following identity holds:
\beqa
(L_i\circ L_j \circ -L_j \circ L_i\circ)|jr\rangle\langle j'r'|
=i\epsilon_{ijk}L_k\circ |jr\rangle\langle j'r'|.
\eeqa

For later convenience, we introduce a positive integer constant, $N_0$,
and reparameterize the dimensions of $V_{j}$ and $V_{j'}$ as
\beqa
2j+1=N_0 + \zeta, \;\;\; 2j'+1=N_0 + \zeta',
\eeqa
where $\zeta$ and $\zeta'$ are integers which are greater than $-N_0$.
We will take the $N_0\rightarrow \infty$ limit shortly.
It will turn out that the fuzzy sphere harmonics defined below are identified
with the monopole harmonics in this limit.
We make a change of basis from the above basis to a new basis, 
\beqa
\hat{Y}^{(jj')}_{Jm}
=\sqrt{N_0}
\sum_{r,r'}(-1)^{-j+r'}C^{Jm}_{jr \; j'-r'}
|jr\rangle\langle j'r'|,
\eeqa
where $J$ takes $|j-j'|, |j-j'|+1,\cdots,j+j'$ and
$m$ takes $-J, -J+1, \cdots, J-1,J$.
In other words, $J$ takes $\frac{1}{2}|\zeta-\zeta'|, 
\frac{1}{2}|\zeta-\zeta'|+1, \cdots,
\frac{1}{2}(\zeta+\zeta')+N_0-1$. 
$N_0$ plays a role of an ultraviolet cut-off for the angular momentum.
For a fixed $J$, $\hat{Y}^{(jj')}_{Jm}$ is the basis of 
the spin $J$ irreducible
representation of $SU(2)$. Namely, using (\ref{relation between CG}), 
one can show
\beqa
&&L_{\pm}\circ \hat{Y}^{(jj')}_{Jm}
=\sqrt{(J\mp m)(J\pm m+1)}\hat{Y}^{(jj')}_{Jm\pm 1}, \n
&&L_3\circ \hat{Y}^{(jj')}_{Jm}=m \hat{Y}^{(jj')}_{Jm}.
\label{Yhat 1}
\eeqa
These relations also imply
\beqa
L_i\circ L_i\circ \hat{Y}^{(jj')}_{Jm}=J(J+1)\hat{Y}^{(jj')}_{Jm}.
\label{Yhat 2}
\eeqa
$\hat{Y}^{(jj')}_{Jm}$ satisfies the orthonormality condition
under the following normalized trace:
\beqa
\frac{1}{N_0}\tr (\hat{Y}^{(jj')\dagger}_{J_1m_1}\hat{Y}^{(jj')}_{J_2m_2})
=\delta_{J_1J_2}\delta_{m_1m_2},
\label{Yhat 3}
\eeqa
where $\tr$ stands for the trace over $(2j'+1)\times (2j'+1)$ matrices.
The hermitian conjugate of $\hat{Y}^{(jj')}_{Jm}$ is evaluated as
\beqa
\hat{Y}^{(jj')\dagger}_{Jm}=(-1)^{m-(j-j')}\hat{Y}^{(j'j)}_{J-m}.
\label{Yhat 4}
\eeqa
Using (\ref{6j}) yields
\beqa
&&\frac{1}{N_0}\tr(\hat{Y}^{(j'j)\dagger}_{J_1m_1}\hat{Y}^{(j'j'')}_{J_2m_2}
\hat{Y}^{(j''j)}_{J_3m_3})  \n
&&=(-1)^{J_1+j+j'}
\sqrt{N_0(2J_2+1)(2J_3+1)}C^{J_1m_1}_{J_2m_2\;J_3m_3}
\left\{\begin{array}{ccc}
J_1 & J_2 & J_3 \\
j'' & j   & j'  
\end{array} \right\}.
\label{Yhat 5}
\eeqa
One can see from (\ref{asymtotic relation}) that 
in the $N_0 \rightarrow \infty$ limit this equality reduces to
\beqa
\frac{1}{N_0}\tr(\hat{Y}^{(j'j)\dagger}_{J_1m_1}\hat{Y}^{(j'j'')}_{J_2m_2}
\hat{Y}^{(j''j)}_{J_3m_3}) 
=\sqrt{\frac{(2J_2+1)(2J_3+1)}{2J_1+1}}C^{J_1m_1}_{J_2m_2\;J_3m_3}
C^{J_1(j'-j)}_{J_2(j'-j'')\;J_3(j''-j)}.
\label{Yhat 6}
\eeqa
Comparing the relations (\ref{Yhat 1}), (\ref{Yhat 2}), (\ref{Yhat 3}), 
(\ref{Yhat 4}) and (\ref{Yhat 6}) with the relations in 
(\ref{monopole harmonic function properties}), one can see that 
$\hat{Y}^{(jj')}_{Jm}$ is identified with $\tilde{Y}_{Jmq}$ 
in the $N_0\rightarrow\infty$ limit through the following correspondence:
\beqa
j-j'                 &\leftrightarrow &q \n
L_i\circ             &\leftrightarrow & L_i^{(q)} \n
\frac{1}{N_0}\tr     &\leftrightarrow & \int \frac{d\Omega_2}{4\pi}.
\label{correspondence}
\eeqa
In this limit, the lower bound of $J$ in $\hat{Y}_{Jm}^{(jj')}$, $|j-j'|$,
remains
finite and indeed corresponds to the monopole charge $q$ while the 
upper bound of $J$ goes to infinity, namely, the ultraviolet cut-off is 
removed.

The analogue of (\ref{spin S monopole spherical harmonics}) is defined by
\beqa
\hat{{\cal Y}}^{Sn}_{Jm,\tilde{J}(jj')}
=C^{Jm}_{\tilde{J}p \; Sn}\hat{Y}_{\tilde{J}p}^{(jj')},
\label{spin S fuzzy sphere harmonics}
\eeqa
which we call the spin $S$ fuzzy sphere harmonics.
$\hat{{\cal Y}}^{Sn}_{Jm,\tilde{J}(jj')}$ shares all the properties except
the integral of the product of three harmonics with
$\tilde{{\cal Y}}^{Sn}_{Jm,\tilde{J}q}$ under the correspondence
(\ref{correspondence}). In the $N_0\rightarrow\infty$ limit, 
the trace of the product of three fuzzy sphere harmonics also coincides with
the integral of the product of three monopole harmonics. 
The spin $S$ fuzzy sphere harmonics
is, therefore, considered as a matrix regularization 
of the spin $S$ monopole harmonics.
The counterparts of (\ref{orthonormality and complex conjugate}) are
\beqa
&&\sum_n\frac{1}{N_0}\tr(\hat{{\cal Y}}^{Sn\dagger}_{J_1m_1,\tilde{J}_1(jj')} 
\tilde{{\cal Y}}^{Sn}_{J_2m_2,\tilde{J}_2(jj')})
=\delta_{J_1J_2}\delta_{\tilde{J}_1\tilde{J}_2}\delta_{m_1m_2},
\n
&&\hat{{\cal Y}}^{Sn\dagger}_{Jm,\tilde{J}(jj')}
=(-1)^{-J+\tilde{J}-S+m-(j-j')+n} \:
 \hat{{\cal Y}}^{S\:-n}_{J\:-m,\tilde{J}\:(j'j)}.
\label{orthonormality and complex conjugate 2}
\eeqa
The counterpart of (\ref{linear action 2}) is 
\beqa
&&C^{1r}_{S'n' \; Sn}\check{L}_r \circ
\hat{{\cal Y}}^{Sn}_{Jm,\tilde{J}(jj')}
=(-1)^{-J-\tilde{J}+2S+n'+1}\sqrt{3\tilde{J}(\tilde{J}+1)(2\tilde{J}+1)}
\left\{\begin{array}{ccc}
       S         & S'        & 1 \\
       \tilde{J} & \tilde{J} & J 
       \end{array} \right\}
  \hat{{\cal Y}}^{S'-n'}_{Jm,\tilde{J}(jj')}, \n
\label{linear action 3}
\eeqa
where $\check{L}_{\pm}\circ=\mp\frac{1}{\sqrt{2}}(L_1\pm iL_2)\circ, 
\;\;\check{L}_0\circ=L_3\circ$.
Using (\ref{Yhat 5}) and (\ref{9j}), it is easy to prove the following formula,
which is the counterpart of (\ref{integral of three harmonics 2}),
\beqa
&&\sum_{n_1n_2n_3}\frac{1}{N_0}\tr
(\hat{{\cal Y}}^{S_1n_1\dagger}_{J_1m_1,\tilde{J}_1(j'j)} 
\hat{{\cal Y}}^{S_2n_2}_{J_2m_2,\tilde{J}_2(j'j'')}
\hat{{\cal Y}}^{S_3n_3}_{J_3m_3,\tilde{J}_3(j''j)}) \: 
C^{S_1n_1}_{S_2n_2 \; S_3n_3} \nonumber\\
&&=(-1)^{\tilde{J}_1+j+j'}
\sqrt{N_0(2S_1+1)(2\tilde{J}_1+1)(2J_2+1)(2\tilde{J}_2+1)
(2J_3+1)(2\tilde{J}_3+1)} \n
&&\;\;\;\;\;\;\;\times\left\{   \begin{array}{ccc}
             J_1 & \tilde{J}_1 & S_1 \\
             J_2 & \tilde{J}_2 & S_2 \\
             J_3 & \tilde{J}_3 & S_3 
             \end{array}     \right\} \:
             C^{J_1m_1}_{J_2m_2 \; J_3m_3}
   \left\{\begin{array}{ccc}
    \tilde{J}_1 & \tilde{J}_2 & \tilde{J}_3 \\
    j''         & j           & j'  
\end{array} \right\}.  
\label{integral of three harmonics 3}
\eeqa
One can see from (\ref{asymtotic relation}) that 
in the $N_0\rightarrow\infty$ limit, this formula reduces to
\newpage
\begin{align}
\hspace{-0.5cm}
&\sum_{n_1n_2n_3}\frac{1}{N_0}\tr
(\hat{{\cal Y}}^{S_1n_1\dagger}_{J_1m_1,\tilde{J}_1(j'j)} 
\hat{{\cal Y}}^{S_2n_2}_{J_2m_2,\tilde{J}_2(j'j'')}
\hat{{\cal Y}}^{S_3n_3}_{J_3m_3,\tilde{J}_3(j''j)})\: 
C^{S_1n_1}_{S_2n_2 \; S_3n_3} \nonumber\\
&=\sqrt{(2S_1+1)(2J_2+1)(2\tilde{J}_2+1)(2J_3+1)(2\tilde{J}_3+1)}\:
   \left\{   \begin{array}{ccc}
             J_1 & \tilde{J}_1 & S_1 \\
             J_2 & \tilde{J}_2 & S_2 \\
             J_3 & \tilde{J}_3 & S_3 
             \end{array}     \right\} \:
             C^{J_1m_1}_{J_2m_2 \; J_3m_3}
             C^{\tilde{J}_1j'-j}_{\tilde{J}_2j'-j'' \; \tilde{J}_3j''-j}, \n
\label{integral of three harmonics 4}
\end{align}
which is equivalent to (\ref{integral of three harmonics 2}) with 
the identification $j-j'=q$, as anticipated.

The fuzzy sphere scalar harmonics, the fuzzy sphere vector  harmonics
and the fuzzy sphere spinor harmonics are defined similarly: 
\beqa
&&\hat{Y}_{Jm(jj')}=\hat{{\cal Y}}^{00}_{Jm,\tilde{J}(jj')}
=\hat{Y}_{Jm}^{(jj')}, \n
&&\hat{Y}^{\rho=1}_{Jm(jj')i}=i\hat{{\cal Y}}^{i}_{J+1\:m,J(jj')}, \;\;\;
\hat{Y}^{\rho=-1}_{Jm(jj')i}=-i\hat{{\cal Y}}^{i}_{Jm,J+1\:(jj')}, \;\;\;
\hat{Y}^{\rho=0}_{Jm(jj')i}=\hat{{\cal Y}}^{i}_{Jm,J(jj')}, \n
&&\hat{Y}^{\kappa=1}_{Jm(jj')\alpha}
=\hat{{\cal Y}}^{S=\frac{1}{2},\alpha}_{J+\frac{1}{2}\:m,J(jj')}, \;\;\;
\hat{Y}^{\kappa=-1}_{Jm(jj')\alpha}
=\hat{{\cal Y}}^{S=\frac{1}{2},\alpha}_{Jm,J+\frac{1}{2}\:(jj')},
\eeqa
where $\hat{{\cal Y}}^{i}_{Jm,\tilde{J}(jj')}$ is 
an analogue of
$\tilde{{\cal Y}}^{i}_{Jm,\tilde{J}q}$ and is expressed in terms of 
$\hat{{\cal Y}}_{Jm,\tilde{J}(jj')}^{1n}$'s.
These harmonics are also orthonormal:
\beqa
&&\frac{1}{N_0}\tr(\hat{Y}_{J_1m_1(jj')}^{\dagger} \hat{Y}_{J_2m_2(jj')})
=\delta_{J_1J_2}\delta_{m_1m_2}, \n
&&\frac{1}{N_0}\tr(\hat{Y}^{\rho_1\dagger}_{J_1m_1(jj')i} 
\hat{Y}^{\rho_2}_{J_2m_2(jj')i})
=\delta_{\rho_1\rho_2}\delta_{J_1J_2}\delta_{m_1m_2}, \n
&&\frac{1}{N_0}\tr(\hat{Y}^{\kappa_1\dagger}_{J_1m_1(jj')\alpha}
\hat{Y}^{\kappa_2}_{J_2m_2(jj')\alpha})
=\delta_{\kappa_1\kappa_2}\delta_{J_1J_2}\delta_{m_1m_2}.
\label{orthnormality4}
\eeqa
Their hermitian conjugates are analogous to the complex conjugates
of the monopole  harmonics:
\beqa
&&\hat{Y}_{Jm(jj')}^{\dagger}=(-1)^{m-(j-j')}\hat{Y}_{J-m(j'j)}, \n
&&\hat{Y}^{\rho\dagger}_{Jm(jj')i}
=(-1)^{m-(j-j')+1}\hat{Y}^{\rho}_{J-m(j'j)i}, \n
&&\hat{Y}^{\kappa\dagger}_{Jm(jj')\alpha}
=(-1)^{m-(j-j')+\kappa\alpha+1}\hat{Y}^{\kappa}_{J-m(j'j)-\alpha}.
\label{complexconjugate4}
\eeqa 
Using the formula (\ref{linear action 3}) yields the identities analogous
to (\ref{identities}):
\beqa
&&\vec{L}\circ\hat{Y}_{Jm(jj')}=\sqrt{J(J+1)}\vec{\hat{Y}}{}^{0}_{Jm(jj')},\n
&&\vec{L}\circ\cdot\vec{\hat{Y}}{}^{\rho}_{Jm(jj')}
=\sqrt{J(J+1)}\delta_{\rho 0}\hat{Y}_{Jm(jj')}, \n
&&i\vec{L}\circ\times \vec{\hat{Y}}{}^{\rho}_{Jm(jj')}
+\vec{\hat{Y}}{}^{\rho}_{Jm(jj')}
=\rho (J+1)\vec{\hat{Y}}{}^{\rho}_{Jm(jj')}, \n
&&\left(\vec{\sigma}\cdot\vec{L}\circ
+\frac{3}{4}\right)\hat{Y}^{\kappa}_{Jm(jj')}
=\kappa (J+\frac{3}{4})\hat{Y}^{\kappa}_{Jm(jj')}.
\label{identities3}
\eeqa
We define the traces of various three fuzzy sphere harmonics,
which are analogous to the vertex coefficients:
\beqa
&&\hat{{\cal C}}^{J_1m_1(j'j)}_{J_2m_2(j'j'')\;J_3m_3(j''j)}
\equiv\frac{1}{N_0}\tr(\hat{Y}_{J_1m_1(j'j)}^{\dagger}
\hat{Y}_{J_2m_2(j'j'')}
\hat{Y}_{J_3m_3(j''j)}). \n
&&\hat{{\cal D}}^{Jm(j'j)}_{J_1m_1(j'j'')\rho_1\;J_2m_2(j''j)\rho_2}
\equiv\frac{1}{N_0}\tr(\hat{Y}_{Jm(j'j)}^{\dagger}
\hat{Y}^{\rho_1}_{J_1m_1(j'j'')i}
\hat{Y}^{\rho_2}_{J_2m_2(j''j)i}). \n
&&\hat{{\cal E}}_{J_1m_1(jj')\rho_1\;J_2m_2(j'j'')\rho_2\;J_3m_3(j''j)\rho_3}
\equiv\epsilon_{ijk}\frac{1}{N_0}\tr
(\hat{Y}^{\rho_1}_{J_1m_1(jj')i} 
\hat{Y}^{\rho_2}_{J_2m_2(j'j'')j} 
\hat{Y}^{\rho_3}_{J_3m_3(j''j)k}). \n
&&\hat{{\cal F}}^{J_1m_1(j'j)\kappa_1}_{J_2m_2(j'j'')\kappa_2\;Jm(j''j)}
\equiv\frac{1}{N_0}\tr
(\hat{Y}^{\kappa_1\dagger}_{J_1m_1(j'j)\alpha}
\hat{Y}^{\kappa_2}_{J_2m_2(j'j'')\alpha}
\hat{Y}_{Jm(j''j)}). \n
&&\hat{{\cal G}}^{J_1m_1(j'j)\kappa_1}_{J_2m_2(j'j'')\kappa_2\;Jm(j''j)\rho}
\equiv\frac{1}{N_0}\tr
(\hat{Y}^{\kappa_1\dagger}_{J_1m_1(j'j)\alpha}
\sigma^i_{\alpha\beta}
\hat{Y}^{\kappa_2}_{J_2m_2(j'j'')\beta}
\hat{Y}^{\rho}_{Jm(j''j)i}). 
\label{hat vertex coefficients}
\eeqa
These can be evaluated using (\ref{integral of three harmonics 3}) and
the explicit expression are given in appendix F.
We see from (\ref{integral of three harmonics 4}) that these reduce to
the corresponding quantities without the hat, namely
the vertex coefficients, with the identification
$j-j'=q$ in the $N_0\rightarrow\infty$ limit.

\section{$2+1$ SYM on $R\times S^2$ vs the plane wave matrix model}
\setcounter{equation}{0}
\subsection{Embedding of $\mbox{SYM}_{R\times S^2}$ into PWMM}
In this subsection, we prove the prediction 1). Namely, we show that
in the $N_0\rightarrow 0$ limit
the theory around the vacuum (\ref{vacuum for PWMM}) in PWMM is equivalent
to the one around the vacuum (\ref{vacuum for S^2}) with the 
identification 
\beqa
j_s-j_t=\frac{1}{2}(\alpha_s-\alpha_t)
\label{j_s-j_t}
\eeqa
and the relation between the coupling constants 
in (\ref{coupling relation between S^2 and PWMM}).

We expand the action (\ref{S^2SU(4)form}) around the background
\beqa
\hat{\vec{Y}}=\vec{e}_r\hat{\Phi}+\vec{e}_{\phi}\hat{A}_1
-\vec{e}_{\theta}\hat{A}_2.
\eeqa 
We make a substitution $\vec{Y}\rightarrow \hat{\vec{Y}}+\vec{Y}$ in
(\ref{S^2SU(4)form}).
The terms including $\vec{Y}$ in (\ref{S^2SU(4)form})
are evaluated as
\beqa
 &&(\vec{\cL}X_{AB})^{(s,t)}\rightarrow \mu\vec{L}^{(q_{st})}X_{AB}^{(s,t)}
 -[\vec{Y},X_{AB}]^{(s,t)}, \n
  &&\vec{\cZ}^{(s,t)}\rightarrow \mu\vec{Y}^{(s,t)}
  +i\mu\vec{L}^{(q_{st})}\times \vec{Y}^{(s,t)}
  -i(\vec{Y}\times\vec{Y})^{(s,t)}, \n
  &&(D_0\vec{Y}-i\mu\vec{L}^{(0)}A_0)^{(s,t)}\rightarrow
  (D_0\vec{Y})^{(s,t)}-i\mu\vec{L}^{(q_{st})}A_0^{(s,t)},
\label{terms including vecY}
\eeqa
where the suffix $(s,t)$ stands for the $(s,t)$ block of an 
$\tilde{N}\times\tilde{N}$ matrix, 
which is an $N_s\times N_t$ rectangular matrix, and $s,t$ run from 1 to $T$.
The monopole charge $q_{st}$ is given by
\beqa
q_{st}=\frac{1}{2}(\alpha_s-\alpha_t).
\eeqa
By using (\ref{terms including vecY}), we obtain the theory around
the vacuum (\ref{vacuum for S^2}):
\begin{align}
&S_{R\times S^2}=S_{R\times S^2}^{free}+S_{R\times S^2}^{int}, \n
&S_{R\times S^2}^{free}=\frac{1}{g_{R\times S^2}^2}
\int dt\frac{d\Omega_2}{\mu^2}\sum_{s,t}\tr\Biggl(
\frac{1}{2}\partial_0X^{AB(t,s)}\partial_0X_{AB}^{(s,t)} \n
&\hspace{3cm}+\frac{\mu^2}{2}\vec{L}^{(q_{ts})}X^{AB(t,s)}
\cdot\vec{L}^{(q_{st})}X_{AB}^{(s,t)}
-\frac{\mu^2}{8}X^{AB(t,s)}X_{AB}^{(s,t)} \n
& \hspace{3cm}
+\frac{1}{2}
\partial_0\vec{Y}^{(t,s)}\cdot\partial_0\vec{Y}^{(s,t)}
-\frac{1}{2}(i\mu\vec{L}^{(q_{ts})}\times \vec{Y}^{(t,s)}+\mu\vec{Y}^{(t,s)})
\cdot(i\mu\vec{L}^{(q_{st})}\times \vec{Y}^{(s,t)}+\mu\vec{Y}^{(s,t)}) \n
& \hspace{3cm}
-\frac{\mu^2}{2}\vec{L}^{(q_{ts})}A_0^{(t,s)}\cdot\vec{L}^{(q_{st})}A_0^{(s,t)}
-i\mu\partial_0\vec{Y}^{(t,s)}\cdot\vec{L}^{(q_{st})}A_0^{(s,t)} \n
& \hspace{3cm}
+i\psi_A^{\dagger(t,s)}\partial_0\psi^{A(s,t)} 
-\mu\psi_A^{\dagger(t,s)}\vec{\sigma}\cdot\vec{L}^{(q_{st})}\psi^{A(s,t)}
-\frac{3\mu}{4}\psi_A^{\dagger(t,s)}\psi^{A(s,t)}
\Biggr), \n
&S_{R\times S^2}^{int}
=\frac{1}{g_{R\times S^2}^2}\int dt\frac{d\Omega_2}{\mu^2}\sum_{s,t}\tr
\Biggl(
-i\partial_0X_{AB}^{(t,s)}[A_0,X^{AB}]^{(s,t)}
-\frac{1}{2}[A_0,X_{AB}]^{(t,s)}[A_0,X^{AB}]^{(s,t)} \n
& \hspace{3cm}
-\mu\vec{L}^{(q_{ts})}X_{AB}^{(t,s)}\cdot[\vec{Y},X^{AB}]^{(s,t)}
+\frac{1}{2}[\vec{Y},X_{AB}]^{(t,s)}\cdot[\vec{Y},X^{AB}]^{(s,t)} \n
& \hspace{3cm}
+\frac{1}{4}[X_{AB},X_{CD}]^{(t,s)}[X^{AB},X^{CD}]^{(s,t)}
-\frac{1}{2}[\vec{Y},A_0]^{(t,s)}\cdot[\vec{Y},A_0]^{(s,t)} \n
& \hspace{3cm}
-i\partial_0\vec{Y}^{(t,s)}\cdot[A_0,\vec{Y}]^{(s,t)}
-\mu[A_0,\vec{Y}]^{(t,s)}\cdot\vec{L}^{(q_{st})}A_0^{(s,t)} \n
& \hspace{3cm}
+i(i\mu\vec{L}^{(q_{ts})}\times\vec{Y}^{(t,s)}+\mu\vec{Y}^{(t,s)})
\cdot(\vec{Y}\times\vec{Y})^{(s,t)}
+\frac{1}{2}(\vec{Y}\times\vec{Y})^{(t,s)}\cdot(\vec{Y}\times\vec{Y})^{(s,t)}
\n 
& \hspace{3cm}
+\psi_A^{\dagger(t,s)}[A_0,\psi^A]^{(s,t)}
+\psi_A^{\dagger(t,s)}\vec{\sigma}\cdot[\vec{Y},\psi^A]^{(s,t)} \n
& \hspace{3cm}
-\psi^{AT(t,s)}\sigma^2[X_{AB},\psi^B]^{(s,t)}
+\psi_A^{\dagger(t,s)}\sigma^2[X^{AB},\psi_B^*]^{(s,t)}
\Biggr), 
\label{S^2 action around monopole vacua}
\end{align}
where tr should be understood as the trace over  
square matrices with a certain size which are  
the products of some rectangular matrices.

Moreover, we make the mode expansion for the fields in terms of 
the monopole harmonics as 
\begin{align}
 A_0^{(s,t)}
  &=\sum_{J\geq|q_{st}|}\sum_{m=-J}^{J}b^{(s,t)}_{Jm}\tilde{Y}_{Jmq_{st}}, 
 \qquad
 X_{AB}^{(s,t)}
  =\sum_{J\geq|q_{st}|}\sum_{m=-J}^{J}x^{(s,t)}_{ABJm}\tilde{Y}_{Jmq_{st}},
  \n
 \psi^{A(s,t)}
  &=\sum_{\kappa=\pm1}\sum_{\tilde{U}\geq|q_{st}|}\sum_{m=-U}^{U}
 \psi_{Jm\kappa}^{A(s,t)}\tilde{Y}_{Jmq_{st}}^\kappa  \n
  &=\sum_{J\geq|q_{st}|}\sum_{m=-J-\frac{1}{2}}^{J+\frac{1}{2}}
  \psi_{Jm1}^{A(s,t)}\tilde{Y}_{Jmq_{st}}^1 
  +\sum_{J\geq|q_{st}|-\frac{1}{2}}\sum_{m=-J}^{J}
  \psi_{Jm-1}^{A(s,t)}\tilde{Y}_{Jmq_{st}}^{-1},  \n
 \vec{Y}^{(s,t)}
 &=\sum_{\rho=-1}^{1}\sum_{\tilde{Q}\geq|q_{st}|}\sum_{m=-Q}^{Q}
 y_{Jm\rho}^{(s,t)}\vec{\tilde{Y}}{}_{Jmq_{st}}^\rho,  \n
 &=\sum_{J\geq|q_{st}|}\sum_{m=-J-1}^{J+1}
  y_{Jm1}^{(s,t)}\vec{\tilde{Y}}{}_{Jmq_{st}}^1 
  +\sum_{J\geq|q_{st}|}\sum_{m=-J}^{J}
  y_{Jm0}^{(s,t)}\vec{\tilde{Y}}{}_{Jmq_{st}}^0
  +\sum_{J\geq|q_{st}|-1}\sum_{m=-J}^{J}
 y_{Jm-1}^{(s,t)}\vec{\tilde{Y}}{}_{Jmq_{st}}^{-1},
\label{mode expansion in 2+1 SYM}
\end{align}
where 
$U\equiv J+\frac{1+\kappa}{4}, \tilde{U}\equiv J+\frac{1-\kappa}{4},
Q\equiv J+\frac{(1+\rho)\rho}{2}$ and 
$\tilde{Q}\equiv J-\frac{(1-\rho)\rho}{2}$. Due to (\ref{complexconjugate3}), 
the conditions
$A_0^{(s,t)\dagger}=A_0^{(t,s)}, X_{AB}^{(s,t)\dagger}=X^{AB(t,s)}$
and $\vec{Y}^{(s,t)\dagger}=\vec{Y}^{(t,s)}$ imply
\begin{align}
 b_{Jm}^{(s,t)\dagger}&=(-1)^{m-q_{st}}b_{J-m}^{(t,s)},
  \qquad x_{ABJm}^{(s,t)\dagger}=(-1)^{m-q_{st}}x_{J-m}^{AB(t,s)}, \n
  y_{Jm\rho}^{(s,t)\dagger}&=(-1)^{m-q_{st}+1}y_{J-m\rho}^{(t,s)}.
\label{hermitian conjugate of mode in 2+1 SYM}
\end{align}
By substituting (\ref{mode expansion in 2+1 SYM})
into (\ref{S^2 action around monopole vacua}) and 
using (\ref{orthnormality3}), (\ref{identities2}) and 
(\ref{vertex coefficients 2}),
we obtain the mode-expanded form of the theory:
\begin{align}
&S_{R\times S^2}^{free}
=\frac{4\pi}{g_{R\times S^2}^2}\int \frac{dt}{\mu^2}
\mbox{tr} \left[
\frac{1}{2}\partial_0x^{(s,t)\dagger}_{AB\omega}\partial_0x^{(s,t)}_{AB\omega}
-\frac{\mu^2}{2}\left(J+\frac{1}{2}\right)^2
x^{(s,t)\dagger}_{AB\omega}x^{(s,t)}_{AB\omega} \right.
\n
&\hspace{4cm}+\frac{1}{2}\partial_0y^{(s,t)\dagger}_{\omega\rho}
\partial_0y^{(s,t)}_{\omega\rho}
-\frac{\mu^2}{2}\rho^2\left(J+1\right)^2
y^{(s,t)\dagger}_{\omega\rho}y^{(s,t)}_{\omega\rho}
\n
&\hspace{4cm}+\frac{\mu^2}{2}J(J+1)b^{(s,t)\dagger}_{\omega}b^{(s,t)}_{\omega}
-i\mu\sqrt{J(J+1)}\partial_0y^{(s,t)\dagger}_{\omega 0}b^{(s,t)}_{\omega}
\n
&\hspace{4cm}\left.+i\psi^{(s,t)\dagger}_{A\omega\kappa}
\partial_0\psi^{A(s,t)}_{\omega\kappa}-\mu\kappa\left(J+\frac{3}{4}\right)
\psi^{(s,t)\dagger}_{A\omega\kappa}\psi^{A(s,t)}_{\omega\kappa}
\right], \n
&S_{R\times S^2}^{int}
=\frac{4\pi}{g_{R\times S^2}^2}\int \frac{dt}{\mu^2}
\mbox{tr} \biggl[-i {\cal C}_{\omega_1q_{st}\;\omega_2q_{tu}\;\omega_3q_{us}}
\partial_0
x^{(s,t)}_{AB,\omega_1}\Bigl(b^{(t,u)}_{\omega_2}x^{AB(u,s)}_{\omega_3}-
x^{AB(t,u)}_{\omega_2}b^{(u,s)}_{\omega_3}\Bigr) \biggr. 
\n
&\hspace{-5mm}-\frac{1}{2}{\cal C}^{\omega q}_{\omega_1q_{st}\;
\omega_2q_{tu}}
{\cal C}_{\omega q \;\omega_3 q_{uv}\; \omega_4 q_{vs}}
\left(b^{(s,t)}_{\omega_1}x^{(t,u)}_{AB,\omega_2}-
x^{(s,t)}_{AB,\omega_1}b^{(t,u)}_{\omega_2}\right)
\Bigl( b^{(u,v)}_{\omega_3}x^{AB(v,s)}_{\omega_4}-
x^{AB(u,v)}_{\omega_3}b^{(v,s)}_{\omega_4} \Bigr)
\n
&\hspace{-5mm}-\mu \sqrt{J_1(J_1+1)}\left( {\cal D}_{\omega_2q_{us}
\; \omega_1q_{st}0 \; \omega q_{tu}\rho}
x_{AB\omega_1}^{(s,t)}y_{\omega\rho}^{(t,u)}
x_{\omega_2}^{AB(u,s)} 
-{\cal D}_{\omega q_{tu} \;  \omega_2 q_{us}\rho_2 \;\omega_1 q_{st}0}
x_{AB\omega_1}^{(s,t)}x_{\omega}^{AB(t,u)}y_{\omega_2\rho_2}^{(u,s)}\right)
\n
&\hspace{-5mm}+(-1)^{m-q_{su}+1}
{\cal D}_{\omega_4 q_{vs} \; \omega q_{su}\rho \; \omega_3q_{uv}\rho_3}
{\cal D}_{\omega_2 q_{tu} \; J-mq_{us}\rho \; \omega_1q_{st}\rho_1}
y^{(s,t)}_{\omega_1\rho_1}x^{(t,u)}_{AB\omega_2}
y^{(u,v)}_{\omega_3\rho_3}x^{AB(v,s)}_{\omega_4}
\n
&\hspace{-5mm}
-(-1)^{m-q_{su}+1}
{\cal D}_{\omega_4 q_{uv} \; \omega_3q_{vs}\rho_3 \; \omega q_{su}\rho}
{\cal D}_{\omega_2 q_{tu} \; J-mq_{us}\rho \; \omega_1q_{st}\rho_1}
y^{(s,t)}_{\omega_1\rho_1}x^{(t,u)}_{AB\omega_2}
x^{AB(u,v)}_{\omega_4}y^{(v,s)}_{\omega_3\rho_3}
\n
&\hspace{-5mm}+\frac{1}{4}{\cal C}^{\omega q}_{\omega_1 q_{st} \;
\omega_2 q_{tu}}{\cal C}_{\omega q  \; \omega_3 q_{uv} \; \omega_4 q_{vs}}
\left(x^{(s,t)}_{AB\omega_1}x^{(t,u)}_{CD\omega_2}-
x^{(s,t)}_{CD\omega_1}x^{(t,u)}_{AB\omega_2}\right)
\Bigl(x^{AB(u,v)}_{\omega_3}x^{CD(v,s)}_{\omega_4}-
x^{CD(u,v)}_{\omega_3}x^{AB(v,s)}_{\omega_4}\Bigr)
\n
&\hspace{-5mm}-i\Bigl({\cal D}_{\omega q_{tu}  \;
\omega_2 q_{us}\rho_2  \; \omega_1 q_{st}\rho_1}
\partial_0 y^{(s,t)}_{\omega_1\rho_1} 
b^{(t,u)}_{\omega} y^{(u,s)}_{\omega_2\rho_2}
-{\cal D}_{\omega_2 q_{us}  \;\omega_1 q_{st}\rho_1  \;\omega q_{tu}\rho}
\partial_0 y^{(s,t)}_{\omega_1\rho_1}  y^{(t,u)}_{\omega\rho}
b^{(u,s)}_{\omega_2} \Bigr)
\n
&\hspace{-5mm}+\mu \sqrt{J_1(J_1+1)}\left({\cal D}_{\omega_2 q_{us} \;
\omega_1 q_{st}0 \; \omega q_{tu}\rho}
b_{\omega_1}^{(s,t)}y_{\omega\rho}^{(t,u)}b_{\omega_2}^{(u,s)} 
-{\cal D}_{\omega q_{tu}  \; \omega_2 q_{us}\rho_2 \; \omega_1 q_{st}0}
b_{\omega_1}^{(s,t)}b_{\omega}^{(t,u)}y_{\omega_2\rho_2}^{(u,s)}\right)
\n
&\hspace{-5mm}-(-1)^{m-q_{su}+1}
{\cal D}_{\omega_4 q_{vs} \; \omega q_{su}\rho \; \omega_3q_{uv}\rho_3}
{\cal D}_{\omega_2 q_{tu} \; J-mq_{us}\rho \; \omega_1q_{st}\rho_1}
y^{(s,t)}_{\omega_1\rho_1}b^{(t,u)}_{\omega_2}
y^{(u,v)}_{\omega_3\rho_3}b^{(v,s)}_{\omega_4}
\n
&\hspace{-5mm}+(-1)^{m-q_{su}+1}
{\cal D}_{\omega_4 q_{uv} \; \omega_3q_{vs}\rho_3 \; \omega q_{su}\rho}
{\cal D}_{\omega_2 q_{tu} \; J-mq_{us}\rho \; \omega_1q_{st}\rho_1}
y^{(s,t)}_{\omega_1\rho_1}b^{(t,u)}_{\omega_2}
b^{(u,v)}_{\omega_4}y^{(v,s)}_{\omega_3\rho_3}
\n
&\hspace{-5mm}+i\mu \rho_1(J_1+1){\cal E}_{\omega_1 q_{st}\rho_1  \;
\omega_2 q_{tu}\rho_2 \; \omega_3 q_{us}\rho_3}
y^{(s,t)}_{\omega_1\rho_1}y^{(t,u)}_{\omega_2\rho_2}y^{(u,s)}_{\omega_3\rho_3}
\n
&\hspace{-5mm}+\frac{1}{2}(-1)^{m-q_{su}+1}
{\cal E}_{J-mq_{us}\rho \; \omega_1q_{st}\rho_1 \; \omega_2q_{tu}\rho_2}
{\cal E}_{\omega q_{su}\rho \; \omega_3q_{uv}\rho_3 \; \omega_4q_{vs}\rho_4}
y^{(s,t)}_{\omega_1\rho_1}y^{(t,u)}_{\omega_2\rho_2}
y^{(u,v)}_{\omega_3\rho_3}y^{(v,s)}_{\omega_4\rho_4}
\n
&\hspace{-5mm}
+(-1)^{m-q_{su}+\frac{\kappa_1-\kappa_2}{2}}
{\cal F}^{J_2-m2-q_{ut}\kappa_2}_{J_1-m_1-q_{st}\kappa_1 \;\omega q_{su}}
\psi^{(s,t)\dagger}_{A\omega_1 \kappa_1}b^{(s,u)}_{\omega}
\psi^{A(u,t)}_{\omega_2\kappa_2}
-{\cal F}^{\omega_1q_{st}\kappa_1}_{\omega q_{su}\kappa \;\omega_2 q_{ut}}
\psi^{(s,t)\dagger}_{A\omega_1\kappa_1}\psi^{A(s,u)}_{\omega\kappa}
b^{(u,t)}_{\omega_2}
\n
&\hspace{-5mm}
-(-1)^{m-q_{su}+\frac{\kappa_1-\kappa_2}{2}}
{\cal G}^{J_2-m_2 -q_{ut}\kappa_2}_{J_1-m_1 -q_{st}\kappa_1 
\; \omega q_{su}\rho}
\psi^{(s,t)\dagger}_{A\omega_1\kappa_1}
y^{(s,u)}_{\omega\rho}\psi^{A(u,t)}_{\omega_2\kappa_2}
-{\cal G}^{\omega_1q_{st}\kappa_1}_{\omega q_{su}\kappa\;\omega_2 q_{ut}\rho_2}
\psi^{(s,t)\dagger}_{A\omega\kappa_1}\psi^{A(s,u)}_{\omega\kappa}
y^{(u,t)}_{\omega_2\rho_2}
\n
&\hspace{-5mm}-i(-1)^{m_2-q_{ut}-\frac{\kappa_2}{2}}
{\cal F}^{J_2 -m_2 -q_{ut} \kappa_2}_{\omega_1 q_{ts}\kappa_1 \; \omega q_{su}}
\psi^{A(t,s)}_{\omega_1\kappa_1}x^{(s,u)}_{AB\omega}
\psi^{B(u,t)}_{\omega_2\kappa_2}
\n
&\hspace{-5mm}+i(-1)^{m_1-q_{ts}+\frac{\kappa_1}{2}}
{\cal F}^{J_1 -m_1 -q_{ts} \kappa_1}_{\omega q_{su}\kappa \; \omega_2 q_{ut}}
\psi^{A(t,s)}_{\omega_1\kappa_1}\psi^{B(s,u)}_{\omega\kappa}
x^{(u,t)}_{AB\omega_2}
\n
&\hspace{-5mm}-i(-1)^{m_1-q_{st}-\frac{\kappa_1}{2}}
{\cal F}^{\omega_2 q_{tu}\kappa_2}_{J_1 -m_1 -q_{st} \kappa_1  \;\omega q_{su}}
\psi^{(s,t)\dagger}_{A\omega_1\kappa_1}x^{AB(s,u)}_{\omega}
\psi^{(t,u)\dagger}_{B\omega_2\kappa_2}
\n
&\hspace{-5mm}-i(-1)^{m-q_{us}-\frac{\kappa}{2}}
{\cal F}^{\omega_1 q_{st}\kappa_1}_{J -m -q_{us} \kappa \; \omega_2 q_{ut}}
\psi^{(s,t)\dagger}_{A\omega_1\kappa_1}\psi^{(u,s)\dagger}_{B\omega\kappa}
 x^{AB(u,t)}_{\omega_2}
\biggl.\biggr], \label{mode expansion of RxS^2}
\end{align}
where the summation over the indices that appear twice or more than
twice is assumed and we have introduced the abbreviated notations:
$\omega$ represents
a pair, $(J,m)$. 

Similarly, we expand the action (\ref{PWMMSU(4)form})
around the vacuum (\ref{vacuum for PWMM}). We make a substitution
$\vec{Y}\rightarrow \hat{\vec{Y}}+\vec{Y}$ in (\ref{PWMMSU(4)form}), where
$\hat{Y}_i=-\mu L_i$ and $L_i$ is given in (\ref{vacuum for PWMM}).
The result is 
\begin{align}
 &S_{PW}=S_{PW}^{free}+S_{PW}^{int}, \n
 &S_{PW}^{free}=\frac{1}{g_{PW}^2} \int \frac{dt}{\mu^2} \sum_{s,t}\tr
  \Biggl(
  \frac{1}{2}\partial_0X^{AB(t,s)}\partial_0X_{AB}^{(s,t)}
  +\frac{\mu^2}{2}\vec{L}\circ X^{AB(t,s)}\cdot\vec{L}\circ X_{AB}^{(s,t)}
  -\frac{\mu^2}{8}X^{AB(t,s)}X_{AB}^{(s,t)} \n
  & \hspace{4cm}
  +\frac{1}{2}
  \partial_0\vec{Y}^{(t,s)}\cdot\partial_0\vec{Y}^{(s,t)}
  -\frac{1}{2}(i\mu\vec{L}\circ \times \vec{Y}^{(t,s)}+\mu\vec{Y}^{(t,s)})
  \cdot(i\mu\vec{L}\circ \times \vec{Y}^{(s,t)}+\mu\vec{Y}^{(s,t)}) \n
  & \hspace{4cm}
  -\frac{\mu^2}{2}\vec{L}\circ A_0^{(t,s)}\cdot\vec{L}\circ A_0^{(s,t)}
  -i\mu\partial_0\vec{Y}^{(t,s)}\cdot\vec{L}\circ A_0^{(s,t)} \n
  & \hspace{4cm}
  +i\psi_A^{\dagger(t,s)}\partial_0\psi^{A(s,t)} 
  -\mu\psi_A^{\dagger(t,s)}\vec{\sigma}\cdot\vec{L}\circ \psi^{A(s,t)}
  -\frac{3\mu}{4}\psi_A^{\dagger(t,s)}\psi^{A(s,t)}
  \Biggr), \n
 &S_{PW}^{int}
  =\frac{1}{g_{PW}^2}\int \frac{dt}{\mu^2} \sum_{s,t}\tr
  \Biggl(
  -i\partial_0X_{AB}^{(t,s)}[A_0,X^{AB}]^{(s,t)}
  -\frac{1}{2}[A_0,X_{AB}]^{(t,s)}[A_0,X^{AB}]^{(s,t)} \n
  & \hspace{4cm}
  -\mu\vec{L}\circ X_{AB}^{(t,s)}\cdot[\vec{Y},X^{AB}]^{(s,t)}
  +\frac{1}{2}[\vec{Y},X_{AB}]^{(t,s)}\cdot[\vec{Y},X^{AB}]^{(s,t)} \n
  & \hspace{4cm}
  +\frac{1}{4}[X_{AB},X_{CD}]^{(t,s)}[X^{AB},X^{CD}]^{(s,t)}
  -\frac{1}{2}[\vec{Y},A_0]^{(t,s)}\cdot[\vec{Y},A_0]^{(s,t)} \n
  & \hspace{4cm}
  -i(\partial_0\vec{Y})^{(t,s)}\cdot[A_0,\vec{Y}]^{(s,t)}
  -\mu[A_0,\vec{Y}]^{(t,s)}\cdot(\vec{L}\circ A_0)^{(s,t)} \n
  & \hspace{4cm}
  +i(i\mu\vec{L}\circ \times\vec{Y}^{(t,s)}+\mu\vec{Y}^{(t,s)})
  \cdot(\vec{Y}\times\vec{Y})^{(s,t)}
  +\frac{1}{2}(\vec{Y}\times\vec{Y})^{(t,s)}\cdot(\vec{Y}\times\vec{Y})^{(s,t)}
  \n 
  & \hspace{4cm}
  +\psi_A^{\dagger(t,s)}[A_0,\psi^A]^{(s,t)}
  +\psi_A^{\dagger(t,s)}\vec{\sigma}\cdot[\vec{Y},\psi^A]^{(s,t)} \n
  & \hspace{4cm}
  -\psi^{AT(t,s)}\sigma^2[X_{AB},\psi^B]^{(s,t)}
  +\psi_A^{\dagger(t,s)}\sigma^2[X^{AB},\psi_B^*]^{(s,t)}
  \Biggr).
\label{PWMM action around fuzzy sphere vacua}
\end{align}
Here the suffix $(s,t)$ stands for the $(s,t)$ `large' block of an 
$\hat{N}\times\hat{N}$ matrix, which is an $N_s(2j_s+1)\times N_t(2j_t+1)$
rectangular matrix, and $s,t$ run from $1$ to $T$.
The reader would notice resemblance between 
(\ref{S^2 action around monopole vacua}) and
(\ref{PWMM action around fuzzy sphere vacua}).
We make a mode expansion analogous to (\ref{mode expansion in 2+1 SYM}):
\begin{align}
 A_0^{(s,t)}
  &=\sum_{J=|j_s-j_t|}^{j_s+j_t}\sum_{m=-J}^{J}b^{(s,t)}_{Jm}
  \otimes\hat{Y}_{Jm(j_sj_t)}, 
 \qquad
 X_{AB}^{(s,t)}
  =\sum_{J=|j_s-j_t|}^{j_s+j_t}\sum_{m=-J}^{J}x^{(s,t)}_{ABJm}\otimes
  \hat{Y}_{Jm(j_sj_t)},
  \n
 \psi^{A(s,t)}
  &=\sum_{\kappa=\pm1}\sum_{\tilde{U}=|j_s-j_t|}^{j_s+j_t}\sum_{m=-U}^{U}
 \psi_{Jm\kappa}^{A(s,t)}\otimes\hat{Y}_{Jm(j_sj_t)}^\kappa  \n
  &=\sum_{J=|j_s-j_t|}^{j_s+j_t}\sum_{m=-J-\frac{1}{2}}^{J+\frac{1}{2}}
  \psi_{Jm1}^{A(s,t)}\otimes\hat{Y}_{Jm(j_sj_t)}^1 
  +\sum_{J=|j_s-j_t|-\frac{1}{2}}^{j_s+j_t-\frac{1}{2}}\sum_{m=-J}^{J}
  \psi_{Jm-1}^{A(s,t)}\otimes\hat{Y}_{Jm(j_sj_t)}^{-1},  \n
 \vec{Y}^{(s,t)}
 &=\sum_{\rho=-1}^{1}\sum_{\tilde{Q}=|j_s-j_t|}^{j_s+j_t}\sum_{m=-Q}^{Q}
 y_{Jm\rho}^{(s,t)}\otimes\vec{\hat{Y}}{}_{Jm(j_sj_t)}^\rho  \n
 &=\sum_{J=|j_s-j_t|}^{j_s+j_t}\sum_{m=-J-1}^{J+1}
  y_{Jm1}^{(s,t)}\otimes\vec{\hat{Y}}{}_{Jm(j_sj_t)}^1 
  +\sum_{J=|j_s-j_t|}^{j_s+j_t}\sum_{m=-J}^{J}
  y_{Jm0}^{(s,t)}\otimes\vec{\hat{Y}}{}_{Jm(j_sj_t)}^0 \n
  &\;\;\;\;\;+\sum_{J=|j_s-j_t|-1}^{j_s+j_t-1}\sum_{m=-J}^{J}
  y_{Jm-1}^{(s,t)}\otimes\vec{\hat{Y}}{}_{Jm(j_sj_t)}^{-1},
\label{mode expansion in PWMM}
\end{align}
In the above expressions, the both sides are
$N_s(2j_s+1)\times N_t(2j_t+1)$ matrices and the modes in the righthand sides
such as $x_{ABJm}^{(s,t)}$ are $N_s\times N_t$ matrices.
Due to (\ref{complexconjugate4}), 
(\ref{hermitian conjugate of mode in 2+1 SYM}) also holds for this case.

By substituting (\ref{mode expansion in PWMM}) into
(\ref{PWMM action around fuzzy sphere vacua}) and using (\ref{orthnormality4}),
(\ref{identities3}) and (\ref{hat vertex coefficients}),
we obtain the mode-expanded form of the theory around 
the vacuum (\ref{vacuum for PWMM}). By setting
\beqa
\frac{4\pi}{g_{R\times S^2}^2}=\frac{N_0}{g_{PW}^2}
\label{coupling relation between S^2 and PWMM 2}
\eeqa
and 
\beqa
q_{st}=j_s-j_t,
\label{q_st=j_s-j_t}
\eeqa
it is easy to see that the free part
completely coincides with $S_{R\times S^2}^{free}$ in 
(\ref{mode expansion of RxS^2}) while the interaction part is obtained by
attaching the hat to the vertex coefficients in $S_{R\times S^2}^{int}$ and
replacing $q_{st}$ in the vertex coefficients with $(j_sj_t)$.
As seen in section 4.3, the vertex coefficients with the hat 
reduce to the vertex coefficients with the identification $q=j-j'$ in 
the $N_0\rightarrow\infty$ limit.  Thus, in the $N_0\rightarrow\infty$ limit,
the interaction part also coincides with $S_{R\times S^2}^{int}$
in (\ref{mode expansion of RxS^2}).
Furthermore, the relation (\ref{q_st=j_s-j_t}) is equivalent to
(\ref{j_s-j_t}), and the relation 
(\ref{coupling relation between S^2 and PWMM 2})
is consistent with (\ref{coupling relation between S^2 and PWMM}).
Thus we have completed the proof of the prediction 1).

\subsection{Topologically nontrivial configurations on fuzzy spheres}
In this subsection, we comment on a relation of our results in the
previous subsection with the works 
\cite{Aoki:2003ye,Aoki:2004sd}.

The authors of \cite{Aoki:2003ye,Aoki:2004sd} considered a configuration
\beqa
Y_i=-\mu L_i=-\mu\left(\begin{array}{cc}
           L_i^{[j_1]} & 0        \\
           0           & L_i^{[j_2]}
           \end{array} \right) 
\label{SU(2)fuzzy}
\eeqa
as a topologically nontrivial gauge configuration,
where $\zeta_1-\zeta_2=2\alpha\;\;(2j_1+1=N_0+\zeta_1,\;2j_2+1=N_0+\zeta_2)$
with $\alpha$ an integer.
They introduced the topological index on a fuzzy sphere which 
can be defined for the configuration (\ref{SU(2)fuzzy}).
Their topological index for (\ref{SU(2)fuzzy})
is equal to $\frac{1}{2}|\zeta_1-\zeta_2|=|\alpha|$, and they claimed that it
coincides with 
the winding number $\pi_2(SU(2)/U(1))$ in the continuum limit 
($N_0\rightarrow\infty$ limit).
Actually, in the case in which $\alpha=1$,  
they directly
obtained from (\ref{SU(2)fuzzy}) the 't Hooft-Polyakov monopole
solution, which has the winding number one. 

According to our result in the previous subsection, 
the vacuum configuration of $\mbox{SYM}_{R\times S^2}$ corresponding to 
(\ref{SU(2)fuzzy}) in the $N_0\rightarrow\infty$ limit is
\beqa
&&\hat{\Phi}=\frac{\mu}{2}\left(
           \begin{array}{cc}
           \alpha & 0        \\
           0      & -\alpha
           \end{array} \right), \n
&&\hat{A}_1=0, \n
&&\hat{A}_2=\left\{\begin{array}{ll}
            \tan\frac{\theta}{2}\:\hat{\Phi}  & \mbox{in region I} \\
            -\cot\frac{\theta}{2}\:\hat{\Phi} & \mbox{in region II}
           \end{array} \right., 
\label{SU(2)configuration}
\eeqa
where we have extracted the $SU(2)$ part separating the decoupled $U(1)$ part.
Namely, for generic $\alpha$,
we found the gauge configuration on $S^2$ to which
(\ref{SU(2)fuzzy}) reduces in the $N_0\rightarrow\infty$ limit.
In the following, we check a consistency that the configuration
(\ref{SU(2)configuration}) has the winding number $|\alpha|$.
 
We define a gauge invariant quantity by
\beqa
{\cal F}_{a'b'}&=&\mbox{Tr}(\tilde{\Phi}F_{a'b'}
-\tilde{\Phi}[D_{a'}\tilde{\Phi},D_{b'}\tilde{\Phi}]) \n
&=&\mbox{Tr}(\nabla_{a'}(\tilde{\Phi}A_{b'})-\nabla_{b'}(\tilde{\Phi}A_{a'})
-\tilde{\Phi}[\nabla_{a'}\tilde{\Phi},\nabla_{b'}\tilde{\Phi}]),
\label{gauge invariant F}
\eeqa
where 
\beqa
\tilde{\Phi}=\frac{\Phi}{\sqrt{2\mbox{Tr}\Phi^2}}.
\eeqa
Then the topological charge is given by
\beqa
Q=\frac{1}{8\pi}\int d\theta d\phi \sin\theta {\cal F}_{12}
\label{Q}
\eeqa
Actually, for configurations where 
$f_{a'b'}=
\mbox{Tr}(\nabla_{a'}(\tilde{\Phi}A_{b'})-\nabla_{b'}(\tilde{\Phi}A_{a'}))$
is total derivative, (\ref{Q}) reduces to
\beqa
Q=-\frac{1}{8\pi}\int d\theta d\phi \sin\theta 
\mbox{Tr}(\tilde{\Phi}[\nabla_{1}\tilde{\Phi},\nabla_{2}\tilde{\Phi}]),
\label{winding number}
\eeqa
which is the winding number $\pi_2(SU(2)/U(1))$.
For the configuration (\ref{SU(2)configuration}), $f_{a'b'}$ is not
total derivative while 
$\mbox{Tr}(\tilde{\Phi}[\nabla_{a'}\tilde{\Phi},\nabla_{b'}\tilde{\Phi}])$
vanishes. $Q$ is evaluated from (\ref{Q}) as $Q=|\alpha|$.
One can also obtain the same value for $Q$ 
from (\ref{winding number}) by applying 
a singular gauge transformation to (\ref{SU(2)configuration}).
In  the region II, it takes the form
\beqa
V=\left(\begin{array}{cc}
  \cos\frac{\theta}{2}e^{-i\alpha\phi} & \sin\frac{\theta}{2} \\
  -\sin\frac{\theta}{2}                & \cos\frac{\theta}{2}e^{i\alpha\phi}
                          \end{array} \right).
\label{gauge transf}
\eeqa
The resultant gauge transformed configuration is 
\beqa 
&&\hat{\Phi}\;\rightarrow\;V^{\dagger}\hat{\Phi}V 
=\frac{\mu\alpha}{2}\left(\begin{array}{cc}
                   \cos\theta                  & \sin\theta e^{i\alpha\phi} \\
                   \sin\theta e^{-i\alpha\phi} & -\cos\theta
                          \end{array} \right), \n
&&{\hat{A}_1}\;\rightarrow\;V^{\dagger}\hat{A}_1V+iV^{\dagger}\nabla_1V
=\frac{i\mu}{2}\left(\begin{array}{cc}
                   0                  &  e^{i\alpha\phi} \\
                   -e^{-i\alpha\phi}  &  0
                          \end{array} \right), \n
&&{\hat{A}_2}\;\rightarrow\;V^{\dagger}\hat{A}_2V+iV^{\dagger}\nabla_2V
=\frac{\mu\alpha}{2}\left(\begin{array}{cc}
                   \sin\theta                & -\cos\theta e^{i\alpha\phi} \\
                -\cos\theta e^{-i\alpha\phi} & -\sin\theta
                          \end{array} \right).
\label{'t Hooft-Polyakov}
\eeqa
In the region I, the same configuration of the fields are obtained by
the gauge transformation $V_{I\rightarrow II}V$, where $V_{I\rightarrow II}$
is given in (\ref{V from I to II}).
Note that the single-valuedness of $V$ and the gauge transformed fields
requires $\alpha$ to be an integer. For the gauge transformed configuration
(\ref{'t Hooft-Polyakov}), $f_{a'b'}$ vanishes and 
(\ref{winding number}) indeed gives $Q=|\alpha|$.
Thus, for the configuration (\ref{SU(2)configuration}) with 
generic $\alpha$, $|\alpha|$ is interpreted as the winding number.
For $\alpha=\pm 1$, it is easy to check
that (\ref{'t Hooft-Polyakov}) is nothing but 
the 't Hooft-Polyakov monopole solution, which is smooth everywhere on $S^2$.
For $\alpha \neq \pm 1$, 
although the gauge fields in (\ref{'t Hooft-Polyakov}) are not smooth 
everywhere, $\Phi$ is smooth everywhere and 
$Q$ is given by (\ref{winding number}).

When $\zeta_1-\zeta_2$ in (\ref{SU(2)fuzzy}) is an odd integer,
one can also consider the corresponding configuration on $S^2$  
(\ref{SU(2)configuration}) in which $2\alpha$ is equal to
the odd integer $\zeta_1-\zeta_2$.
This configuration indeed gives $Q=|\alpha|$ which is a half odd integer.
However, in this case, the gauge transformation (\ref{gauge transf}) does
not exist, so that one cannot interpret this $Q$ as the winding number.

\section{${\cal N}=4$ SYM on $R\times S^3/Z_k$ vs $2+1$ SYM on $R\times S^2$}
\setcounter{equation}{0}
\subsection{Embedding of $\mbox{SYM}_{R\times S^3/Z_k}$ into 
$\mbox{SYM}_{R\times S^2}$}
In this subsection, we prove the prediction 2) for the trivial vacuum
of $\mbox{SYM}_{R\times S^3/Z_k}$. 
According to the prediction 2), the theory around the trivial vacuum of 
$\mbox{SYM}_{R\times S^3/Z_k}$ with $U(N)$ gauge group is equivalent to 
the theory around the vacuum
(\ref{periodic Phi}) of $\mbox{SYM}_{R\times S^2}$ 
with the relation (\ref{relation for coupling constant})
if a single period
is extracted after the periodicity is imposed.

In (\ref{S^2 action around monopole vacua}), by setting 
$\alpha_s=sk,\; N_s= N$ and 
making $s$ run from $-\infty$ to $\infty$,
we obtain the theory around the vacuum (\ref{periodic Phi}) 
of $\mbox{SYM}_{R\times S^2}$. Then, the monopole charge $q_{st}$ takes
the form
\beqa
q_{st}=\frac{k}{2}(s-t),
\label{q_st}
\eeqa
which depends only on $s-t$. This fact enables us to impose the following 
condition on the blocks of the fields 
in (\ref{S^2 action around monopole vacua}):
\begin{eqnarray}
 &&X^{(s+1,t+1)}=X^{(s,t)}, \qquad A_0^{(s+1,t+1)}=A_0^{(s,t)}, \n
 &&\vec{Y}^{(s+1,t+1)}=\vec{Y}^{(s,t)}, \qquad \psi^{A(s+1,t+1)}=\psi^{A(s,t)}.
\label{condition}
\end{eqnarray}
Namely, the $(s,t)$ blocks of the fields depends only on $s-t$.
It is natural to consider that this condition corresponds to
the periodicity on the gravity side.
We show below that this is indeed the case.
 
The condition for the modes of these fields follows from (\ref{condition}):
\begin{eqnarray}
 &&x_{ABJm}^{(s+1,t+1)}=x_{ABJm}^{(s,t)}, \qquad 
 b_{Jm}^{(s+1,t+1)}=b_{Jm}^{(s,t)},
  \n
 &&y_{Jm\rho}^{(s+1,t+1)}=y_{Jm\rho}^{(s,t)}, 
  \qquad \psi_{Jm\kappa}^{A(s,t)}=\psi_{Jmq\kappa}^{A(s,t)}.
\end{eqnarray}
This condition allows us to rewrite the modes as
\begin{eqnarray}
 &&x_{ABJm}^{(s,t)}=x_{ABJmq_{st}}, \qquad b_{Jm}^{(s,t)}=b_{Jmq_{st}},
  \n
 &&y_{Jm\rho}^{(s,t)}=y_{Jmq_{st}\rho}, 
  \qquad \psi_{Jm\kappa}^{A(s,t)}=\psi_{Jmq_{st}\kappa}^A,
  \label{S^3 form of the modes}
\end{eqnarray}
Note that every mode is an $N\times N$ matrix.

By using (\ref{q_st}) and (\ref{S^3 form of the modes}), we rewrite
(\ref{mode expansion of RxS^2}). Here we show calculation of 
some terms in (\ref{mode expansion of RxS^2}) as examples.
We first consider in $S_{R\times S^2}^{free}$ 
\beqa
\sum_{s,t}\sum_{J\geq |q_{st}|}\sum_{m=-J}^J
\left(J+\frac{1}{2}\right)^2x_{ABJm}^{(s,t)\dagger}x_{ABJm}^{(s,t)}.
\label{term in free part}
\eeqa
We set $s-t=n,\;\;s=l$ so that $n,\;l$ take integers.  
We can rewrite (\ref{term in free part}) as
\beqa
\sum_l\sum_n\sum_{J\geq |\frac{k}{2}n|}\sum_{m=-J}^J
\left(J+\frac{1}{2}\right)^2
x_{Jm\frac{k}{2}n}^{AB\dagger}x_{Jm\frac{k}{2}n}^{AB}.
\eeqa
Moreover, by setting $\frac{k}{2}n=\tilde{m}$, we obtain
\beqa
\left.\sum_l\sum_{J=0}^{\infty}\sum_{m=-J}^J
\sum_{\tilde{m}=-J}^J\right|_{\tilde{m}\in\frac{k}{2}{\bm Z}}
\left(J+\frac{1}{2}\right)^2
x_{Jm\tilde{m}}^{AB\dagger}x_{Jm\tilde{m}}^{AB}.
\label{term in free part 2}
\eeqa
We next consider in $S_{R\times S^2}^{int}$ 
\begin{align}
&\sum_{s,t,u}\sum_{J_1\geq |q_{st}|,m_1} 
\sum_{J_2\geq |q_{tu}|,m_2}\sum_{J_3\geq |q_{us}|,m_3} \nonumber\\[2mm]
&\qquad {\cal C}_{J_1m_1q_{st}\;J_2m_2q_{tu}\;J_3m_3q_{us}}
\partial_0x_{ABJ_1m_1q_{st}}
(b_{J_2m_2q_{tu}}x_{J_3m_3q_{us}}^{AB}-x_{J_2m_2q_{tu}}^{AB}b_{J_3m_3q_{us}}).
\label{term in interaction part}
\end{align}
In (\ref{term in interaction part}), we set 
$s-t=n,\;t-u=p,\;t=l$ in the first term and $s-t=n,\;u-s=p,\;s=l$ in the
second term, so that $n,p,l$ take integers. We also  
make exchanges for dummy variables in the second term as
$J_2\leftrightarrow J_3,\;m_2 \leftrightarrow m_3$. Then we can rewrite
(\ref{term in interaction part}) as 
\begin{align}
\sum_{l,n,p}\sum_{J_1\geq |\frac{k}{2}n|,m_1}
\sum_{J_2\geq |\frac{k}{2}p|,m_2}\sum_{J_3\geq |\frac{k}{2}(n+p)|,m_3}
\!{\cal C}_{J_1m_1\frac{k}{2}n\;J_2m_2\frac{k}{2}p\;J_3m_3\frac{k}{2}(-n-p)}
\partial_0x_{ABJ_1m_1\frac{k}{2}n}
[b_{J_2m_2\frac{k}{2}p},x_{J_3m_3\frac{k}{2}(-n-p)}^{AB}].
\n
\end{align}
We further set 
$\frac{k}{2}n=\tilde{m}_1,\;\frac{k}{2}p=\tilde{m}_2,\;
\frac{k}{2}(-n-p)=\tilde{m}_3$, and obtain
\beqa
&&\sum_l
\left.\sum_{J_1=0}^{\infty}\sum_{m_1,\tilde{m}_1=-J_1}^{J_1}
\sum_{J_2=0}^{\infty}\sum_{m_2,\tilde{m}_2=-J_2}^{J_2}
\sum_{J_3=0}^{\infty}\sum_{m_3,\tilde{m}_3=-J_3}^{J_3}
\right|_{\tilde{m}_1,\tilde{m}_2,\tilde{m}_3\in\frac{k}{2}{\bm Z}} \nonumber
\\[2mm]
&&\qquad{\cal C}_{J_1m_1\tilde{m}_1\;J_2m_2\tilde{m}_2\;J_3m_3\tilde{m}_3}
\partial_0x_{ABJ_1m_1\tilde{m}_1}
[b_{J_2m_2\tilde{m}_2},x_{J_3m_3\tilde{m}_3}^{AB}].
\label{term in interaction part 2}
\eeqa
We can easily rewrite the other terms in (\ref{mode expansion of RxS^2})
in the same way. There appears in common the overall factor $\sum_l$ 
in all the terms of 
the rewritten form of (\ref{mode expansion of RxS^2}).

In appendix G, we give the mode expansion of the theory around the
trivial vacuum of $\mbox{SYM}_{R\times S^3/Z_k}$
(\ref{mode expansion of S^3/Z_k}),
which we obtained in our previous publication \cite{ITT}.
In the rewritten form of  
(\ref{mode expansion of RxS^2}) obtained above, we make the following 
identifications 
\beqa
&&b_{Jm\tilde{m}}=B_{Jm\tilde{m}},\;\;\;
y_{Jm\tilde{m}\rho}=A_{Jm\tilde{m}\rho}, \n
&&x_{Jm\tilde{m}}^{AB}=X_{Jm\tilde{m}}^{AB}, \;\;\;
\psi_{Jm\tilde{m}\kappa}^A=\Psi_{Jm\tilde{m}\kappa}^A
\eeqa
and input the relation (\ref{relation for coupling constant}).
Moreover, we divide this rewritten form by the overall factor $\sum_l$.
This procedure corresponds to extracting a single period.
Then, it is easy to see that this rewritten form of 
(\ref{mode expansion of RxS^2}) coincides with
(\ref{mode expansion of S^3/Z_k}).\footnote{More precisely, the terms 
proportional to $\mu$ differ in signature. However, this difference can be
compensated by the parity transformation, so that it does not matter.}
Thus we have completed the proof of the prediction 2) for the trivial vacuum
of $\mbox{SYM}_{R\times S^3/Z_k}$.

The configuration (\ref{periodic Phi}), the condition (\ref{condition})
and the procedure of dividing by $\sum_l$
physically mean that a circle with the radius$\sim k$ is constructed 
in the $\Phi$ direction and 
the $(s,t)$ block of the fields is interpreted as the winding mode 
around the circle with the winding number $s-t$. 
We have reinterpreted the winding number $s-t$ as
the Kaluza-Klein momentum $\frac{k}{2}(s-t)$ 
on a circle with the radius$\sim\frac{1}{k}$.
This is similar to Taylor's prescription for 
the compactification (the T-duality) 
in matrix models \cite{Taylor:1996ik}. The difference between
our prescription and Taylor's is the existence of the nontrivial
gauge fields in (\ref{periodic Phi}), 
which makes a nontrivial fibration of the circle
over $S^2$ rather than a direct product $S^2\times S^1$ so that 
$S^3/Z_k$ is realized.

\subsection{$S^3$ from three matrices}
\begin{figure}[htbp]
\begin{center}
\hspace{2.5cm}
\includegraphics[height=7cm, keepaspectratio, clip]{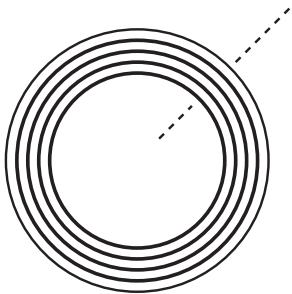}
\end{center}
\vspace{-3cm}
\caption{$S^3/Z_k$ is realized through a stack of fuzzy spheres.
Each circle represents $N$ coincident fuzzy spheres.}
\label{stack of fuzzy spheres}
\end{figure}

Combining the result in section 5.1 with that in section 6.1 leads us to
conclude that the trivial vacuum of $\mbox{SYM}_{R\times S^3/Z_k}$ with
gauge group $U(N)$ is embedded in PWMM. The corresponding vacuum configuration
of PWMM is $Y_i=-\mu L_i$, where
\begin{align}
 L_i=
 \begin{pmatrix}
  \rotatebox[origin=tl]{-35}
  {$\cdots \;\;\;
  \overbrace{\rotatebox[origin=c]{35}{$L_{i}^{[j_{s-1}]}$} \;
  \cdots \;
  \rotatebox[origin=c]{35}{$L_{i}^{[j_{s-1}]}$}}^{\rotatebox{35}{$N$}}
  \;\;\;
  \overbrace{\rotatebox[origin=c]{35}{$L_i^{[j_s]}$} \;
  \cdots \;
  \rotatebox[origin=c]{35}{$L_i^{[j_s]}$}}^{\rotatebox{35}{$N$}}
  \;\;\;
  \overbrace{\rotatebox[origin=c]{35}{$L_{i}^{[j_{s+1}]}$} \;
  \cdots \;
  \rotatebox[origin=c]{35}{$L_{i}^{[j_{s+1}]}$}}^{\rotatebox{35}{$N$}}
  \;\;\; \cdots$}
 \end{pmatrix}
\label{from S^2 to S^3/Z_k}
\end{align}
with $2j_s+1=N_0+ks$. $s$ runs from $-\infty$ to $\infty$ and the following
periodicity for the fluctuations of the fields around the vacuum 
(\ref{from S^2 to S^3/Z_k}) is imposed:
\beqa
\vec{Y}^{(s+1,t+1)}=\vec{Y}^{(s,t)},\;\;\; X_m^{(s+1,t+1)}=X_m^{(s,t)},\;\;\;
\lambda^{(s+1,t+1)}=\lambda^{(s,t)}.
\eeqa
The vacuum (\ref{from S^2 to S^3/Z_k}) is interpreted as 
a stack of infinitely many sets of
$N$ coincident fuzzy spheres
(See Fig.\ref{stack of fuzzy spheres} ). 
Note that the $N_0\rightarrow\infty$ limit must be taken from the 
beginning in order for the configuration (\ref{from S^2 to S^3/Z_k}) to
be realized.

It is interesting that $S^3/Z_k$ is realized by the three matrices,
$Y_1,\:Y_2,\:Y_3$. 
It is well-known that fuzzy sphere is realized by three
matrices through the $SU(2)$ algebra and in the continuum limit an ordinary
$S^2$ is realized with one of three directions remained on $S^2$ 
as a Higgs field.
In the present case, the Higgs field is utilized 
to make the $U(1)$ bundle on $S^2$.
In particular, in the $k=1$ case, one realizes $S^3$ by the three matrices and
obtains from PWMM ${\cal N}=4$ SYM on $R\times S^3$, 
which is important in the AdS/CFT
context, namely, dual to $AdS_5\times S^5$ in the global coordinates.
In this case, the $SU(2|4)$ symmetry is enhanced to the $SU(2,2|4)$ symmetry.

\section{Summary and outlook}
\setcounter{equation}{0}
In this paper, we show that every vacuum of 
$\mbox{SYM}_{R\times S^2}$ is embedded in PWMM and the trivial vacuum
of $\mbox{SYM}_{R\times S^3/Z_k}$ is embedded in $\mbox{SYM}_{R\times S^2}$.
This is predicted from the gravity duals through Lin-Maldacena's method.
Our results serve as a nontrivial check of the gauge/gravity correspondence
for the theories with $SU(2|4)$ symmetry. As by-products,
we reveal the relationships among the spherical harmonics on $S^3$,
the monopole harmonics and the fuzzy sphere harmonics, and extend
an extension of the compactification (T-duality) in matrix models a la Taylor
to that on spheres.

We treated only embedding of the trivial vacuum of 
$\mbox{SYM}_{R\times S^3/Z_k}$ into $\mbox{SYM}_{R\times S^2}$.
Indeed, we have the vacuum configurations in $\mbox{SYM}_{R\times S^2}$
that would give the theories around the nontrivial vacua of 
$\mbox{SYM}_{R\times S^3/Z_k}$.
It is important to prove the prediction 2) for the nontrivial vacua. 

It is interesting to extend the T-duality in matrix models in this paper, 
which realizes $S^3/Z_k$ as an $S^1$ fibration over $S^2$,
to other fiber bundles and to obtain
a general recipe for such T-duality in matrix models.

$\mbox{SYM}_{R\times S^3/Z_k}$ with $k=1$ is nothing but ${\cal N}=4$ 
SYM on $R\times S^3$, which has the unique trivial vacuum and whose
symmetry group is enhanced to $SU(2,2|4)$. The gravity dual of this
theory is $AdS_5\times S^5$.  
Hence as mentioned in section 6.2, our results tell
that ${\cal N}=4$ SYM on $R\times S^3$ which is a gauge theory in
a typical example of the AdS/CFT correspondence is 
embedded in PWMM.
However, this does not mean that we have obtained 
a matrix model that regularizes
${\cal N}=4$ SYM on $R\times S^3$ preserving gauge symmetry and supersymmetry
and in principle enables us to perform a numerical simulation
for the AdS/CFT correspondence. Indeed, in the T-duality, we need to
consider matrices with infinite size. Presumably, by referring to the work 
\cite{Kaplan:2002wv}, we can  
make the size of matrices finite with a part of supersymmetry preserved
and obtain a lattice gauge theory with few parameters to be fine-tuned
for ${\cal N}=4$ SYM on $R\times S^3$.

We hope to report progress in the above projects in the near future.

\section*{Note added}
While we are writing the manuscript, we are informed that Aoki et al. are
preparing for a publication \cite{AIM}, 
which has some overlap with section 4.3 of the present paper.

\section*{Acknowledgements}
We would like to thank H. Aoki for discussions and informing us of 
the results in \cite{AIM} before publication. We are also grateful to
H. Kawai for discussions.
We thank the Yukawa Institute for Theoretical Physics at Kyoto University.
Discussions during the YITP workshop YITP-W-06-11 on ``String Theory and
Quantum Field Theory" were useful to complete this work.
The work of Y.T. is supported in part by The 21st Century COE Program
``Towards a New Basic Science; Depth and Synthesis."
The work of A.T. is supported in part by Grant-in-Aid for Scientific
Research (No.16740144) from the Ministry of Education, Culture, Sports,
Science and Technology.

\appendix

\section*{Appendices}

\section{Some conventions}
\setcounter{equation}{0}
\renewcommand{\theequation}{A.\arabic{equation}}
In this appendix, we describe some conventions which we follow
in the present paper.

We use the following metric for $R\times S^3$:
\beqa
ds_{R\times S^3}^2=-dt^2+\frac{1}{\mu^2}(d\theta^2+\sin^2\theta d\phi^2
+(d\psi+\cos\theta d\phi)^2),
\eeqa
where $0 \leq \theta \leq \pi$, $0 \leq \phi <2\pi$, $0 \leq \psi <4\pi$, 
and the radius of $S^3$ is $\frac{2}{\mu}$.
The nonvanishing components of the vierbeins and the spin connections are 
\begin{align}
\hspace{-1cm}
&e^1_{\theta}=\mu^{-1},\;\;\; e^2_{\phi}=\mu^{-1}\sin\theta,\;\;\; 
e^3_{\phi}=\mu^{-1}\cos\theta,\;\;\; e^3_{\psi}=\mu^{-1}, \n
&\omega_{12}=-\omega_{21}=-\frac{1}{2}\cos\theta d\phi+\frac{1}{2}d\psi,\;\;\;
\omega_{23}=-\omega_{32}=-\frac{1}{2}d\theta,\;\;\;
\omega_{31}=-\omega_{13}=-\frac{1}{2}\sin\theta d\phi. 
\end{align}
We use the following metric for $R\times S^2$:
\beqa
ds_{R\times S^2}^2=-dt^2+\frac{1}{\mu^2}(d\theta^2+\sin^2\theta d\phi^2).
\eeqa
Here the radius of $S^2$ is $\frac{1}{\mu}$.
The nonvanishing components of the dreibeins and the spin connections are
\beqa
b^1_{\theta}=\mu^{-1},\;\;\; b^2_{\phi}=\mu^{-1}\sin\theta,\;\;\;
k_{12}=-k_{21}=-\cos\theta d\phi.
\eeqa

It is convenient for the mode expansions to rewrite the actions 
in the $SU(4)$ symmetric form.  
The 10-dimensional Lorentz group has been decomposed as
$SO(9,1) \supset SO(3,1)\times SO(6)$. We identify $SO(6)$ with $SU(4)$.
We use $A,B=1,2,3,4$ as the indices of $\mbox{\boldmath $4$}$
in $SU(4)$ while we have used $m,n=4,\cdots,9$ as the indices of
$\mbox{\boldmath $6$}$ in $SO(6)$. The $SO(6)$ vector, $\mbox{\boldmath $6$}$,
corresponds to the antisymmetric tensor of $\mbox{\boldmath $4$}$ in $SU(4)$.
The $SO(6)$ and $SU(4)$ basis are related as
\beqa
&&X_{i4}=\frac{1}{2}(X_{i+3}+iX_{i+6}) \;\;\; (i=1,2,3), \n
&&X_{AB}=-X_{BA},\;\;\;X^{AB}=-X^{BA}=X_{AB}^{\dagger},\;\;\;
X^{AB}=\frac{1}{2}\epsilon^{ABCD}X_{CD}.
\eeqa
Similar identities hold for the gamma matrices:
\beqa
\Gamma^{i4}=\frac{1}{2}(\Gamma^{i+3}-i\Gamma^{i+6}), \;\;\;\mbox{etc.}
\eeqa
The 10-dimensional gamma matrices are decomposed as
\beqa
\Gamma^a=\gamma^a\otimes 1_8,\;\;\;
\Gamma^{AB}=\gamma_5\otimes \left( \begin{array}{cc}
                                   0          &  -\tilde{\rho}^{AB} \\
                                   \rho^{AB}  &  0
                                   \end{array}  \right)
=-\Gamma^{BA},
\eeqa
where $\gamma^a$ is the 4-dimensional gamma matrix, satisfying
$\{\gamma^a,\gamma^b\}=2\eta^{ab}$, and 
$\gamma_5=i\gamma^0\gamma^1\gamma^2\gamma^3$. $\Gamma^{AB}$ satisfies
$\{\Gamma^{AB},\Gamma^{CD}\}=\epsilon^{ABCD}$, and $\rho^{AB}$ and
$\tilde{\rho}^{AB}$ are defined by
\beqa
(\rho^{AB})_{CD}=\delta^A_C\delta^B_D-\delta^A_D\delta^B_C,\;\;\;
(\tilde{\rho}^{AB})^{CD}=\epsilon^{ABCD}.
\eeqa
The charge conjugation matrix and the chirality matrix are given by
\beqa
C_{10}=C_4 \otimes \left( \begin{array}{cc}
                          0   &  1_4 \\
                          1_4 &  0   
                          \end{array} \right), \;\;\;\;\;
\Gamma^{11}=\Gamma^0\cdots\Gamma^9=\gamma_5\otimes 
            \left( \begin{array}{cc}
                     1_4   &  0    \\
                     0     &  -1_4    
                   \end{array} \right),
\eeqa
where $(\Gamma^{a,m})^T=-C_{10}^{-1}\Gamma^{a,m}C_{10}$ and 
$C_4$ is the charge conjugation matrix in 4 dimensions.
The Majorana-Weyl spinor in 10 dimensions is decomposed as
\beqa
\lambda=\Gamma_{11}\lambda
=\left(\begin{array}{c} \lambda_+^A \\ \lambda_{-A} \end{array}\right),
\label{10to4}
\eeqa
where $\lambda_{-A}$ is the charge conjugation of $\lambda_+^A$:
\beqa
\lambda_{-A}=(\lambda_+^A)^c=C_4(\bar{\lambda}_{+A})^T,\;\;\;\;\;
\gamma_5\lambda_{\pm}=\pm\lambda_{\pm}.
\eeqa
We further fix the forms of 4-dimensional gamma matrices:
\beqa
\gamma^a=\left(\begin{array}{cc} 
               0                & i\sigma^a \\
               i\bar{\sigma}^a  & 0
               \end{array}\right),
\eeqa
where
$\sigma^0=-1_2$ and $\sigma^i\;\;\;(i=1,2,3)$ are the Pauli matrices.
$\bar{\sigma}^0=\sigma^0$ and $\bar{\sigma}^i=-\sigma^i$.
In this convention,
\beqa
\gamma_5=\left(\begin{array}{cc} 
               1_2 & 0 \\
               0   & -1_2
               \end{array}\right), \;\;\;\;\;
C_4=\left(\begin{array}{cc} 
          -\sigma^2 & 0 \\
          0         & \sigma^2
          \end{array}\right).
\eeqa
We introduce a two-component spinor:
\beqa
\lambda_+^A=\left(\begin{array}{c} \psi^A \\ 0 \end{array}\right). 
\label{4to2}
\eeqa
Using the $SU(4)$ symmetric notation, 
one can rewrite the actions (\ref{action of N=4 SYM on R times S^3}),
(\ref{2+1 SYM convenient form}) and (\ref{action of PWMM}) as follows:
\beqa
&&S_{R\times S^3}=\frac{1}{g_{R\times S^3}^2}\int dt\frac{d\Omega_3}{(\mu/2)^3}
\: \mbox{Tr}\left(
-\frac{1}{4}F_{ab}F^{ab}-\frac{1}{2}D_aX_{AB}D^aX^{AB}
-\frac{1}{2}X_{AB}X^{AB} \right. \n
&&\hspace{3.7cm}+i\psi_A^{\dagger}D_0\psi^A
+i\psi_A^{\dagger}\sigma^iD_i\psi^A 
+\psi_A^{\dagger}\sigma^2[X^{AB},(\psi_B^{\dagger})^T]
-\psi^{AT}\sigma^2[X_{AB},\psi^B] \n
&&\hspace{3.7cm}\left.
+\frac{1}{4}[X_{AB},X_{CD}][X^{AB},X^{CD}]\right), \n
\label{actiontwo-component}
\eeqa
\newpage
\begin{align}
 &S_{R\times S^2}
=\frac{1}{g_{R\times S^2}^2}\int dt\frac{d\Omega_2}{\mu^2} \mbox{Tr}\biggl(
\frac{1}{2}(D_0\vec{Y}-i\mu\vec{L}^{(0)}A_0)^2
 -\frac{1}{2}\vec{{\cal Z}}^2+\frac{1}{2}D_0X_{AB}D_0X^{AB}\nonumber \\
 &\hspace{2.5cm}
 +\frac{1}{2}\vec{{\cal L}}X_{AB}\cdot\vec{{\cal L}}X^{AB}
 -\frac{\mu^2}{8}X_{AB}X^{AB}+\frac{1}{4}[X_{AB},X_{CD}][X^{AB},X^{CD}]
  \nonumber \\
 &\hspace{2.5cm}
 +i\psi_{A}^\dagger D_0\psi^A
 -\psi_A^\dagger\vec{\sigma}\cdot \vec{\cal L}\psi^A
 -\frac{3\mu}{4}\psi_A^\dagger\psi^A
 +\psi_A^\dagger\sigma^2[X^{AB},(\psi_B^\dagger)^T]
 -\psi^{AT}\sigma^2[X_{AB},\psi^B]\biggr),
\label{S^2SU(4)form}
\end{align}

\begin{align}
 &S_{PW}=\frac{1}{g_{PW}^2}\int \frac{dt}{\mu^2}\:  \mbox{Tr}\biggl(
\frac{1}{2}(D_0Y_i)^2
 -\frac{1}{2}(\mu Y_i-\frac{i}{2}\epsilon_{ijk}[Y_j,Y_k])^2
 +\frac{1}{2}D_0X_{AB}D_0X^{AB}
 \nonumber \\
 &\hspace{2.5cm}
 -\frac{\mu^2}{8}X_{AB}X^{AB}
 +\frac{1}{2}[Y_i,X_{AB}][Y_i,X^{AB}]
 +\frac{1}{4}[X_{AB},X_{CD}][X^{AB},X^{CD}] \nonumber \\
 &\hspace{2.5cm}
 +i\psi_{A}^\dagger D_0\psi^A
-\frac{3\mu}{4}\psi_A^\dagger\psi^A
+\psi_A^\dagger\sigma^i [Y_i,\psi^A]
+\psi_A^\dagger\sigma^2[X^{AB},(\psi_B^\dagger)^T]
-\psi^{AT}\sigma^2[X_{AB},\psi^B]\biggr).
\label{PWMMSU(4)form}
\end{align}
\section{The plane wave matrix model}
\setcounter{equation}{0}
\renewcommand{\theequation}{B.\arabic{equation}}
In this appendix, we give the relationship between the action 
(\ref{action of PWMM}) 
and the conventional form 
of the action of the plane wave matrix model in the literature.
We introduce another representation of the 10-dimensional gamma matrices as
follows:
\beqa
\Gamma^0=1_{16}\otimes (-i)\sigma^2,\;\;\;
\Gamma^{\hat{M}}=\gamma^{\hat{M}}\otimes \sigma^3,
\eeqa
where $\gamma^{\hat{M}}$ is the $SO(9)$ gamma
matrix, which is a $16\times 16$ real symmetric matrix, and $\hat{M}=(i,m)$.
In this representation, the charge conjugation matrix is $C_{10}=\Gamma^0$,
and $\Gamma^{11}=1_{16}\otimes \sigma^1$.
Then the Majorana-Weyl spinor $\lambda$ is represented as 
\beqa
\lambda=\frac{1}{\sqrt{2}}\left( 
\begin{array}{c} \Psi \\ \Psi \end{array} \right),
\eeqa
where $\Psi$ is a real 16-components spinor.
We make a redefinition, $Y^i \: \rightarrow \: X^i$.
We also rescale the fields, the coupling constant and the time as follows:
\beqa
&&A_0\:\rightarrow \: -3\mu g A_0,\;\;\; 
X^{\hat{M}} \:\rightarrow \: -\mu g X^{\hat{M}}, \;\;\;
\Psi \:\rightarrow \: -\sqrt{3}\mu^{\frac{3}{2}}g \Psi,\;\;\; \n
&& g \:\rightarrow \:\sqrt{3\mu}g,\;\;\; t \:\rightarrow \: 3\mu t.
\eeqa
We finally obtain from (\ref{action of PWMM}) 
\beqa
&&S_{PWMM}=\int dt\: \Tr \left(
\frac{1}{2}D_0X^{\hat{M}}D_0X^{\hat{M}}-\frac{1}{18}X^iX^i
-\frac{1}{72}X^mX^m-\frac{ig}{18}\epsilon_{ijk}X^i[X^j,X^k] \right.\n
&&\qquad\qquad\qquad\qquad\!\left. 
+\frac{g^2}{36}[X^{\hat{M}},X^{\hat{N}}]^2 
+\frac{i}{2}\Psi^{\dagger}D_0\Psi
-\frac{i}{8}\Psi^{\dagger}\gamma^{123}\Psi
+\frac{g}{6}\Psi^{\dagger}\gamma^{\hat{M}}[X^{\hat{M}},\Psi] \right),
\eeqa
where $D_0=\partial_t+ig[A_0,\:]$. This is the conventional form
of the action of the plane wave matrix model seen in the literature.

\section{Supersymmetry transformations}
\setcounter{equation}{0}
\renewcommand{\theequation}{C.\arabic{equation}}

In this appendix, we give the supersymmetry 
transformation rules for the theories with $SU(2|4)$ symmetry.

First, the action of PWMM (\ref{action of PWMM}) is invariant under the 
following supersymmetry transformations:
\beqa
&&\delta A^0=-i\bar{\eta} \Gamma^0 \lambda,
\n
&&\delta \vec{Y}=-i\bar{\eta} \vec{\Gamma} \lambda,
\n
&&\delta X^m=-i\bar{\eta} \Gamma^m \lambda,
\n
&&\delta \lambda=D_0Y^i \Gamma^{0i}\eta
+D_0X^m \Gamma^{0m}\eta+\mu Y^i \Gamma^{i123}\eta
-\frac{\mu}{2}X^m \Gamma^{m123}\eta 
\n
&&\qquad\;
-\frac{i}{2}[Y^i,Y^j]\Gamma^{ij}\eta
-i[Y^i,X^m]\Gamma^{im}\eta
-\frac{i}{2}[X^m,X^n]\Gamma^{mn}\eta,
\label{susy transf for PWMM}
\eeqa
where the parameter $\eta$ is a 10-dimensional 
Majorana-Weyl spinor which satisfies 
$\partial_0\eta=-\frac{\mu}{4}\Gamma^{0123}\eta$.
Then, the theory has 16 supercharges.

Next, the action of $\mbox{SYM}_{R\times S^2}$ 
(\ref{2+1 SYM convenient form}) is 
invariant under the following transformations:
\beqa
&&\delta A^0=-i\bar{\eta} \Gamma^0 \lambda,
\n
&&\delta \vec{Y}=-i\bar{\eta} \vec{\Gamma} \lambda,
\n
&&\delta X^m=-i\bar{\eta} \Gamma^m \lambda,
\n
&&\delta \lambda=D_0Y^i \Gamma^{0i}\eta
+D_0X^m \Gamma^{0m}\eta
-\frac{\mu}{2}X^m \Gamma^{m123}\eta
+i{\cal L}_iX^m \Gamma^{im}
\n
&&\qquad\;
-\frac{i}{2}[X^m,X^n]\Gamma^{mn}\eta
+\frac{1}{2}\epsilon_{ijk}{\cal Z}_i\Gamma^{jk}\eta
-i\mu L^{(0)}_i A_0 \Gamma^{0i}\eta.
\label{susy transf for S^2}
\eeqa
Again, $\eta$ is a 10-dimensional 
Majorana-Weyl spinor which satisfies 
$\partial_0\eta=-\frac{\mu}{4}\Gamma^{0123}\eta$.
The theory also has 16 supercharges.

Finally, the transformation rule for the original
${\cal N}=4$ SYM on $R \times S^3$ (\ref{action of N=4 SYM on R times S^3}) 
is as follows:
\beqa
&&\delta A_a =i\bar{\lambda}\Gamma_a\epsilon,
\n
&&\delta X_m =i\bar{\lambda}\Gamma_m\epsilon,
\n
&&\delta \lambda =\left[\frac{1}{2}F_{ab}\Gamma^{ab}
+D_aX_m\Gamma^{am}-\frac{1}{2}X_m\Gamma^{ma}\nabla_a
-\frac{i}{2}[X_m,X_n]\Gamma^{mn} \right]\epsilon.
\label{susy transf for S^3}
\eeqa
In this case, the parameter $\epsilon$ is a conformal
Killing spinor on $R\times S^3$. In order to write down
the conformal Killing spinor equation, we decompose 
$\epsilon$ into the 4-dimensional Majorana-Weyl spinors as 
\beqa
\epsilon = \left(
\begin{array}{c}
\epsilon^A_+ \\
\epsilon_{-A}
\end{array}
\right),
\eeqa 
where $\epsilon^A_+$ and $\epsilon_{-A}$ are the 4-dimensional
Majorana-Weyl spinors, and $\epsilon_{-A}$ is the charge 
conjugation of $\epsilon^A_+$ (see Appendix A).
Then, the conformal Killing spinor equation on $R\times S^3$ is
written as
\beqa
\nabla_a \epsilon^A_+=\pm \frac{i}{2}\gamma_a\gamma^0
\epsilon^A_+,\hspace{3mm}\gamma_5\epsilon^A_+=\epsilon^A_+.
\label{conformal Killing equation}
\eeqa
A general solution of above equation has four real degrees
of freedom for each sign, and there are four SU(4) indices,
so that the original 10-dimensional parameter $\epsilon$ 
possess 32 real degrees of freedom.
In $\mbox{SYM}_{R\times S^3/Z_k}$, there remain only supersymmetries caused by
the conformal Killing spinors that satisfy the lower sign of 
(\ref{conformal Killing equation}), so that only 16 supercharges survive.

\section{Useful formulae for representations of $SU(2)$}
\setcounter{equation}{0}
\renewcommand{\theequation}{D.\arabic{equation}}

In this appendix, we gather some useful formulae concerning
the representations of $SU(2)$, most of which are found in \cite{vmk}.
The relationship between the Clebsch-Gordan coefficient and the $3-j$ symbol is
\beqa
\left( \begin{array}{ccc}
        J_1 & J_2 & J_3 \\
        m_1 & m_2 & m_3  
       \end{array}  \right)
=(-1)^{J_3+m_3+2J_1}\frac{1}{\sqrt{2J_3+1}}\:C^{J_3m_3}_{J_1\:-m_1\;J_2\:-m_2}.
\label{C-Gand3-j}
\eeqa
The Clebsch-Gordan coefficient possesses the following symmetries:
\newpage
\begin{align}
 \cg{J_3m_3}{J_1m_1}{J_2m_2}
  &=(-1)^{J_1+J_2-J_3}\cg{J_3m_3}{J_2m_2}{J_1m_1} \n
  &=(-1)^{J_1-m_1}\sqrt{\frac{2J_3+1}{2J_2+1}}\cg{J_2\:-m_2}{J_1m_1}{J_3\:-m_3}
  =(-1)^{J_1-m_1}\sqrt{\frac{2J_3+1}{2J_2+1}}\cg{J_2\:m_2}{J_3m_3}{J_1\:-m_1}
 \n
  &=(-1)^{J_2+m_2}\sqrt{\frac{2J_3+1}{2J_1+1}}\cg{J_1\:-m_1}{J_3\:-m_3}{J_2m_2}
  =(-1)^{J_2+m_2}\sqrt{\frac{2J_3+1}{2J_1+1}}\cg{J_1\:m_1}{J_2\:-m_2}{J_3m_3},
  \n[2mm]
  \cg{J_3m_3}{J_1m_1}{J_2m_2}
 &=(-1)^{J_1+J_2-J_3}\cg{J_3\:-m_3}{J_1\:-m_1}{J_2\:-m_2}.
\label{CG symmetry}
\end{align}
The recursion relation for the Clebsch-Gordan coefficient is 
\begin{align}
 \sqrt{(c\pm \gamma)(c\mp \gamma+1)}
  \cg{c\gamma\mp 1}{a\alpha}{b\beta}
  =
  \sqrt{(a\mp \alpha)(a\pm \alpha+1)}\cg{c\gamma}{a\alpha\pm 1}{b\beta}
  +\sqrt{(b\mp \beta)(b\pm \beta+1)}\cg{c\gamma}{a\alpha}{b\beta\pm 1}.
\label{relation between CG}
\end{align}
In sections 4, 
we frequently use summation formulae for the Clebsch-Gordan coefficient,
\begin{align}
 &\sum_{\alpha\beta}\cg{c\gamma}{a\alpha}{b\beta}
 \cg{c'\gamma'}{a\alpha}{b\beta}
  =\delta_{cc'}\delta_{\gamma\gamma'}, \\
  &\sum_{\alpha\beta\delta}
  \cg{c\gamma}{a\alpha}{b\beta}\cg{e\epsilon}{d\delta}{b\beta}
  \cg{d\delta}{a\alpha}{f\varphi}
  =(-1)^{b+c+d+f}\sqrt{(2c+1)(2d+1)}\cg{e\epsilon}{c\gamma}{f\varphi}
  \begin{Bmatrix}
   a & b & c \\
   e & f & d
  \end{Bmatrix}, \label{6j}\\
  &\sum_{\beta\gamma\epsilon\varphi}
  \cg{a\alpha}{b\beta}{c\gamma}\cg{d\delta}{e\epsilon}{f\varphi}
  \cg{b\beta}{e\epsilon}{g\eta}\cg{c\gamma}{f\varphi}{j\mu}
  =\sum_{k\kappa}\sqrt{(2b+1)(2c+1)(2d+1)(2k+1)}
  \cg{k\kappa}{g\eta}{j\mu}\cg{a\alpha}{d\delta}{k\kappa}
  \begin{Bmatrix}
   a & b & c \\
   d & e & f \\
   k & g & j
  \end{Bmatrix}.
  \label{9j}
\end{align}
In section 4, the following identity is often used:
\beqa
\langle Jm|e^{i\theta J_1} | Jn\rangle^*=(-1)^{-m+n}\langle J-\!m|
e^{i\theta J_1}|J-\!n\rangle.
\label{complex conjugate of matrix element}
\eeqa
In section 5, we use a formula for the asymptotic relations between the
$6-j$ symbols and the $3-j$ symbols. If $R\gg 1$, one obtains
\begin{align}
 \begin{Bmatrix}
  a & b & c \\
  d+R & e+R & f+R
 \end{Bmatrix}
 \approx \frac{(-1)^{a+b+c+2(d+e+f+R)}}{\sqrt{2R}}
 \begin{pmatrix}
  a & b & c \\
  e-f & f-d & d-e
 \end{pmatrix}.
\label{asymtotic relation}
\end{align}

\section{Vertex coefficients}
\setcounter{equation}{0}
\renewcommand{\theequation}{E.\arabic{equation}}
In this appendix, we give expressions for
the vertex coefficients we defined in section 4.
These expressions are obtained by using the formula 
(\ref{integralofthreeharmonics}). In the following,
$Q\equiv J+\frac{(1+\rho)\rho}{2}$, 
$\tilde{Q}\equiv J-\frac{(1-\rho)\rho}{2}$, $U\equiv J+\frac{1+\kappa}{4}$
and $\tilde{U}\equiv J+\frac{1-\kappa}{4}$. Suffices on these variables must
be understood appropriately.
\begin{align}
 &{\cal C}^{J_1m_1\tilde{m}_1}_{J_2m_2\tilde{m}_2\;J_3m_3\tilde{m}_3}
 =\sqrt{\frac{(2J_2+1)(2J_3+1)}{2J_1+1}}
 C^{J_1m_1}_{J_2m_2\;J_3m_3}
 C^{J_1\tilde{m}_1}_{J_2\tilde{m}_2\;J_3\tilde{m}_3}, \\
 &{\cal D}^{Jm\tilde{m}}_{J_1m_1\tilde{m}_1\rho_1\;J_2m_2\tilde{m}_2\rho_2}
 =(-1)^{\frac{\rho_1+\rho_2}{2}+1}
 \sqrt{3(2J_1+1)(2J_1+2\rho_1^2+1)(2J_2+1)(2J_2+2\rho_2^2+1)} \n
 &\qquad\qquad\qquad\qquad
 \times\left\{ \begin{array}{ccc}
	Q_1 & \tilde{Q}_1 &1 \\
                Q_2 & \tilde{Q}_2 &1 \\
                J   & J           &0
	       \end{array}   \right\}
 C^{Jm}_{Q_1m_1\; Q_2m_2}
 C^{J\tilde{m}}_{\tilde{Q}_1\tilde{m}_1\;\tilde{Q}_2\tilde{m}_2}, \\
 &{\cal E}
 _{J_1m_1\tilde{m}_1\rho_1\;J_2m_2\tilde{m}_2\rho_2\;J_3m_3\tilde{m}_3\rho_3} 
 \n
 &=
 \sqrt{6(2J_1+1)(2J_1+2\rho_1^2+1)(2J_2+1)(2J_2+2\rho_2^2+1)
 (2J_3+1)(2J_3+2\rho_3^2+1)} \n
 &\;\;\;\;\;\times (-1)^{-\frac{\rho_1+\rho_2+\rho_3+1}{2}}
 \left\{ \begin{array}{ccc}
  Q_1 & \tilde{Q}_1 &1 \\
	  Q_2 & \tilde{Q}_2 &1 \\
	  Q_3 & \tilde{Q}_3 &1        
	 \end{array}   \right\}
 \left( \begin{array}{ccc}
  Q_1 & Q_2 & Q_3 \\
	 m_1 & m_2 & m_3  
	\end{array}  \right)
 \left( \begin{array}{ccc}
  \tilde{Q}_1 & \tilde{Q}_2 & \tilde{Q}_3 \\
	 \tilde{m}_1 & \tilde{m}_2 & \tilde{m}_3  
	\end{array}  \right), \\
 &{\cal F}^{J_1m_1\tilde{m}_1\kappa_1}_{J_2m_2\tilde{m}_2\kappa_2\;Jm\tilde{m}}
 =\sqrt{2(2J+1)^2(2J_2+1)(2J_2+2)} 
 \left\{ \begin{array}{ccc}
  U_1 & \tilde{U}_1 &\frac{1}{2} \\
	  U_2 & \tilde{U}_2 &\frac{1}{2} \\
	  J   & J           &0        
	 \end{array}   \right\}
 C^{U_1m_1}_{U_2m_2\;Jm}
 C^{\tilde{U}_1\tilde{m}_1}_{\tilde{U}_2\tilde{m}_2\;J\tilde{m}}, \\
 &{\cal G}^{J_1m_1\tilde{m}_1\kappa_1}
 _{J_2m_2\tilde{m}_2\kappa_2\;Jm\tilde{m}\rho}
 =(-1)^{\frac{\rho}{2}}\sqrt{6(2J_2+1)(2J_2+2)(2J+1)(2J+2\rho^2+1)} \n
 &\qquad\qquad\qquad\;\;\;
 \times\left\{ \begin{array}{ccc}
	U_1 & \tilde{U}_1 &\frac{1}{2} \\
		U_2 & \tilde{U}_2 &\frac{1}{2} \\
		Q   & \tilde{Q}   &1        
	       \end{array}   \right\}
 C^{U_1m_1}_{U_2m_2\;Qm}
 C^{\tilde{U}_1\tilde{m}_1}_{\tilde{U}_2\tilde{m}_2\;\tilde{Q}\tilde{m}}.
 \label{vertexfunctions}
\end{align}
                   
 \section{Vertex coefficients of the fuzzy sphere harmonics}
\setcounter{equation}{0}
\renewcommand{\theequation}{F.\arabic{equation}}

In this appendix, we give expressions for the traces of various three
fuzzy sphere harmonics which are defined in section 4.3.
\begin{align}
 &\hat{{\cal C}}^{J_1m_1(j'j)}_{J_2m_2(j'j'')\;J_3m_3(j''j)} \n
 &=(-1)^{J_1+j+j'}\sqrt{N_0(2J_2+1)(2J_3+1)}
 C^{J_1m_1}_{J_2m_2\;J_3m_3}
 \begin{Bmatrix}
  J_1 & J_2 & J_3 \\
  j'' & j & j'
 \end{Bmatrix}, \\
 &\hat{{\cal D}}^{Jm(j'j)}
 _{J_1m_1(j'j'')\rho_1\;J_2m_2(j''j)\rho_2} \n
 &=
 \sqrt{3N_0(2J+1)(2J_1+1)(2J_1+2\rho_1^2+1)(2J_2+1)(2J_2+2\rho_2^2+1)} \n
 &\qquad
 \times
 (-1)^{\frac{\rho_1+\rho_2}{2}+1+J+j+j'}
 \left\{ \begin{array}{ccc}
  Q_1 & \tilde{Q}_1 &1 \\
	  Q_2 & \tilde{Q}_2 &1 \\
	  J   & J           &0
	       \end{array}   \right\}
 C^{Jm}_{Q_1m_1\; Q_2m_2}
 \begin{Bmatrix}
  J & \tilde{Q}_1 & \tilde{Q}_2 \\
  j'' & j & j'
 \end{Bmatrix}, \\
 &\hat{{\cal E}}
 _{J_1m_1(jj')\rho_1\;J_2m_2(j'j'')\rho_2\;J_3m_3(j''j)\rho_3} 
 \n
 &=
 \sqrt{6N_0(2J_1+1)(2J_1+2\rho_1^2+1)(2J_2+1)(2J_2+2\rho_2^2+1)
 (2J_3+1)(2J_3+2\rho_3^2+1)} \n
 &\;\;\;\;\;
 \times
 (-1)^{-\frac{\rho_1+\rho_2+\rho_3+1}{2}
 -\tilde{Q}_1-\tilde{Q}_2-\tilde{Q}_3+2j+2j'+2j''}
 \left\{ \begin{array}{ccc}
  Q_1 & \tilde{Q}_1 &1 \\
	  Q_2 & \tilde{Q}_2 &1 \\
	  Q_3 & \tilde{Q}_3 &1        
	 \end{array}   \right\}
 \begin{pmatrix}
  Q_1 & Q_2 & Q_3 \\
  m_1 & m_2 & m_3  
 \end{pmatrix}
 \begin{Bmatrix}
  \tilde{Q}_1 & \tilde{Q}_2 & \tilde{Q}_3 \\
  j'' & j & j'  
 \end{Bmatrix}, \\
 &\hat{{\cal F}}^{J_1m_1(j'j)\kappa_1}
 _{J_2m_2(j'j'')\kappa_2\;Jm(j''j)} \n
 &=\sqrt{2N_0(2\tilde{U}_1+1)(2J+1)^2(2J_2+1)(2J_2+2)} \n
 & \hspace{3cm}
 \times (-1)^{\tilde{U}_1+2J+j+j'}
 \left\{ \begin{array}{ccc}
  U_1 & \tilde{U}_1 &\frac{1}{2} \\
	  U_2 & \tilde{U}_2 &\frac{1}{2} \\
	  J   & J           &0        
	 \end{array}   \right\}
 C^{U_1m_1}_{U_2m_2\;Jm}
 \begin{Bmatrix}
  \tilde{U}_1 & \tilde{U}_2 & J \\
  j'' & j & j'
 \end{Bmatrix},
\end{align}
\begin{align}
 &\hat{{\cal G}}^{J_1m_1(j'j)\kappa_1}
 _{J_2m_2(j'j'')\kappa_2\;Jm(j''j)\rho} \n
 &=\sqrt{6N_0(2\tilde{U}_1+1)(2J_2+1)(2J_2+2)(2J+1)(2J+2\rho^2+1)} \n
 &\qquad\qquad\qquad\;\;\;
 \times 
 (-1)^{\frac{\rho}{2}+\tilde{U}_1+j+j'}
 \left\{ \begin{array}{ccc}
	U_1 & \tilde{U}_1 &\frac{1}{2} \\
                U_2 & \tilde{U}_2 &\frac{1}{2} \\
                Q   & \tilde{Q}   &1        
	       \end{array}   \right\}
 C^{U_1m_1}_{U_2m_2\;Qm}
 \begin{Bmatrix}
  \tilde{U}_1 & \tilde{U}_2 & \tilde{Q} \\
  j'' & j & j'
 \end{Bmatrix}.
\end{align}
 As mentioned in section 4.3, 
 In the $N_0\rightarrow \infty$, these reduce to the vertex
 coefficients in appendix E.

\section{Mode expansion of $\mbox{SYM}_{R\times S^3/Z_k}$}
\setcounter{equation}{0}
\renewcommand{\theequation}{G.\arabic{equation}}
In this appendix, we describe the mode expansion of the theory around 
the trivial vacuum of $\mbox{SYM}_{R\times S^3/Z_k}$, which was obtained in
our previous publication \cite{ITT}. 
The result is 
\begin{eqnarray}
&&S_{R\times S^3/Z_k}=S_{R\times S^3/Z_k}^{free}+S_{R\times S^3/Z_k}^{int}, \n
&&S_{R\times S^3/Z_k}^{free}=\frac{16\pi^2}{g_{R\times S^3/Z_k}^2k\mu^3}\int dt \mbox{Tr}\biggl\{
\sum_{Jm\tilde{m}}\frac{1}{2}
(\partial_0X_{Jm\tilde{m}}^{AB\dagger}\partial_0X_{Jm\tilde{m}}^{AB}
-\mu^2(J+\frac{1}{2})^2X_{Jm\tilde{m}}^{AB\dagger}X_{Jm\tilde{m}}^{AB})
\n
&&\hspace{1.8cm}
+\sum_{\rho=-1}^{1}\sum_{Jm\tilde{m}}\frac{1}{2}
(\partial_0A_{Jm\tilde{m}\rho}^{\dagger}\partial_0A_{Jm\tilde{m}\rho}
-\mu^2\rho^2(J+1)^2A_{Jm\tilde{m}\rho}^{\dagger}A_{Jm\tilde{m}\rho}) 
\n
&&\hspace{1.8cm}
 +\sum_{Jm\tilde{m}}\left(
 \frac{\mu^2}{2}J(J+1)B_{Jm\tilde{m}}^{\dagger}B_{Jm\tilde{m}}
 +i\mu\sqrt{J(J+1)}\partial_0A_{Jm\tilde{m}0}^{\dagger}B_{Jm\tilde{m}}\right)
\n
&&\hspace{1.8cm}
+\sum_{\kappa=\pm 1}\sum_{Jm\tilde{m}}
\left(i\Psi_{AJm\tilde{m}\kappa}^{\dagger}\partial_0\Psi_{Jm\tilde{m}\kappa}^A
+\kappa\mu(J+\frac{3}{4})
\Psi_{AJm\tilde{m}\kappa}^{\dagger}\Psi^A_{Jm\tilde{m}\kappa}\right)\biggr\}, 
\n
&&S_{R\times S^3/Z_k}^{int}
=\frac{16\pi^2}{g_{R\times S^3/Z_k}^2k\mu^3}\int dt\mbox{Tr}\biggl\{
-i{\cal C}_{Jm\tilde{m}\;J_1m_1\tilde{m}_1\;J_2m_2\tilde{m}_2}
\partial_0X_{AB}^{J_1m_1\tilde{m}_1}
[B_{Jm\tilde{m}},X^{AB}_{J_2m_2\tilde{m}_2}] \n
&&\qquad\; 
-\frac{1}{2}
{\cal C}^{Jm\tilde{m}}_{J_1m_1\tilde{m}_1\;J_2m_2\tilde{m}_2}
{\cal C}_{Jm\tilde{m}\;J_2m_3\tilde{m}_3\;J_4m_4\tilde{m}_4}
[B_{J_1m_1\tilde{m}_1},X_{AB}^{J_2m_2\tilde{m}_2}]
[B_{J_3m_3\tilde{m}_3},X^{AB}_{J_4m_4\tilde{m}_4}]) \n
&&\qquad\;
+\mu\sqrt{J_1(J_1+1)}
{\cal D}_{J_2m_2\tilde{m}_2\;J_1m_1\tilde{m}_10\;Jm\tilde{m}\rho}
X_{AB}^{J_1m_1\tilde{m}_1}[A_{Jm\tilde{m}\rho},X^{AB}_{J_2m_2\tilde{m}_2}] \n
&&\qquad\;
+\frac{1}{2}(-1)^{m-\tilde{m}+1}
{\cal D}_{J_1m_1\tilde{m}_1\;J_2m_2\tilde{m}_2\rho_2\;Jm\tilde{m}\rho}
{\cal D}_{J_3m_3\tilde{m}_3\;J_4m_4\tilde{m}_4\rho_4\;J-m-\tilde{m}\rho}
\n
&&\qquad\qquad\qquad\qquad\qquad\qquad 
\times[X_{AB}^{J_1m_1\tilde{m}_1},A_{J_2m_2\tilde{m}_2\rho_2}]
[X^{AB}_{J_3m_3\tilde{m}_3},A_{J_4m_4\tilde{m}_4\rho_4}] \n
&&\qquad\;
+\frac{1}{4}{\cal C}^{Jm\tilde{m}}_{J_1m_1\tilde{m}_1\;J_2m_2\tilde{m}_2}
{\cal C}_{Jm\tilde{m}\;J_3m_3\tilde{m}_3\;J_4m_4\tilde{m}_4}
[X_{AB}^{J_1m_1\tilde{m}_1},X_{CD}^{J_2m_2\tilde{m}_2}]
[X_{J_3m_3\tilde{m}_3}^{AB},X_{J_4m_4\tilde{m}_4}^{CD}] \n 
&&\qquad\;
-i{\cal D}_{Jm\tilde{m}\;J_1m_1\tilde{m}_1\rho_1\;J_2m_2\tilde{m}_2\rho_2}
\partial_0A_{J_1m_1\tilde{m}_1\rho_1}
[B_{Jm\tilde{m}},A_{J_2m_2\tilde{m}_2\rho_2}] \n
&&\qquad\;
-\mu\sqrt{J_1(J_1+1)}
{\cal D}_{J_2m_2\tilde{m}_2\;J_1m_1\tilde{m}_10\;Jm\tilde{m}\rho}
B_{J_1m_1\tilde{m}_1}[A_{Jm\tilde{m}\rho},B_{J_2m_2\tilde{m}_2}]  \n
&&\qquad\;
-\frac{1}{2}(-1)^{m-\tilde{m}+1}
{\cal D}_{J_1m_1\tilde{m}_1\;J_2m_2\tilde{m}_2\rho_2\;Jm\tilde{m}\rho}
{\cal D}_{J_3m_3\tilde{m}_3\;J_4m_4\tilde{m}_4\rho_4\;J-m-\tilde{m}\rho}
\n
&&\qquad\qquad\qquad\qquad\qquad\qquad 
\times [B_{J_1m_1\tilde{m}_1},A_{J_2m_2\tilde{m}_2\rho_2}]
[B_{J_3m_3\tilde{m}_3},A_{J_4m_4\tilde{m}_4\rho_4}] \n
&&\qquad\;
-i\frac{\mu}{2}\rho_1(J_1+1)
{\cal E}_{J_1m_1\tilde{m}_1\rho_1\;J_2m_2\tilde{m}_2\rho_2
\;J_3m_3\tilde{m}_3\rho_3}
A_{J_1m_1\tilde{m}_1\rho_1}
[A_{J_2m_2\tilde{m}_2\rho_2},A_{J_3m_3\tilde{m}_3\rho_3}] \n
&&\qquad\;
+\frac{1}{8}(-1)^{m-\tilde{m}+1}
{\cal E}_{J-m-\tilde{m}\rho\; J_1m_1\tilde{m}_1\rho_1\;J_2m_2\tilde{m}_2\rho_2}
{\cal E}_{Jm\tilde{m}\rho\;J_3m_3\tilde{m}_3\rho_3\;J_4m_4\tilde{m}_4\rho_4}
\n
&&\qquad\qquad\qquad\qquad\qquad\qquad 
\times [A_{J_1m_1\tilde{m}_1\rho_1},A_{J_2m_2\tilde{m}_2\rho_2}]
[A_{J_3m_3\tilde{m}_3\rho_3},A_{J_4m_4\tilde{m}_4\rho_4}] \n
&&\qquad\;
+{\cal F}^{J_1m_1\tilde{m}_1\kappa_1}_{J_2m_2\tilde{m}_2\kappa_2\;Jm\tilde{m}}
\Psi_{AJ_1m_1\tilde{m}_1\kappa_1}^{\dagger}
[B_{Jm\tilde{m}},\Psi_{J_2m_2\tilde{m}_2\kappa_2}^A] \n
&&\qquad\;
+{\cal G}^{J_1m_1\tilde{m}_1\kappa_1}
_{J_2m_2\tilde{m}_2\kappa_2\;Jm\tilde{m}\rho}
\Psi_{AJ_1m_1\tilde{m}_1\kappa_1}^{\dagger}
[A_{Jm\tilde{m}\rho},\Psi_{J_2m_2\tilde{m}_2\kappa_2}^A] \n
&&\qquad\;
-i(-1)^{m_2-\tilde{m}_2+\frac{\kappa_2}{2}}
{\cal F}^{J_1m_1\tilde{m}_1\kappa_1}_{J_2-m_2-\tilde{m}_2\kappa_2\;Jm\tilde{m}}
\Psi_{AJ_1m_1\tilde{m}_1\kappa_1}^{\dagger}
[X^{AB}_{Jm\tilde{m}},\Psi_{BJ_2m_2\tilde{m}_2\kappa_2}^{\dagger}] 
\n
&&\qquad\;
+i(-1)^{-m_1+\tilde{m}_1+\frac{\kappa_1}{2}}
{\cal F}^{J_1-m_1-\tilde{m}_1\kappa_1}_{J_2m_2\tilde{m}_2\kappa_2\;Jm\tilde{m}} 
\Psi_{J_1m_1\tilde{m}_1\kappa_1}^A 
[X_{AB}^{Jm\tilde{m}},\Psi_{J_2m_2\tilde{m}_2\kappa_2}^B]
\biggr\}, \label{mode expansion of S^3/Z_k}
\end{eqnarray}
where the summation over the indices that appear twice or more than twice
in $S_{R\times S^3/Z_k}^{int}$ is
assumed and $\tilde{m}$ only takes $\frac{k}{2}n\;\;(n\in{\bm Z})$. 
In comparison of (\ref{mode expansion of S^3/Z_k}) with 
(\ref{mode expansion of RxS^2}) in section 6.1, we use the identity 
\begin{align}
&\sum_{Jm\tilde{m}\rho}(-1)^{m-\tilde{m}+1}
{\cal D}_{J_1m_1\tilde{m}_1\; J_2m_2 \tilde{m}_2\rho_2\; Jm\tilde{m}\rho}
{\cal D}_{J_3m_3\tilde{m}_3\; J_4m_4 \tilde{m}_4\rho_4\; J-m-\tilde{m}\rho}\n
&=\sum_{Jm\tilde{m}\rho}(-1)^{m-\tilde{m}+1}
{\cal D}_{J_1m_1\tilde{m}_1\; J_4m_4 \tilde{m}_4\rho_4\; Jm\tilde{m}\rho}
{\cal D}_{J_3m_3\tilde{m}_3\; J_2m_2 \tilde{m}_2\rho_2\; J-m-\tilde{m}\rho}.
\end{align}

\end{document}